\documentclass[11pt]{article}
\global\arraycolsep=1pt
\usepackage{subfig}

\usepackage{amsmath,amssymb,graphicx}
\usepackage{hyperref}
\usepackage{multicol,color,longtable}
\definecolor{darkred}{rgb}{0.65,0.15,0}
\hypersetup{pdfborder={0 0 0},colorlinks=true,urlcolor=blue,citecolor=blue,linkcolor=darkred,linktocpage=true}

\usepackage[shadow,textwidth=2.7cm]{todonotes}
\usepackage{ifthen}
\reversemarginpar

\usepackage{cite}

\usepackage{tikz}
\usetikzlibrary{arrows}
\usepackage{pgfplots}

\usepackage{amsmath}
\usepackage{amsthm}
\usepackage{amssymb}
\usepackage{mathrsfs}
\usepackage[Symbol]{upgreek}
\usepackage{dsfont}
\usepackage[vcentermath]{youngtab}
\usepackage{textcomp}

\usepackage{latexsym}
\usepackage{graphicx}

\numberwithin{equation}{section}
\newcommand{\be}{\begin{equation}}
\newcommand{\ee}{\end{equation}}
\def\bea{\begin{eqnarray}}\def\eea{\end{eqnarray}}
\newcommand{\CR}{\nonumber \\*}

\newcommand{\y}{y}
\newcommand{\lP}{\ell_{\scalebox{0.6}{P}}}

\newcommand{\II}{I\hspace{-0.79mm}I}
\newcommand{\gra}[2]{{\scriptscriptstyle (#1 , #2 )}}
\newcommand{\ord}[1]{{\scriptscriptstyle (#1)}}
\def\Ep{{\rm Ep}}
\newcommand{\EpN}[1]{{\rm \phantom{\hat{E}}\hspace{-2.9mm} {E}p}^{\hspace{-1.1mm} #1}}
\def\Epreg{{\rm \widehat{E}p}}
\newcommand{\tr}{\, \mbox{tr}\, }
\newcommand{\dd}{{\rm d}}

\begin{document}

\begin{flushright} CPHT-RR053.082023 \end{flushright} 
 \vspace{8mm}

\begin{center}

{\LARGE \bf \sc Saturating unitarity bounds at U-duality symmetric points}

\vspace{6mm}
\normalsize
{\large  Guillaume Bossard and Adrien Loty}

\vspace{10mm}

{\it Centre de Physique Th\'eorique, CNRS,  Institut Polytechnique de Paris\\
91128 Palaiseau cedex, France \footnote{email: guillaume.bossard@polytechnique.edu,  adrien.loty@polytechnique.edu}}
\vskip 1 em
\vspace{20mm}

\hrule

\vspace{5mm}

 \begin{tabular}{p{14cm}}
It has recently been shown that the leading Wilson coefficient in type II string theory can take (almost) all values allowed by unitarity, crossing symmetry and maximal supersymmetry in $D=10$ and $D=9$ dimensions. This suggests that string theory might define the unique consistent quantum theory of gravity with maximal supersymmetry. We study  the minima of the leading  Wilson coefficient in $D=6$, $7$ and $8$ dimensions and find the global minimum at the point in moduli space with maximal symmetry. The minimum value turns out to always be negative for $D\le 7$. 
\end{tabular}

\vspace{5mm}
\hrule
\end{center}

\vspace{5mm} 

\thispagestyle{empty}

\newpage

\setcounter{page}{1}

\setcounter{tocdepth}{2}
\tableofcontents

\section{Introduction}
Superstring theory on asymptotically flat spacetimes defines a large number of scattering theories including gravitons. In the low energy limit, the scattering amplitudes behave as quantum field theory amplitudes. In spacetime dimensions larger than four, the corresponding S-matrices are well defined and must therefore satisfy the usual quantum field theory conditions of analyticity, crossing symmetry and unitarity. A string theorist may wish that all consistent quantum gravity S-matrices could be obtained in superstring theory. Assuming this is the case, one can in principle derive constraints on consistent effective field theories that would not be visible in perturbative quantum field theory. This is often used in the swampland conjectures, see \cite{Palti:2019pca} for a review. A more humble conjecture, and maybe more realistic, is that all  consistent quantum gravity theories with extended supersymmetry could be formulated as superstring theories. In this paper, we wish to study this question for maximally supersymmetric theories. 

Type II superstring theory on the Cartesian product of $D$-dimensional Minkowski spacetime and a compact torus admits the maximal number of supersymmetries. The low energy effective theory is then maximal supergravity in $D$ spacetime dimensions and the scattering amplitudes of massless states can in principle be described in supergravity using the Wilsonian effective action obtained by integrating out massive string states. In practice one first computes the perturbative string amplitudes and compare them with the supergravity amplitudes with operator insertions to deduce the Wilsonian effective action \cite{Gross:1986mw,Green:2008uj,Green:2008bf}.  The effective action is highly constrained from supersymmetry and U-duality \cite{Green:1981yb,Green:1997tv,Green:1997di,Berkovits:1997pj,Pioline:1998mn,Green:1998by,Obers:1999um,Green:1999pv,Kazhdan:2001nx,Basu:2008cf,Green:2005ba,Pioline:2010kb,Green:2011vz,Bossard:2014lra,Bossard:2014aea,Gustafsson:2014iva,Bossard:2015uga,Gourevitch:2019knu}. The leading Wilson coefficient is completely determined by supersymmetry, U-duality and anomaly cancelations, provided one requires consistency in the decompactification limits. 

\vskip 3mm

It is commonly believed  that type II superstring theory on a torus is the unique consistent quantum theory with maximal supersymmetry. One may therefore expect that the superstring S-matrix of massless states in a maximally supersymmetric vacuum covers all possible S-matrices satisfying analyticity, crossing symmetry, unitarity and maximal supersymmetry, as well as all the required anomaly cancelations. These consistency conditions can be analysed within the S-matrix bootstrap initiated in \cite{Paulos:2016but,Paulos:2017fhb,EliasMiro:2019kyf}. A lower bound on the leading Wilson coefficient was computed in \cite{Guerrieri:2021ivu,Guerrieri:2022sod} in maximal supergravity in $D=9,10,11$ dimensions, using the constraints from the two-to-two S-matrix. The unitarity bound was found to be close below the minimal value these Wilson coefficients can take in string theory and eleven-dimensional supergravity. The unitary bounds derived in  \cite{Guerrieri:2021ivu,Guerrieri:2022sod} are not sharp since they neglect non-elastic contributions to the optical theorem. Only integrability in two dimensions provides non-trivial examples of purely elastic S-matrices \cite{Coleman:1967ad,Aks:1965}, and taking particle production into account is expected to raise the bound on the leading Wilson coefficient \cite{Antunes:2023irg}. So one may conclude from \cite{Guerrieri:2021ivu,Guerrieri:2022sod} that string theory does indeed seem to saturate the sharp unitarity bound. 

The case of $D=11$ is particular because there is no moduli and the leading Wilson coefficient is a fixed number in M-theory. It is determined by the cancelation of the M5-brane anomaly \cite{Duff:1995wd,Witten:1996hc}. One may argue that this anomaly inflow argument is  independent of string theory \cite{Kim:2019vuc}, so that there should not be any consistent theory in eleven dimensions with a different value of the leading Wilson coefficient. Supersymmetry also fixes the next-to-leading Wilson coefficient and the first Wilson coefficient to be determined by unitarity and crossing symmetry multiplies $\nabla^8 R^4$.

In $D\le 10$ the leading Wilson coefficient is a function of the moduli and can take arbitrary large values, such that all the values consistent with the S-matrix unitarity bound seem to be covered by the string theory amplitude in $D=9$ and $D=10$ dimensions. 

The analysis of \cite{Guerrieri:2022sod} can in principle be generalised to all spacetime dimensions $5\le D\le 11$. However, one must modify the amplitude ansatz in dimension $D\le 8$ to include the contribution from the supergravity one-loop amplitude in the low energy limit. In $D=5$ one  would furthermore need to include the contribution from the two-loop amplitude. Although it is technically challenging to include the one-loop correction in the S-matrix bootstrap method, it is a priori doable, see for example \cite{Guerrieri:2020bto}. 

\vskip 3mm

On the string theory side one needs to find the minimum value of the leading Wilson coefficient. It is a maximal parabolic Eisenstein series of the U-duality group in $D$ dimensions \cite{Green:1997tv,Kiritsis:1997em,Obers:1999um,Green:2010wi,Green:2011vz}. In particular for the type IIB superstring amplitude in ten dimensions it is a real analytic Eisenstein series $2\zeta(3) E_{3/2}(S)$ on the upper complex half-plane \cite{Green:1997tv}.  The minimum is known to be at the $\mathds{Z}_3$-symmetric point  $S =  \frac{1+i \sqrt{3}}{2}$  \cite{MinimaSL2R}.  In dimension $D \le 8$  the leading Wilson coefficient is again a specific Eisenstein series of higher rank groups associated to their minimal automorphic representation \cite{Pioline:2010kb,Green:2011vz}. The $SL(3)$ Epstein series relevant for the leading Wilson coefficient in $D=8$ dimensions has been studied numerically in \cite{MinimaSL3}. The global minimum was found to be at the point in moduli space defined by the unimodular symmetric matrix proportional to the Gram matrix of the lattice $A_3$. Finding minima of Eisenstein series is generally an open problem, and is the subject of this paper. 

\vskip 3mm

We provide strong evidence that the global minimum of the $SL(N)$ Epstein series $\Ep_s^N(H)$ is obtained at the unimodular symmetric matrix $H=H_{\rm dlp}$ proportional to the Gram matrix of the densest lattice sphere packing  in $N$ dimensions for all $s>\frac{N}{4}$. It is proved for asymptotically large $s$ in \cite{MinimaLatticePacking}. We prove that the densest lattice sphere packing Gram matrix $H_{\rm dlp}$ is a local minimum of the Epstein series for all $s$ and $N\le 8$. We identify candidates for local minima as symmetric points and we checked that the lowest minimum is at $H_{\rm dlp}$ numerically. For $N=5$ we study the leading Wilson coefficient in $D=7$ dimensions on several surfaces containing $H_{\rm dlp}$ in moduli space and find each time that it is a global minimum on these surfaces. 

We generalise this analysis to the Spin$(5,5)$ Eisenstein series appearing as the leading Wilson coefficient in $D=6$ dimensions and find strong evidence that the global minimum is at the $W(D_5)\times W(D_5)$ symmetric point. This point is the analogue of the point of enhanced Spin$(10)$ symmetry in perturbative heterotic string theory on $T^5$.  

\vskip 3mm

One striking feature is that the leading Wilson coefficient can always be negative in dimension $D<8$. This is not in contradiction with unitarity because the leading Wilson coefficient is subleading with respect to the supergravity one-loop correction for $D\le 8$. One therefore expects the Wilson coefficient to possibly be negative, and comparable to the one-loop correction at Planck scale \cite{Caron-Huot:2022ugt}. 

The  Eisenstein series appearing in the Wilson coefficients are absolutely convergent when they are dominant compare to the loop corrections, and defined by analytic continuation when they are subleading \cite{Green:2010sp,Bossard:2015oxa}. It follows that the Wilson coefficient are necessary positive when they are dominant, and can always be possibly negative when they are subleading, consistently with unitarity. We will show indeed that the global minimum of an Eisenstein series is always negative in the critical strip, where it cannot be defined as an absolutely convergent sum. 

\vskip 3mm

The paper is organised as follows. In the second section we give some notations and summarise our results. We define a fundamental domain for the various moduli spaces of interest in Section 3. In Section 4 we give the main results leading to the conjectured minimum of Epstein series at symmetric points and in particular prove they are local minima. We discuss the specific case of dimension 8 in Section 5, where the splitting of the string amplitude into analytic and non-analytic pieces is ambiguous due to a logarithmic divergence. In Section 6 we expose numerical checks of our conjecture. 

\section{Notations and summary of the results}
The four-graviton superstring amplitude on $\mathds{R}^{1,9-d}\times T^d$ factorises in the form
\be \mathcal{M}_{4} = - i \frac{\kappa_{\scalebox{0.6}{$D$}}^2}{2^{10}} t_8 t_8 \prod_{a=1}^4 R(k_a,\epsilon_a) \mathcal{A}(s,t,u,\varphi)  \ee
where $\mathcal{A}(s,t,u,\varphi)$ is invariant under permutations of the three Mandelstam variables and is a function of the moduli $\varphi$ in  $K(E_{d{+}1}) \backslash E_{d{+}1(d{+}1)} / E_{d{+}1}(\mathds{Z})$ \cite{Hull:1994ys}. We define the Planck length in $D=10-d$ spacetime dimensions as
\be \kappa_{\scalebox{0.6}{$D$}}^2 = \frac12 (2\pi)^{7-d} \lP^{8-d} \; , \ee
and the $D$-dimensional effective string coupling $g_{\scalebox{0.6}{$D$}} = e^{\phi} /\sqrt{\upsilon_d}$ in terms of the ten-dimensional dilaton $\phi$ and the volume $V_d = (2\pi \sqrt{\alpha^\prime} )^d \upsilon_d $ of $T^d$, such that 
\be \alpha^\prime = g_{\scalebox{0.6}{$D$}}^{- \frac{4}{8-d}} \lP^2 \; . \ee
In the low energy limit, one can write 
\begin{multline} \mathcal{A}(s,t,u,\varphi)  = \frac{64}{s t u} + 32 (2\pi)^{7-d} \lP^{8-d} \bigl( I_4(s,t) + I_4(t,u) + I_4(u,s)\bigr) + \lP^6 \mathcal{E}_\gra{0}{0}(\varphi) \\
+ \frac{ \lP^{10}}{16}  \mathcal{E}_\gra{1}{0}(\varphi)(s^2+t^2+u^2)   + o( \lP^{15-2d})   \label{LowEnergyExpansion} \end{multline} 
so that the leading Wilson coefficient $\lP^6 \mathcal{E}_\gra{0}{0}(\varphi)$ is between the one-loop and the two-loop supergravity corrections for $D=6,7,8$. It is equal to the maximal parabolic Eisenstein series 
\be
 \mathcal{E}_\gra{0}{0}(\varphi) =4\pi\xi(d-2)E^{E_{d+1}}_{\frac{d-2}{2}\Lambda_{d{+}1}}
\ee
where $\Lambda_{d{+}1}$ is the fundamental weight associated to the electric charges representation  in $D$ dimensions and $\xi(s) = \pi^{-s/2} \Gamma(s/2) \zeta(s)$ is the completed zeta function. In $D\ge 7$ it can be written in terms of $SL(N)$ Epstein series  defined by analytic continuation of the  sum \footnote{In this paper we use $\Ep^N_s$ to distinguish the Epstein series normalisation used in the original papers \cite{Green:1997tv,Kiritsis:1997em,Obers:1999um}  from the Langlands Eisenstein series normalisation $E^{\scalebox{0.65}{$SL(N)$}}_{s\Lambda_1}$ more commonly used since \cite{Green:2010wi,Green:2011vz}.}
\be \Ep^{N}_s(H)=\sum_{n\in \mathds{Z}^N \smallsetminus \{0\} } \frac{1}{H[n]^s} =  2\zeta(2s)E^{\scalebox{0.65}{$SL(N)$}}_{s\Lambda_1}(H) \label{EpsteinN} \; . \ee
With this definition one has 
\be  \mathcal{E}_\gra{0}{0}(\varphi)  =  \EpN{5}_\frac32(H) \; , \ee
in seven dimensions \cite{Obers:1999um}. We conjecture that the global minimum is at the densest sphere lattice packing point $D_5$ 
\be   \EpN{5}_\frac32(H_{\rm dlp}) \approx -9.50663\; .  \ee

In eight dimensions one must take into account that the one-loop supergravity amplitude is logarithmically divergent, and so are the Epstein series appearing in $\mathcal{E}_\gra{0}{0}(\varphi)$. One must therefore introduce an appropriate renormalisation  
\be  \mathcal{E}_{\gra{0}{0},\mu }(\varphi)  = \Epreg^{\!\! 3}_\frac32(H) + 2 \Epreg^{\!\! 2}_1(U)  + \frac{22\pi}{3} - 4\pi \ln (2\pi \lP \mu) \; , \ee
where the renormalised Epstein series $ \Epreg^{\!\! 3}_\frac32$ and $\Epreg^{\!\! 2}_1$ are defined as in \cite{Green:2010wi}. The renormalisation scale $\mu$ cancels in the complete amplitude, and the specific finite number is determined by our definition of the 1-loop box integral, as we explain in Section \ref{Renormalisation}. Choosing $\mu = 1/ \lP$, we find the  minimum value
\be  \mathcal{E}_{\gra{0}{0},\frac{1}{\ell_{\scalebox{0.4}{P}}}}(\varphi_{\rm dlp}) \approx 15.2363\; .   \ee
In eight dimensions it also makes sense to include the next-to-leading Wilson coefficient
\be   \mathcal{E}_\gra{1}{0}(\varphi)  = \frac12 \EpN{3}_\frac52(H) -4  \EpN{3}_{-\frac12}(H)\EpN{2}_2(U)\; , \ee
while neglecting the two-loop supergravity integral. One computes the value at the minimum 
\be \mathcal{E}_\gra{1}{0}(\varphi_{\rm dlp}) \approx 10.7196 \; . \ee
 In six dimensions the leading Wilson coefficient is written similarly as the vector representation Eisenstein series of $SO(5,5)$ \cite{Obers:1999um}
 \be   \mathcal{E}_\gra{0}{0}(\varphi)  =  \EpN{5,5}_\frac32(H) \; . \ee
We conjecture the global minimum to be at the point of enhancement Spin$(10)$  symmetry 
\be  \EpN{5,5}_\frac32(H_{D_5}) \approx - 3.445\; .  \ee

\section{Fundamental domain of $K \backslash G / G(\mathds{Z})$}
In order to determine the minium of an automorphic function on $K \backslash G / G(\mathds{Z})$, it is useful to find an appropriate fundamental domain $\mathcal{F}$ for the action of the arithmetic  subgroup $G(\mathds{Z})$ on the symmetric space $\mathcal{M} = K \backslash G$.  Here we assume that $G$ is a simple group of real split form  different from $E_8$, $F_4$ or $G_2$, and $G(\mathds{Z})$ is its Chevalley subgroup associated to the weight lattice. This includes in particular the locally symmetric spaces relevant for the type II string theory effective action in dimensions greater than four. 

\vskip 3mm

Let us first recall the definition of a fundamental domain. One defines a free regular set $\mathring{\mathcal{F}} \subset \mathcal{M}$ as an open set in $\mathcal{M}$ such that any point in $\mathcal{M}$ can be mapped under the action of $G(\mathds{Z})$ to its closure $\mathcal{F}$ in $\mathcal{M}$
\be \mathcal{F} = \overline{\mathring{\mathcal{F}}} \subset \mathcal{M} \; , \ee
and for any element $\gamma\in G(\mathds{Z})$ acting non-trivially on $\mathcal{M}$ one has 
\be \gamma \mathcal{F}  \cap \mathring{\mathcal{F}}  = \emptyset \; . \ee
It is required that $\partial \mathcal{F} = \mathcal{F} \smallsetminus \mathring{\mathcal{F}}$ is of measure zero in $\mathcal{M}$. 
We then call $\mathcal{F}$ a fundamental domain of  $G(\mathds{Z})$ in $\mathcal{M}$. Equivalently one can define $\mathcal{F}$ to satisfy 
\be \bigcup_{\gamma \in G(\mathds{Z})} \gamma \mathcal{F} = \mathcal{M}\; , \qquad  \gamma \mathcal{F}  \cap \mathcal{F}  \subset \partial   \mathcal{F} \; . \ee

There is generally no canonical fundamental domain. A first fundamental domain was introduced by Minkowski for $G(\mathds{Z}) = GL(N,\mathds{Z})$ acting on $SO(N) \backslash SL(N) $ \cite{zbMATH02628371}. Grenier then defined a different fundamental domain for $GL(N,\mathds{Z})$ that is easier to generalise to arbitrary simple groups \cite{zbMATH04146017}.  In this section we shall first review Grenier's construction in the case of $SL(N,\mathds{Z})$. Then we will show that there is a natural generalisation of Grenier's fundamental domain for all simple groups of split real form but $E_8$, $F_4$ or $G_2$. We will finally discuss $SO(N,N,\mathds{Z})$ in more detail as an example.

\subsection{Fundamental domain of $SL(N,\mathds{Z})$}
\label{GrenierDomain}

Following Grenier, we parametrise the symmetric space $\mathcal{M} =SO(N) \backslash  SL(N)$ in the Iwasawa decomposition with $N-1$ variables $y_i \in \mathds{R}_+$ with $1\le i\le N-1$ and $\frac{N(N-1)}{2}$ variables $x_{ij} \in \mathds{R}$ with $1\le i<j\le N$.  The group representative in $ SL(N)$ can then be written as the upper triangular matrix 
\be
\mathcal{V}=\frac{1}{y}\begin{pmatrix}1&&&&\\&y_1&&&\\&&y_1y_2&&\\&&&\ddots&\\&&&&y_1y_2\dots y_{N-1}\end{pmatrix}\begin{pmatrix}1&&x_{ij}\\&\ddots&\\&&1\end{pmatrix}
 \label{Vyx}\; ,  \ee
with
\be
y=\left(\prod_{i=1}^{N-1}y_i^{N-i}\right)^\frac{1}{N}\; . 
\ee
For short we do not distinguish the group representative from the corresponding point in $\mathcal{M}$. To any element $\mathcal{V}\in \mathcal{M}$ we can associate a positive definite symmetric bilinear form over $\mathds{Z}^N$
\be  H[n] = n^{\intercal}\mathcal{V}^{\intercal}\mathcal{V}n \; . \ee
The Grenier fundamental domain is defined by induction by writing first  $\mathcal{V}$ in the maximal parabolic subgroup $(GL(1)\times SL(N-1))\ltimes\mathds{R}^{N-1}\subset SL(N)$ as 
\be
\mathcal{V}=\frac{1}{y}\begin{pmatrix}1& 0 \\ 0 &y^\frac{N}{N-1}\mathcal{V}_{1}\end{pmatrix}\begin{pmatrix}1&\; x^{\intercal}\\  0&\; \mathds{1}\end{pmatrix}
\ee
where $x^{\intercal}=(x_{12},\dots,x_{1N})$ and $\mathcal{V}_{1}\in SL(N-1)$, and then recursively for each $\mathcal{V}_i \in SL(N-i)$. To avoid taking care of the factors of $y$, Grenier introduces the non-unimodular symmetric matrix $Y=y^2\mathcal{V}^{\intercal}\mathcal{V}$
\be Y[n]\equiv n^\intercal Y n =\left(n_1+\sum_{k=2}^Nx_{1k}n_k\right)^2+y_1^2Y_1[ n]
\ee
where $Y_1 = y_1^{-2} y^{\frac{2N}{N-1}} \mathcal{V}_{1}^\intercal \mathcal{V}_1 $ is a symmetric bilinear form over $\mathds{Z}^{N-1}$, so its argument $n$ is implicitly $(n_2,\dots,n_N)$. One defines recursively the bilinear forms $Y_i[n] \equiv (n_{i+1}, \dots n_n)^\intercal Y_i (n_{i+1}, \dots n_n)$ for $1\leq k\leq N$  as
\be
Y_i[n]=\biggl(n_{i+1}+\sum_{k=i+2}^Nx_{i+1,k}n_k\biggr)^2+y_{i+1}^2Y_{i+1}[ n]\; , 
\ee
with 
\be
Y_{N-1}[n]=n_N^2\; , 
\ee
and by convention $Y_0[n]=Y[n]$. The condition for $\mathcal{V}$ to be in the Grenier fundamental domain $\mathcal{F}$ of $\mathcal{M}$ is defined recursively by the two conditions:

\begin{enumerate}
    \item $Y_i[n]\geq 1$, for all non-vanishing $n\in\mathds{Z}^{N-i}$ of greatest common divisor gcd$(n)=1$.
    \item $x_{i,i+j}\in\left[-\frac{1}{2},\frac{1}{2}\right]$ for all  $i\in\{1,\dots,N-1\}$ and $j\in\{2,\dots,N-i\}$, $x_{i,i+1}\in\left[0,\frac{1}{2}\right]$ for all  $i\in\{2,\dots,N-1\}$, whereas $x_{1,2} \in \left[0,\frac{1}{2}\right]$ if $N$ is odd and $x_{1,2} \in \left[- \frac12,\frac{1}{2}\right]$ if $N$ is even.
\end{enumerate}
Grenier proved that one can always restrict the first condition to a finite set of vectors such that this gives a finite number of inequalities. This follows from the property that $Y_i[n]$ is positive definite and there is a finite number of lattice points in the ball $Y_i[n]\leq1$. 

\subsubsection*{Fundamental domain of $SL(2,\mathds{Z})$}

For the case $N=2$ the Grenier fundamental domain agrees with the standard $SL(2,\mathds{Z})$ fundamental domain. The only vector for which the first condition is non-trivial is $n=(0,1)^{\intercal}$ and together with the second condition one gets \\

i) $x_{12}^2+y_1^2\geq1$\; , \qquad ii) $-\frac{1}{2}\leq x_{12}\leq\frac{1}{2}$\; .

\subsubsection*{Fundamental domain of $SL(3,\mathds{Z})$}

Another case of interest for us is $N=3$. The fundamental domain of $SO(3)\backslash SL(3)$ is given by\\

i) $x_{23}^2+y_2^2\geq1$\\

ii) $x_{12}^2+y_1^2\geq1$\\

iii) $x_{13}^2+y_1^2(x_{23}^2+y_2^2)\geq1$\\

iv) $(x_{12}-x_{13})^2+y_1^2((1-x_{23})^2+y_2^2)\geq1$\\

v) $(1-x_{12}+x_{13})^2+y_1^2((1-x_{23})^2+y_2^2)\geq1$\\

vi) $0\leq x_{12}\leq\frac{1}{2}$\\

vii) $0 \leq x_{23}\leq\frac{1}{2}$\\

viii) $-\frac{1}{2}\leq x_{13}\leq\frac{1}{2}$\\

The condition i) comes from the condition 1) for $Y_1[n]$ applied to $n=(0,1)^{\intercal}$ while the conditions ii)-v) come from the condition 1) for $Y_0[n]$ applied to $n=(0,1,0)^{\intercal},\,n=(0,0,1)^{\intercal},\,n=(0,-1,1)^{\intercal},\,n=(1,-1,1)^{\intercal}$ respectively. Again vi)-viii) are a direct consequence of condition 2).

\subsection{Generalisation to $G(\mathds{Z})$}

It appears that Grenier's construction of a fundamental domain is based on a sequence of maximal abelian parabolic subgroups. In this section we therefore consider $G$ to be the  split real form of a simple group of rank $r$ admitting an abelian parabolic subgroup, which excludes $E_8$, $F_4$ and $G_2$. We shall discuss the explicit example of $SO(N,N)$ in the next section. One can probably generalise this construction to Heisenberg parabolic subgroups to encompass all exceptional groups, but this requires further analysis and will not be relevant for us. In what follows the group $G$ is defined from its fundamental representations and $G(\mathds{Z})\subset G$ is the Chevalley subgroup associated to the weight lattice. 

\vskip 5mm

Let us first describe an appropriate coordinate system on $\mathcal{M} = K \backslash G$. Let $P_1$ be a maximal abelian parabolic subgroup 
\be
P_1=(GL(1)\times G_1)\ltimes U_1 \subset G \; , 
\ee
where $G_1$ is itself the split real form of a semi-simple group and $U_1$ is an abelian unipotent subgroup. One calls respectively $GL(1)\times G_1$ the Levi subgroup and $U_1$ the unipotent radical of the parabolic subgroup $P_1$. In the last section, for $G=SL(N)$, we had $G_1=SL(N-1)$ and $U_1=\mathds{R}^{N-1}$ the additive group for example.

We can then apply the same decomposition to $G_1$ and all $G_i$ semi-simple subgroups successively 
\be
G_i\supset P_{i+1}=(GL(1)\times G_{i+1})\ltimes U_{i+1}\; , 
\ee
with $U_{i}$ an abelian unipotent group for all $i = 1$ to $r$. By construction $P_r=GL(1)\ltimes U_{r}$ and the Borel subgroup can be defined as 
\be B = GL(1)^r \ltimes U_r \ltimes U_{r-1} \dots \ltimes U_1 \; , \ee
and the Iwasawa decomposition of $G$  is  compatible with this succession of abelian parabolic subgroups
\be B \subset GL(1)^{r-2} \times P_{r-1} \ltimes U_{r-2} \dots \ltimes U_1 \subset \dots\subset GL(1)\times P_2 \ltimes U_1 \subset  P_1 \subset G \; . \label{ParaSucc} \ee
The Iwasawa decompositions provides coordinates on $\mathcal{M}$, with $y_i\in \mathds{R}_+$ for $i=1$ to $r$, parametrising each $GL(1)$ factors of the Cartan torus and a vector ${\bf x}_i\in \mathfrak{u}_i$ the Lie algebra of $U_i$ for  $i=1$ to $r$ parametrising each $U_i$ unipotent group. By construction the Borel subgroup $B_i\subset G_i$ can be decomposed accordingly into 
\be B_i = GL(1)^{r-i} \ltimes U_r \ltimes U_{r-1} \dots \ltimes U_{i+1} \; . \ee
Therefore $\{ y_j , {\bf x}_j\}$ for $i+1\le j\le r$ provide coordinates on the symmetric space  $\mathcal{M}_i = K_i \backslash G_i$, where $K_i$ is the maximal compact subgroup of $G_i$. 

By definition each maximal parabolic subgroup $P_i\subset G_{i-1}$ is determined by a fundamental weight $\Lambda^{\ord{i}}$ of the subgroup $G_{i-1}$. The associated set of roots of $G$ is called an abelian enumeration in \cite{Gourevitch:2019knu}. We write $R^{\ord{i}}=R(\Lambda^{\ord{i}},\mathds{R})$ the associated highest weight representation, which highest weight vector ${\bf e}^{\ord{i}} \in R^{\ord{i}}$ admits as stabiliser  the subgroup $G_i \ltimes U_i \subset G_{i-1}$. The $G_{i-1}$-orbit of ${\bf e}^\ord{i}$ is therefore the coset space 
\be G_{i-1} \cdot {\bf e}^\ord{i} = G_{i-1} /( G_i \ltimes U_i ) \; . \label{CosetSmall} \ee
For example in the case of $SL(N)$, the maximal parabolic subgroup $(GL(1)\times SL(N-1))\ltimes\mathds{R}^{N-1}$ is associated to the fundamental weight $\Lambda_1$ of $SL(N)$ and the corresponding highest weight representation is the fundamental representation $\mathds{R}^N$. The orbit of the highest weight vector is then dense in the  module 
\be \mathds{R}^{N} \smallsetminus \{ 0\} = SL(N) / ( SL(N-1) \ltimes \mathds{R}^{N-1})\; .  \ee
By convention we will call $R^\ord{i}$ the fundamental representation. In general the orbit of the highest weight vector is not necessarily dense in the module $R^\ord{i}$, but is defined instead as the set of non-zero elements  ${\bf v}^{\ord{i}}$ satisfying the quadratic constraint \cite{Peterson:1983}
\be
{\bf v}^{\ord{i}} \times {\bf v}^{\ord{i}} \equiv \kappa_{\alpha\beta}T^\alpha {\bf v}^{\ord{i}}\otimes T^\beta {\bf v}^{\ord{i}} -\bigl( \Lambda^{\ord{i}},\Lambda^{\ord{i}} \bigr) {\bf v}^{\ord{i}}\otimes {\bf v}^{\ord{i}} = 0 \; , \label{xXx}
\ee
where $\kappa_{\alpha\beta}$ denotes the Killing--Cartan form of $\mathfrak{g}_{i-1}$ and $T^\alpha$ its representation matrices in $R^{\ord{i}}$. One can then write \eqref{CosetSmall} as
\be G_{i-1} /( G_i \ltimes U_i )  =  \bigl\{ {\bf v}^{\ord{i}} \in R^\ord{i}  \, |\, {\bf v}^{\ord{i}} \times {\bf v}^{\ord{i}}= 0\, , \; L({\bf v}^{\ord{i}}) > 0  \bigr\} \; , \label{CosetSmallKac} \ee
where some further positivity condition $L({\bf v}^{\ord{i}}) > 0 $ may be necessary for specific groups $G_{i-1} $ and representations $R^\ord{i} $. For example for $SO(N,N)$ one can take the fundamental representation as the vector representation, and ${\bf v}^{\ord{i}}\times {\bf v}^{\ord{i}} = ({\bf v}^{\ord{i}},{\bf v}^{\ord{i}})=0$ is the condition that ${\bf v}^{\ord{i}}$ is light-like. For the connected component $SO_{\scalebox{0.6}{$0$}}(1,1)$ one would moreover demand that the non-zero lightcone coordinate is positive. 

For each point in $\mathcal{M}_i$ we define the associated matrix  $\mathcal{V}_i\in  P_{i+1}$ in the representation $R^{\ord{i+1}}$ of $G_i$.  For short we will write $\mathcal{V}_i= \rho_{\scalebox{0.6}{$\Lambda^{(i+1)}$}}[ \mathcal{V}_i]$ for  both the point $\mathcal{V}_i\in  \mathcal{M}_i$ and the matrix $\rho_{\scalebox{0.6}{$\Lambda^{(i+1)}$}}[ \mathcal{V}_i]$ in the representation $R^{\ord{i+1}}$. We can then define the symmetric bilinear form $H_i=\mathcal{V}_i^{\intercal}\mathcal{V}_i$ in the representation $R^{\ord{i+1}}$
\be H_i[{\bf v}^{\ord{i+1}}] \equiv  {\bf v}^{\ord{i+1} \intercal }H_i{\bf v}^{\ord{i+1}}\; . \label{Hix} \ee
The matrix $\mathcal{V}_i$ can be written as
\be
\mathcal{V}_i=\rho_{\scalebox{0.6}{$\Lambda^{(i+1)}$}} \bigl[ e^{-\ln\y_{i+1}{\bf h}^{(i+1)}}\mathcal{V}_{i+1}(\y_{j>i+1},{\bf  x}_{j>i+1})e^{{\bf x}_{i+1}} \bigr] \; ,
\ee
where $\rho_{\scalebox{0.6}{$\Lambda^{(i+1)}$}} $ is the representation homomorphism and ${\bf h}^{(i+1)}$ the Cartan generator of $G_i$ associated to the weight $\Lambda^\ord{i+1}$ with the normalisation\footnote{${\bf h}^{\ord{i}}$ are related to the Cartan--Weyl basis generators of $G_{i-1}$ by ${\bf h}^{\ord{i}}
=\frac{1}{( \Lambda^{\ord{i}},  \Lambda^{\ord{i}})}{\bf h}_{\Lambda^{\ord{i}}}$ and ${\bf x}_{i}$ decomposes in the basis of root generators ${\bf e}_\alpha$ satisfying $(\Lambda^{\ord{i}}, \alpha)=1$.}
\be \rho_{\scalebox{0.6}{$\Lambda^{(i+1)}$}}[ {\bf h}^{(i+1)}] {\bf e}^{\ord{i+1}}  = {\bf e}^{\ord{i+1}}  \; , \ee
such that
\be
\mathcal{V}_{i} {\bf e}^\ord{i+1} =\frac{1}{y_{i+1}} {\bf e}^\ord{i+1} \; . 
\ee
Note that the coordinates $y_i$ defined above are not the same as the coordinates introduced in \eqref{Vyx}.\footnote{Writing $\tilde y_i$ the Grenier coordinates in \eqref{Vyx}, the coordinates $y_i$ used in this section are $y_i=\left(\prod_{j=i}^{N-1}\tilde y_j^{N-j}\right)^\frac{1}{N+1-i}$.}

In the following we shall use that a point in $\mathcal{M}_i$ can be parametrised equivalently by all $\{ y_j,{\bf x}_j\}$ for $j\ge i+1$ or by the symmetric matrix $H_i$. One can also equivalently use the coordinates $\{ y_{i+1},{\bf x}_{i+1}\} $ and the symmetric matrix $H_{i+1}$, etc...

\vskip 5mm

The idea behind Grenier's construction of a fundamental domain is that one can construct recursively the fundamental domains $\mathcal{F}_{i}$ for $\mathcal{M}_{i}$ under the action of $G_{i}(\mathds{Z})$ from $i=r-1$ to $0$. At each step, one can use the bilinear form \eqref{Hix} to define a set of inequalities  that determines $\mathcal{F}_{i}$. 

Before generalising Grenier's construction, let us give some definitions. We define for $1\le i \le r-1$
\be P_i(\mathds{Z}) = G(\mathds{Z}) \cap P_i = G_i(\mathds{Z}) \ltimes U_i(\mathds{Z}) \; , \quad G_i(\mathds{Z}) =  G(\mathds{Z}) \cap  GL(1) \times G_i\; , \quad U_i(\mathds{Z}) = G(\mathds{Z}) \cap U_i \; .  \ee
Note that with this definition $G_i(\mathds{Z})$ may not be a subgroup of $G_i$ but includes the entire discrete Levi subgroup. For $SL(N,\mathds{Z})$ we would then define $G_i(\mathds{Z}) = GL(N-i,\mathds{Z})$ for example. 

The construction of the fundamental domain is defined by induction. In order to show that $\mathcal{F}_{i}$ is a fundamental domain of $\mathcal{M}_{i}$ under the action of $G_{i}(\mathds{Z})$ we have to show that for any element $H_i\in\mathcal{M}_i$ there exists an element $\gamma\in G_{i}(\mathds{Z})$ such that $\gamma^{\intercal}H_i\gamma\in \mathcal{F}_i$ and conversely that for any non-trivial element $\gamma\in G_{i}(\mathds{Z})$ and any element $H_i\in\mathcal{F}_i$, either $\gamma^{\intercal}H_i\gamma\notin\mathcal{F}_i$ or both $H_i$ and $\gamma^{\intercal}H_i\gamma$ are on the boundary of $\mathcal{F}_i$. 

 We first consider the fundamental domain in $\mathcal{M}_i$ for the action of the parabolic  subgroup $P_{i+1}(\mathds{Z})$.  Let $p$ be an element of $P_{i+1}(\mathds{Z})=  G_{i+1}(\mathds{Z})\ltimes U_{i+1}(\mathds{Z})$, we can decompose it as
\be
p=l\,\exp({\bf b})
\ee
with $l\in G_{i+1}(\mathds{Z})$ and $\exp({\bf b}) \in U_{i+1}(\mathds{Z})$. $P_{i+1}(\mathds{Z})$ acts on $\mathcal{M}_i$ by
\be
p^{\intercal}H_ip=(y_{i+1},l^{\intercal}H_{i+1}l,l^{-1}{\bf x}_{i+1}+{\bf b})\; .
\ee
The adjoint Levi subgroup $G_{i+1}(\mathds{Z}) / Z(G_{i+1}(\mathds{Z}))$, with $Z(G_{i+1}(\mathds{Z}))$ the centre of $G_{i+1}(\mathds{Z})$, acts freely on a dense open set in $\mathcal{M}_{i+1}$, and determines the fundamental domain 
 \be \mathcal{F}_{i+1}= \mathcal{M}_{i+1}/ (G_{i+1}(\mathds{Z}) / Z(G_{i+1}(\mathds{Z}))) \; . \ee
To find a fundamental domain of  $P_{i+1}(\mathds{Z})$  in $\mathcal{M}_i$, it remains to act with $Z(G_{i+1}(\mathds{Z}))\ltimes {U}_{i+1}(\mathds{Z})$ on $\mathfrak{u}_{i+1}$. By construction ${U}_{i+1}(\mathds{Z})$ acts by translation and $Z(G_{i+1}(\mathds{Z}))$ acts either trivially or by multiplying ${\bf x}^\ord{i+1}\in \mathfrak{u}_{i+1}$ by $-1$. 
One gets therefore  
\be  \mathcal{F}^{\scalebox{0.8}{P}}_i =  \mathcal{M}_i / P_{i+1}(\mathds{Z}) = \mathds{R}_+\times \mathcal{F}_{i+1}\times \left[-\tfrac{1}{2},\tfrac{1}{2}\right]^{d_{i+1}} \hspace{-1.5mm} /\mathds{Z}_\mu \; , 
\ee
where $d_{i+1}$ is the dimension of $U_{i+1}$ and $\mu=2$ if $Z(G_{i+1}(\mathds{Z}))$ acts as  $\mathds{Z}_2$ on $\mathfrak{u}_{i+1}$ and $\mu=1$ if it acts trivially.  

To define the fundamental domain of $G_{i}(\mathds{Z})$, we introduce the Chevalley lattice $R^{\ord{i+1}}(\mathds{Z})\subset R^\ord{i+1}$, which is preserved by the action of $G_{i}(\mathds{Z})$. 

The equivalent of \eqref{CosetSmallKac} for the discrete subgroup $G_{i}(\mathds{Z})$ gives 
\be
G_{i}(\mathds{Z})/P_{i+1}(\mathds{Z})={S}_{i}=\{n\in R^{\ord{i}}(\mathds{Z})\ \lvert\ n\times n=0, \; L(n) > 0   ,\, \text{gcd}\,n =1\}
\label{Sset}
\ee
where $n\times n$ is defined as in \eqref{xXx} and an additional positivity condition $L(n)$ may be required when the only elements in $G_{i}(\mathds{Z})$ that change the sign of $n$ are trivial in $G_{i}(\mathds{Z})/G_{i+1}(\mathds{Z})$. 

One defines the positive set $S_{i\, >}\subset  \mathcal{M}_i$ as
\be
S_{i\, >}=\Bigl\{H_i\in \mathcal{M}_i\ \lvert\ H_i[n]\geq \frac{1}{\y_{i+1}^2},\, \forall n\in S_{i} \Bigr\}
\ee
Where $H_i[n]$ is the positive definite bilinear form \eqref{Hix}. This domain is non-empty. For fixed $(H_{i+1},{\bf x}_{i+1})$ in $\mathcal{F}^P_i$, one finds that $H_i \in S_{i\, >}$ for $y_{i+1}> L_{i+1}(H_{i+1},{\bf x}_{i+1})$ sufficiently large. 

We will show that a fundamental domain $\mathcal{F}_i = \mathcal{M}_i / G_i(\mathds{Z})$ can be defined as the intersection 
\be
\mathcal{F}_i=\mathcal{F}^P_i\cap S_{i\, >}\; . 
\label{fundamental domain}
\ee

By construction, any element $\gamma \in G_i(\mathds{Z})$ can be decomposed as
\be \gamma = h(n) p = h(n) \, l \exp(b) \; , \ee
with $p \in P_{i+1}(\mathds{Z})$, $l\in G_{i+1}(\mathds{Z})$ and $\exp(b) \in U_{i+1}(\mathds{Z})$. For each point in $G_{i}(\mathds{Z})/P_{i+1}(\mathds{Z})$ one can find a representative $h\in G_{i}(\mathds{Z})$ that preserves $\mathcal{F}^P_i$. Using the isomorphism \eqref{Sset}, one has therefore for each $n\in S_{i}$ a unique $h(n) \in G_{i}(\mathds{Z})$ such that 
\be
h(n) {\bf e}^\ord{i} =n\; , \qquad h(n) \mathcal{F}^P_i = \mathcal{F}^P_i\; . 
\ee
We have already checked that for any point  $H_i \in \mathcal{M}_i$, there is an element $p\in P_{i+1}(\mathds{Z})$ such that $p^\intercal H_i p \in \mathcal{F}^P_i$. For any such $  H_i^\prime \in \mathcal{F}^P_i$ there is a smallest norm element $n_0 \in S_{i}$ such that 
\be H^\prime_i[n]\geq H^\prime_i[n_0],\, \qquad \forall n\in S_{i} \; . \label{MinBall} \ee
It may not be unique, but because $R^{\ord{i}}(\mathds{Z})$ is discrete in $R^{\ord{i}}$, there is a finite number of discrete points in any ball $H^\prime_i[{\bf x}]\leq H^\prime_i[n]$ and therefore a finite number of $n_0$ satisfying \eqref{MinBall}. Choosing $h(n_0)$ one obtains 
\be \frac{1}{\y_{i+1}^2(\gamma)}=h^{\intercal}(n_0) H_i^\prime h(n_0)[{\bf e}^\ord{i} ]= H_i^\prime[n_0] \; , 
\ee
and $\gamma^\intercal H_i \gamma \in \mathcal{F}^P_i\cap S_{i\, >}$ for $\gamma = ph(n_0) $. We have therefore proved that for all $H_i\in \mathcal{M}_i$, there exists $\gamma \in G_i(\mathds{Z})$ such that $\gamma^\intercal H_i \gamma \in \mathcal{F}_i$ so that $\mathcal{F}_i$ includes a fundamental domain. 

Now we need to prove that for any point in the interior of $\mathcal{F}_i$
\be
\mathring{\mathcal{F}}_i=\mathring{\mathcal{F}}^P_i\cap \mathring{S}_{i\, >} \; , 
\ee
and any non-trivial element $\gamma \in G(\mathds{Z})/ Z(G(\mathds{Z}))$ we have $\gamma \mathring{\mathcal{F}}_i\cap {\mathcal{F}}_i = \{ 0 \} $. First of all let us show that one can always determine $S_{i\, >}$ with a finite number of vectors $n \in S_{i}$. For $(y_{i+1},H_{i+1},{\bf x}_{i+1})$ in $\mathcal{F}^P_i$, the set $S_{i\, >}$ includes the condition $H_i[ {\bf e}_{-\alpha_{i+1}} {\bf e}^\ord{i+1}] \ge \frac{1}{y_{i+1}^{\, 2}}$ for the simple root $\alpha_{i+1}$ of $G_i$ satisfying $(\Lambda^\ord{i+1},\alpha_{i+1})=1$. In the appropriate coordinate system this reads  
\be \frac{1}{y_{i+1}^{\; 2}} x(\alpha_{i+1})^2 +  \frac{1}{y_{i+1}^{\; 2-\frac{2}{(\Lambda^{\scalebox{0.4}{$(i+1)$}} ,\Lambda^{\scalebox{0.4}{$(i+1)$}})}}} \frac1{y(\alpha_{i+1})^2} \ge \frac{1}{y_{i+1}^{\; 2}}   \quad \Rightarrow \quad y_{i+1}^{\frac{2}{(\Lambda^{\scalebox{0.4}{$(i+1)$}} ,\Lambda^{\scalebox{0.4}{$(i+1)$}})}} \ge \frac{3}{4} y(\alpha_{i+1})^2 \; , \ee
for $y(\alpha_{i+1})$  the associated Cartan torus coordinate in the Levi subgroup $G_{i+1}$. In particular one can always choose the succession of parabolic subgroups \eqref{ParaSucc} such that $y(\alpha_{i+1}) = y_{i+2}$. It follows by induction that there exists $l_{i+1}>0$ independent of $(H_{i+1}, {\bf x}_{i+1})$ such that any point $(y_{i+1},H_{i+1},{\bf x}_{i+1})\in \mathring{\mathcal{F}}^P_i\cap \mathring{S}_{i\, >} $ satisfies $y_{i+1} \ge  l_{i+1}$. One can therefore restrict the conditions defining $S_{i\, >}$ to the vectors $n\in S_{i}$ in the ball $H_i[n]\leq \frac{1}{l_{i+1}
{}^2}$ and there is only a finite number of those. 

As a consequence $S_{i\, >}$ is defined by a finite intersection of closed sets and its interior $\mathring{S}_{i\, >}$ is defined by the finite intersection of open sets 
\be
\mathring{S}_{i\, >}=\Bigl\{H_i\in \mathcal{M}_i\ \lvert\ H_i[n]>\frac{1}{\y_{i+1}^2},\, \forall n\in S_{i},\;  n\neq {\bf e}^\ord{i+1} \Bigr\}\; . 
\ee
It follows directly that for any point $H_i \in \mathring{\mathcal{F}}_i$, the action of a non-trivial $\gamma = h(n) p\in G_i(\mathds{Z})/Z(G(\mathds{Z}))$ gives $\gamma^\intercal H_i \gamma \notin {\mathcal{F}}_i$. To see this, note that any non-trivial $p\in P_{i+1}(\mathds{Z})$ moves $\gamma^\intercal H_i \gamma \notin {\mathcal{F}}^P_i \supset  {\mathcal{F}}_i$. If $p=1$ and $h(n)$ is non-trivial, we have by definition $h(n) {\bf e}^\ord{i+1} = n$ and therefore 
\be \gamma^\intercal H_i[n]\gamma <\frac{1}{\y_{i+1}(\gamma)^2} \ee
for this specific $n$, which shows that $\gamma^\intercal H_i \gamma \notin {\mathcal{F}}_i$. Note that if $H_i \in \partial \mathcal{F}_i$, either it is in the boundary of $\mathcal{F}^P_i$ or in the boundary of $S_{i\, >}$. In the second case it means that there is a finite set of vectors $n\in  S_{i}$ not equal to ${\bf e}^\ord{i+1} $ such that $H_i[n]=\frac{1}{\y_{i+1}^2}$, and the corresponding $h(n)$ map $\partial S_{i\, >} \cap \mathcal{F}^P_i $ to itself. 

This concludes the proof that $\mathcal{F}_i$ is a fundamental domain of $\mathcal{M}_i / G_i(\mathds{Z})$, and by induction that $\mathcal{F}$ is a fundamental domain of $\mathcal{M} / G(\mathds{Z})$.

\subsection{Fundamental domain of $SO(N,N,\mathds{Z})$}

In this section we shall illustrate how this construction applies to the symmetric space $\mathcal{M}=\bigl( SO(N)\times SO(N)\bigr) \backslash SO(N,N)$ and the arithmetic subgroup $SO(N,N,\mathds{Z})$ preserving the even self-dual lattice of split signature $\II_{N,N}$. In string theory, the group of T-duality on the torus $T^d$ is $O(d,d,\mathds{Z})$ and the relevant moduli space is $\mathcal{M} / O(d,d,\mathds{Z})$. The group of U-duality of type II on $T^4$ is Spin$(5,5,\mathds{Z})$ instead and the moduli space is $\mathcal{M} / SO_{\scalebox{0.6}{$0$}}(5,5,\mathds{Z})$. The difference with respect to the different possible duality groups $SO_{\scalebox{0.6}{$0$}}(N,N,\mathds{Z})$, $SO(N,N,\mathds{Z})$ and $O(N,N,\mathds{Z})$ only appear in the first step because $SO_{\scalebox{0.6}{$0$}}(1,1,\mathds{Z}) = \{ 1\} $, $SO(1,1,\mathds{Z}) = \mathds{Z}_2$ and $O(1,1,\mathds{Z}) = \mathds{Z}_2\times \mathds{Z}_2$. 
 
We will use the convention that the split signature metric is 
\be
\eta=\begin{pmatrix}0\, &\, \mathds{1}  \\ \mathds{1} \, &\, 0\end{pmatrix}
\label{Antidiagonal metric}
\ee
and we parametrise points in $\mathcal{M}$ by a symmetric matrix $H$ in the vector representation of $SO(N,N)$, 
which satisfies 
\be
\eta^{\intercal}H \eta=H\; . 
\ee
For short we will use the symbol $H$ for the point in moduli space and for the symmetric matrix in $SO(N,N)$ that represents it. 
In this case it is convenient to choose the succession of parabolic subgroups \eqref{ParaSucc} such that 
\be P_i = GL(1)\times SO(N-i,N-i) \ltimes \mathds{R}^{N-i,N-i} \; . \ee
We decompose accordingly $H= H_0$ into each $H_i$ for $1\le i \le N$ as 
\be
H_{i-1}=\begin{pmatrix}1\, &0&\; \; 0\\ \eta x_i&\mathds{1} &\; \; 0\\\frac{1}{2}(x_i,x_i)\, &x_i^{\intercal}&\;\;  1\end{pmatrix}\begin{pmatrix}\frac{1}{y_i^2}&0&0\\0&H_{i}&0\\0&0&y_i^2\end{pmatrix}\begin{pmatrix}1\;\;  &x_i^{\intercal}\eta&\, \frac{1}{2}(x_i,x_i)\\0\; \; &\mathds{1} &x_i\\0\; \; &0&1\end{pmatrix}
\label{Iwasawa decomp}
\ee
where $y_i \in \mathds{R}_+$, $x_i\in\mathds{R}^{N-i,N-i}$ is a vector of $SO(N-i,N-i)$, $\mathds{1}$ is the identity matrix in $SO(N-i,N-i)$ and $H_{i}$ is a symmetric element in $SO(N-i,N-i)$.\\

For the quotient by $O(N,N,\mathds{Z})$, one derives that 
\be P_i(\mathds{Z}) = O(N-i,N-i,\mathds{Z}) \ltimes \II_{N-i,N-i} \; . \ee
For $SO_{\scalebox{0.6}{$0$}}(N,N,\mathds{Z})$, one simply gets $SO_{\scalebox{0.6}{$0$}}(N-i,N-i,\mathds{Z})$. The sets $S_i$ in \eqref{Sset} can be defined as 
\be
S_{i} = O(N-i,N-i,\mathds{Z})/P_{i+1}(\mathds{Z})=\{ Q \in \II_{N-i,N-i} \ \lvert\ ( Q,Q) =0\, ,  \, \text{gcd}\,Q  =1\} 
\ee
for $0\le i\le N-1$. For $SO_{\scalebox{0.6}{$0$}}(N,N,\mathds{Z})$ one must simply replace $S_{N-1}= \{ (\pm 1,0),(0,\pm 1)\} $ by the trivial set to take into account that $SO_{\scalebox{0.6}{$0$}}(1,1,\mathds{Z})$ is trivial. 

At each step we decompose $Q\in\II_{N+1-i,N+1-i}$ as $Q=(m,q,n)$ with $m,n\in\mathds{Z}$ and $q\in \II_{N-i,N-i}$, such that 
\be \eta[Q] = (q,q)+2mn\; ,\ee
and the bilinear form $H_{i-1}[Q]$ decomposes  as
\be
H_{i-1}[Q]=\frac{1}{y_i^2}\bigl(m+(x_i,q)+\tfrac{1}{2}(x_i,x_i) n \bigr)^2+H_{i}[q+x_i n ]+y_i^2 n^2\; . 
\ee

The fundamental domain $\mathcal{M} / O(N,N,\mathds{Z})$ is defined iteratively  such that $\mathcal{F}_{i-1}$ is defined as
\begin{enumerate}
    \item $H_{i}\in \mathcal{F}_{i}$.
    \item $(x_i)_1\in\left[0,\frac{1}{2}\right]$ and $(x_i)_k\in\left[-\frac{1}{2},\frac{1}{2}\right]$ for all $2\le k\le 2N-2i$.
     \item For any $m,n\in\mathds{Z},\ q\in\II_{N-i,N-i}$ such that  $\text{gcd}(m,q,n)=1$ and $(q,q)+2mn=0$ we have:
    \be
  \bigl(m+(x_i,q)+\tfrac{1}{2}(x_i,x_i)n \bigr)^2+y_i^2 H_{i}[q+x_i n ]+y_i ^4 n ^2\geq 1 
    \label{third condition}
    \ee
\end{enumerate}
The first two conditions ensure that $H_{i-1}\in\mathcal{F}_{i-1}^P$ while  the third imposes  $H_{i-1}\in S_{i-1\, >}$. For the fundamental domain $\mathcal{M} / SO_{\scalebox{0.6}{$0$}}(N,N,\mathds{Z})$ one must further take into account that $SO_{\scalebox{0.6}{$0$}}(1,1)$ is trivial and $(x_{N-1})_1\in\left[-\frac12,\frac{1}{2}\right]$ instead of  $\left[0,\frac{1}{2}\right]$ and the last condition $y_N>1$ is not imposed.

\subsubsection{Inductive proof}
For simplicity we consider the fundamental domain $\mathcal{M} / O(N,N,\mathds{Z})$ in this section. Let us first show that for any $H_{i} \in\mathcal{M}_{i}$ there exists $\gamma\in O(N-i,N-i,\mathds{Z})$ such that $\gamma^{\intercal}H_i \gamma\in \mathcal{M}_i / O(N-i,N-i,\mathds{Z})$ assuming it is true for $i+1$. We can write an element of the discrete parabolic subgroup $P_{i+1}(\mathds{Z}) = O(N-1-i,N-1-i,\mathds{Z})\times\mathds{Z}^{2N-2-2i}$ as 

\be
p =\begin{pmatrix}1\, &\, 0\, &\, 0\\0& l&0\\0&0&1\end{pmatrix}\begin{pmatrix}1\, &b^{\intercal}\eta&\, \frac{1}{2}(b,b)\\0&\mathds{1}&b\\0&0&1\end{pmatrix}
\label{parabolic element}
\ee
with $l\in O(N-1-i,N-1-i,\mathds{Z})$ and $b\in \mathds{Z}^{2N-2-2i}$. Using the decomposition of $H_i$ in \eqref{Iwasawa decomp}, one obtains that such transformation gives 
\be H_{i+1} \rightarrow l^{\intercal}H_{i+1} l\; , \qquad x_{i+1} \rightarrow   l^{-1}x_{i+1}+b\; . \ee
By assumption, there exists $l\in O(N-1-i,N-1-i,\mathds{Z})$ such that $l^{\intercal}H_{i+1}l\in\mathcal{F}_{i+1}$. This only fixes $l$ up to sign, and there is therefore enough symmetry together with the shift in $b$ to fix  $0\leq (x_{i+1})_1\leq1/2$ and $-1/2\leq (x_{i+1})_k\leq 1/2$ for all $k$ between $2$ and $2N-2-2i$. This shows that there exists $p$ such that $p^{\intercal}H_i p \in \mathcal{F}_i^P$.

One can write an element $h$  of $O(N-i,N-i,\mathds{Z})$ as
\be
h=\begin{pmatrix}m\, &\, *\, &\, *\\q&*&*\\n&*&*\end{pmatrix}
\label{coset element}
\ee
for any $m,n\in\mathds{Z}$, $q\in \II_{N-1-i,N-1-i}$ with gcd$(m,n,q)=1$ and $2 mn +(q,q)=0$. These components are determined by the action on a vector $Q=(1,0,0)$ up to right multiplication by an arbitrary element in $P_{i+1}(\mathds{Z})$. Using the result above, one can always determine $h$ such that $h^{\intercal} p^{\intercal}H_i p h \in \mathcal{F}_i^P$. Writing $p^{\intercal}H_i p$ in terms of $y_{i+1}, H_{i+1}, x_{i+1}$ satisfying 1 and 2 in the definition of $\mathcal{F}_i$, one obtains that $y_{i+1}(h)$ of $h^{\intercal} p^{\intercal}H_i p h$ as 
\be 
y_{i+1}(h) =\Bigl( \frac{1}{y_{i+1}^2} \bigl(m+(x_{i+1},q)+\tfrac{1}{2}(x_{i+1},x_{i+1}) n \bigr)^2+H_{i+1}[q+x_{i+1} n ]+y_{i+1}^2 n^2\Bigr)^{ - \frac12} \; . 
\ee 
We then simply choose $Q=(m,q,n)$ such that $y_{i+1}(h) $ is maximal. By construction $\gamma = p h$ is now such that $\gamma^\intercal H_i \gamma \in \mathcal{F}_i$.

Let us now show that any non trivial $\gamma = p h \in O(N-i,N-i,\mathds{Z})$ up to the centre $- \mathds{1} \in O(N-i,N-i,\mathds{Z})$ moves an element $H_i \in\mathring{\mathcal{F}}_i$ outside of the fundamental domain.  By induction we assume that $\gamma = p h = l b h $ and any non-trivial $l \in O(N-1-i,N-1-i,\mathds{Z})$ up to sign moves $H_{i+1} \in\mathring{\mathcal{F}}_{i+1}$ out of $\mathring{\mathcal{F}}_{i+1}$, and it is clear that any non-trivial element $b$ or $\ell = - {\mathds{1}}$ moves $x_{i+1}$ outside of the domain $0\leq (x_{i+1})_1\leq1/2$ and $-1/2\leq (x_{i+1})_k\leq 1/2$. 

Choosing $y_{i+1}$ sufficiently large compared to all eigenvalues of  $H_{i+1}$,, it is clear that 
\be  \bigl(m+(x_i,q)+\tfrac{1}{2}(x_i,x_i)n \bigr)^2+y_i^2 H_{i}[q+x_i n ]+y_i ^4 n ^2>  1 \; , \qquad \forall (m,q,n) \ne (\pm 1,0,0) \; .  \ee 
The open set $\mathring{\mathcal{F}}_i$ is therefore defined by the strict inequality and it follows that any non-trivial $h$ moves $H_i\in\mathring{\mathcal{F}}_i$ outside the fundamental domain. 

Because the action is continuous it follows that any non-trivial element $\gamma \in  O(N-i,N-i,\mathds{Z})$ acts on a point in the boundary of the fundamental domain ${\mathcal{F}}_i$ to give another point in the boundary. It follows by induction if $H_{i+1} $ is in the boundary of $\mathcal{F}_{i+1}$, and it is rather obvious if one of the $(x_{i+1})_k = \pm \frac12$. If there exists a non-trivial vector $Q\ne (\pm 1,0,0)$ such that the inequality \eqref{third condition} is saturated, the corresponding element $h$ also preserves the boundary. 

This establishes the definition of the fundamental domain described in this section for $O(N,N,\mathds{Z})$. To prove the result for $SO_{\scalebox{0.6}{$0$}}(N,N,\mathds{Z})$ only require to study the case of $SO(2,2)$, which we describe now.  

\subsubsection{The example of $SO(2,2)$}
In this case we write explicitly the bilinear form as
\be H[Q] =  \frac{1}{y_1^2}\bigl(m+x_1 q_2 + x_2 q_1 +x_1 x_2 n \bigr)^2+ \frac{1}{y_2^2} (q_1+x_1n )^2 +y_2^2 (q_2+x_2 n)^2   +y_1^2 n^2 \; , \ee
with 
\be \eta[Q] = 2 m n + 2 q_1 q_2 \; . \ee
The induction starts with the condition that $q_1^2 + y_2^4 q_2\ge 1$ for any $(q_1,q_2)$ in the orbit of $(1,0)$. For $O(1,1,\mathds{Z})$ one gets the four vectors $(q_1,q_2) = (\pm 1,\pm 1)$ and so one obtains the condition $y_2 \ge 1$. For $SO(1,1,\mathds{Z})$ or $SO_{\scalebox{0.6}{$0$}}(1,1,\mathds{Z})$ one does not get $q_2 = \pm 1$ and there is no further condition on $y_2>0$. 

At the next step the action of the unipotent subgroup allows to fix both $x_i \in [ - \frac12,\frac12]$. For $O(1,1,\mathds{Z})$ and $SO(1,1,\mathds{Z})$ one can use 
the element $- \mathds{1}$ to constrain $x_1\ge 0$. For $SO_{\scalebox{0.6}{$0$}}(1,1,\mathds{Z})$ the trivial group, we do not get this further restriction. 

Using now the third condition, one obtains for $m=n=0$ the two conditions 
\be x_1^2  + y_1^2 y_2^2 \ge 1\; , \qquad x_2^2  + \frac{y_1^2}{ y_2^2} \ge 1 \; , \ee
and all the other conditions one obtains are consequences of these two. 

We conclude that a fundamental domain of $O(2,2,\mathds{Z})$ is defined by the conditions
\be\text{i) }\,x_1^2  + y_1^2 y_2^2 \ge 1\; ,\quad \text{ii) }\,x_2^2  + \frac{y_1^2}{ y_2^2} \ge 1\; , \quad \text{iii) }\,y_2\ge 1\; , \quad \text{iv) }\,0 \le x_1 \le \frac{1}{2} \; , \quad   \text{v) }\,-1/2\leq x_{2}\leq1/2 \; , \ee
a fundamental domain of $SO(2,2,\mathds{Z})$ by 
\be \text{i) }\,x_1^2  + y_1^2 y_2^2 \ge 1\; , \quad \text{ii) }\,x_2^2  + \frac{y_1^2}{ y_2^2} \ge 1\; ,  \quad \text{iii) }\,0 \le x_1 \le \frac{1}{2} \; , \quad \text{iv) }\,-1/2\leq x_{2}\leq1/2 \; , \ee
and a fundamental domain of $SO_{\scalebox{0.6}{$0$}}(2,2,\mathds{Z})$ by 
\be \text{i) }\,x_1^2  + y_1^2 y_2^2 \ge 1\; , \quad \text{ii) }\,x_2^2  + \frac{y_1^2}{ y_2^2} \ge 1\; ,  \quad \text{iii) }\,-\frac12 \le x_1 \le \frac{1}{2} \; , \quad \text{iv) }\,-1/2\leq x_{2}\leq1/2 \; . \ee

This is consistent with the isomorphism 
\be \bigl( \hspace{-0.3mm} SO(2)\times SO(2) \hspace{-0.3mm} \bigr) \backslash SO(2,2) / SO_{\scalebox{0.6}{$0$}}(2,2,\mathds{Z}) \cong SO(2)\backslash SL(2) / SL(2,\mathds{Z}) \times  SO(2)\backslash SL(2) / SL(2,\mathds{Z}) \; , \ee
and $T = x_1  + i y_1 y_2 $ and $U = x_2  + i \frac{y_1}{y_2}$ in the standard fundamental domain of $SL(2,\mathds{Z})$.  The further condition $x_1\ge 0$ appears for the fundamental domain of $SO(2,2,\mathds{Z})$ which includes the further generator $SO(2,2,\mathds{Z}) = S(GL(2,\mathds{Z})\times_{\mathds{Z}_2} GL(2,\mathds{Z}))$ that changes the signs of $x_1$ and $x_2$. The additional condition $y_1\ge 1$ appears for the fundamental domain of $O(2,2,\mathds{Z})$ which further includes the generator that exchanges $T$ and $U$. 

\section{Minima at symmetric points}
The main purpose of this paper is to find the global minimum of Eisenstein series. The $SL(2)$ real analytic Eisenstein series $2 \zeta(2s) E_s$ is known to have a global minimum at the symmetric point $\tau = \frac{1}{2} +i \frac{\sqrt{3}}{2}$, \cite{MinimaSL2R,MinimaSL2C} for any value of the parameter $s>0$, with the appropriate regularisation at $s=1$.  There is no general result for the global minimum of the Epstein series $ \Ep^N_s(H)$ 
for  $N\ge 4$ and generic $s>0$. The sum \eqref{EpsteinN} is absolutely convergent for $s> \frac{N}{2}$, but admits an analytic continuation to a meromorphic function of $s\in \mathds{C}$ with a single pole at $s= \frac{N}{2}$ \cite{Epstein2}. One can define the regularised value of the Epstein function at the simple pole by minimal subtraction.  For $s\rightarrow \infty$, the global minimum is the solution to the densest lattice sphere packing in $N$ dimensions \cite{MinimaLatticePacking}. One can understand this result intuitively using an expansion of $\Ep^N_s$ in the length of the lattice vectors. Writing $n^{H}_{\scalebox{0.6}{min}}$ a vector of minimal length for the associated quadratic form $H$ and $d^{H}_{\scalebox{0.6}{min}}$ the number of  vectors of minimal length in the lattice, one gets for $s \rightarrow \infty $
\be \Ep^N_s \underset{s\rightarrow \infty}{\gtrsim} \frac{d^{H}_{\scalebox{0.6}{min}}}{ H[n^{H}_{\scalebox{0.6}{min}}]^s}\; .  \ee
Therefore the minimum of $\Ep^N_s$ is obtained when $H[n^{H}_{\scalebox{0.6}{min}}]$ is maximal for large Re$[s]$. The density of spheres of radius $\sqrt{H[n^{H}_{\scalebox{0.6}{min}}]}/2$ in the lattice is 
\be \rho(H) = \frac{\Bigl( \frac{ \scalebox{0.9}{$ \pi H[n^{H}_{\scalebox{0.5}{min}}]$ }\hspace{-2mm}\phantom{\bigr|}}{4}\Bigr)^{\frac{N}{2}} }{ \Gamma(\frac{N}{2}+1)}\; , \label{densityLattice} \ee
and so $H[n^{H}_{\scalebox{0.6}{min}}]$ maximal also gives the densest sphere packing in the lattice. 
For $2\le N\le 8$, the  densest lattice sphere packings  are known to be the following rank $N$ root lattices \cite{ConwaySloane}
\be A_2\; , \quad A_3\; , \quad D_4\; , \quad D_5\; , \quad E_6\; , \quad E_7\; , \quad E_8\; , \label{ADEbest} \ee
such that the minimum of the Epstein series at large $s$ is at $H_{L} =\det C_{L}^{-\frac{1}{N}} C_L$ for $C_L$ the even bilinear form associated to the lattice $L$ (with $L=A_N,\, D_N,\, E_N$). 

However, we are interested in small values of $s$ for the string theory couplings, in particular $s\le \frac{N}{2}$. It was observed in  \cite{MinimaSL3} that the densest lattice sphere packing  bilinear form $H_{\scalebox{0.6}{dlp}}$ cannot be the global minimum of the Epstein series for all $s>0$ when the lattice and its dual do not define the same point in $SO(N) \backslash SL(N) / SL(N,\mathds{Z})$, i.e. when $H_{\scalebox{0.6}{dlp}}^{-1} \not \approx H_{\scalebox{0.6}{dlp}}$ modulo the action of $SL(N,\mathds{Z})$. This is a direct consequence of the functional relation 
\be
\Ep^N_s(H)=\pi^{2s-\frac{N}{2}}\frac{\Gamma(\frac{N}{2}-s)}{\Gamma(s)}\Ep^N_{\frac{N}{2}-s}(H^{-1})\; . \label{FunctionalEpstein} 
\ee
In particular the Epstein series is the same for the two dual points at $s=\frac{N}{4}$
\be \Ep^N_{\frac{N}{4}}(H_{\scalebox{0.6}{dlp}}^{-1} ) = \Ep^N_{\frac{N}{4}}(H_{\scalebox{0.6}{dlp}}) \; .  \ee
One may therefore argue at most that the densest lattice packing $H_{\scalebox{0.6}{dlp}}$ is the global minimum for $s\ge \frac{N}{4}$, which implies by the functional relation above that its dual lattice $H_{\scalebox{0.6}{dlp}}^{-1} $ is the global minimum for $0<s< \frac{N}{4}$. 

The case of $N=3$ was analysed numerically in \cite{MinimaSL3}. It was shown that the densest lattice sphere packing point $H_{A_3} $ is a global minimum for $s>\frac{3}{4}$, and in particular for the regularised series
\be \Epreg^{\!\! 3}_{\frac32} = \lim_{s\rightarrow \frac{3}{2}} \Bigl(\EpN{3}_{s} - \frac{2\pi}{s-\frac{3}{2}}  \Bigr) \; . \ee
They also demonstrate that the minimum of $\Epreg^N_{s} $ must be strictly negative in the critical strip $0<s<\frac{N}{2}$. However, it was proved in \cite{MinimaEpsteinTerras} that $\Epreg^N_{\frac{N}{4}}>0$ if the minimal length vector $n^{H}_{\scalebox{0.6}{min}}$ of the associated quadratic form $H$ has length bounded from above as
\be \sqrt{H[n^{H}_{\scalebox{0.6}{min}}] } < \frac{N}{2\pi e} \quad {\rm or}\quad \sqrt{H^{-1}[n^{H^{-1}}_{\scalebox{0.6}{min}}] }< \frac{N}{2\pi e} \; . \ee
This excludes the possibility that the minimum be at $H_{\mathfrak{g}}$ for an ADE lattice for $N>24$ and suggests instead that the densest lattice sphere packing will be the minimum \cite{MinimaSL3}. The densest lattice sphere packing has been proved to be the global minimum for $N=8$ and $24$ and all values of $s>\frac{N}{2}$ \cite{SpherePacking}. 

It would be very difficult to carry out a complete numerical analysis of the Epstein series for $N\ge 4$. Based on the results above we shall therefore concentrate on symmetric points in moduli space for which the bilinear form $H$ is invariant under a finite subgroup of $SL(N,\mathds{Z})$. In this section we define these symmetric points and determine a criterion for them to be local minima of Eisenstein series. 

We then generalise the same analysis to $SO(N,N,\mathds{Z})$ and prove that symmetric points are local minima of Eisenstein series provided they are negative in the critical strip.

\subsection{Taylor expansion at symmetric points}
\label{TaylorExpansion}
The hyperplanes defining the boundary of the fundamental domain are mapped to themselves under elements  of $G(\mathds{Z})$, and their intersections are invariant under non-trivial finite subgroups of $G(\mathds{Z})$. The maximal intersections define isolated points $\mathcal{V}_0\in \mathcal{M}$ that are invariant under maximal finite subgroups  $G^{\mathcal{V}_0}(\mathds{Z})\subset G(\mathds{Z})$
\be
G^{\mathcal{V}_0}(\mathds{Z})=\bigl\{\gamma\in G(\mathds{Z})\ \lvert\ \gamma^\intercal \mathcal{V}^\intercal_0 \mathcal{V}_0 \gamma = \mathcal{V}^\intercal_0 \mathcal{V}_0 \bigr\} =\bigl\{\gamma\in G(\mathds{Z})\ \lvert\ \mathcal{V}_0\gamma\mathcal{V}_0^{-1}\in K\bigr\}\; . 
\ee
When the $G^{\mathcal{V}_0}(\mathds{Z})$-fixed points in $\mathcal{M}$ are isolated, we call them symmetric points. All the symmetric points are maximal intersections but some maximal intersections may not be symmetric points. 

By construction for all $\gamma\in G^{\mathcal{V}_0}(\mathds{Z})$ there exists a $k_\gamma=\mathcal{V}_0\gamma\mathcal{V}_0^{-1}\in K$ such that
\be
\mathcal{V}_0\gamma=k_\gamma \mathcal{V}_0\; . \label{Kgamma}
\ee
An automorphic form $\Phi$ on $\mathcal{M}$ in a representation $\rho$ of $K$ transforms as
\be \Phi(k \mathcal{V} \gamma) = \rho_\Phi(k) \Phi(\mathcal{V}) \; . \ee
Here $\rho$ generalises the weight for $SL(2)$. By definition, the automorphic form is invariant under the linear action of $\rho_\Phi(k_\gamma)$  at $\mathcal{V}= \mathcal{V}_0$ for any $\gamma\in G^{\mathcal{V}_0}(\mathds{Z})$ and $k_\gamma$  satisfying \eqref{Kgamma}. Indeed, one checks that 
\be  \rho_\Phi(k_\gamma) \Phi(\mathcal{V}_0) = \Phi(k_\gamma \mathcal{V}_0 )=   \Phi( \mathcal{V}_0 \gamma ) = \Phi( \mathcal{V}_0  )\; . \ee
We can use this property to constrain the covariant derivatives of an automorphic function at a symmetric point. We define the covariant derivative from the left-action of $p \in  \mathfrak{p} = \mathfrak{g}\ominus \mathfrak{k}$ as
\be
f(\exp(p)\mathcal{V}_0)\equiv f_{\mathcal{V}_0}(p) = f(\mathcal{V}_0)+p^aD_af(\mathcal{V}_0)+\frac{1}{2}p^ap^bD_aD_bf(\mathcal{V}_0)+\mathcal{O}(p^3)\; , 
\ee
where $a$ labels the components of $p$ in $\mathfrak{p}$. Note that the differential operators $D_a$ can equivalently be defined as the covariant derivative in tangent frame on the symmetric space 
\be D_a = e_a{}^\mu \bigl( \partial_\mu + \rho(\omega_\mu )\bigr) \; . \ee
If $G^{\mathcal{V}_0}(\mathds{Z})$ is non-trivial, we have 
\be f_{\mathcal{V}_0}(p) =  f_{\mathcal{V}_0}(k_\gamma^{-1} pk_\gamma)\ee
for any $k_\gamma = \mathcal{V}_0\gamma\mathcal{V}_0^{-1}\in K$  such that $\gamma \in G^{\mathcal{V}_0}(\mathds{Z})$, because 
\be f_{\mathcal{V}_0}(k_\gamma^{-1} pk_\gamma) =f(k_\gamma^{-1} \exp(p)k_\gamma \mathcal{V}_0) =  f(\exp(p) \mathcal{V}_0\gamma)  =  f_{\mathcal{V}_0}( p)\; . \ee
The Taylor expansion of an automorphic function is therefore highly constrained by the symmetry group $G^{\mathcal{V}_0}(\mathds{Z})$. At order $n$ in the Taylor expansion, one can classify the order $n$ polynomials in $p$ that are invariant under all such elements $k_\gamma = \mathcal{V}_0\gamma\mathcal{V}_0^{-1}\in K$. If $\mathcal{V}_0$ is a symmetric point as we define above, there is no invariant linear polynomial and $\mathcal{V}_0$ is an extremum of any automorphic function. 

For any symmetric point $\mathcal{V}_0$ we must check if the extremum is a minimum. In practice we shall find that the symmetric points  admit a small number of  $G^{\mathcal{V}_0}(\mathds{Z})$-invariant quadratic polynomials, and that it is sufficient to evaluate the automorphic function on a small dimension hypersurface to determine if it is a minimum.

If there is a single invariant quadratic polynomial, by construction it must  be proportional to the quadratic Casimir such that 
\be
f_{\mathcal{V}_0}(p)=f(\mathcal{V}_0)+\frac{1}{2\dim\mathfrak{p}}\kappa_{ab}p^a p^b\Delta f(\mathcal{V}_0)+\mathcal{O}(p^3)
\ee
where $\kappa_{ab}$ is the restriction to $\mathfrak{p}$ of the Killing--Cartan form on $\mathfrak{g}$. The Hessian of the function at $\mathcal{V}_0$ is then completely determined, and the symmetric point is a local minimum provided $\Delta f(\mathcal{V}_0)>0$. In particular the function must be negative at a local minimum if the eigenvalue of the Laplace operator is negative. 

We will be interested in maximal parabolic Eisenstein series $E_{s\Lambda_i}^{G}$ associated to a fundamental weight  $\Lambda_i$,
\be E_{s\Lambda_i}^{G} = \sum_{\gamma \in G(\mathds{Z})/ P_i(\mathds{Z})} \chi(2s \Lambda_i - \varrho,\mathcal{V} \gamma) \; , \ee
which satisfy  the Laplace equation
\be \Delta E_{s\Lambda_i}^{G} = 2  (s  \Lambda_i -\varrho , s \Lambda_i) E_{s\Lambda_i}^{G} \;, \ee
with  $\varrho = \sum_i \Lambda_i$ the Weyl vector and $\chi(\lambda)$ the multiplicative parabolic character of weight $\lambda$. Physically, maximal parabolic Eisenstein series are the sum over the U-duality images of a fixed dilaton to the power $s$. For maximal parabolics, the dilaton is either the string coupling constant or the volume of a torus $T^r$ inside the M-theory torus or the type IIB torus.

Maximal parabolic Eisenstein series are absolutely convergent for  $ {\rm Re}[s] >\frac{( \Lambda_i,\varrho)}{(\Lambda_i,\Lambda_i)} $, and both $ E_{s\Lambda_i}^{G}>0 $ and $\Delta E_{s\Lambda_i}^{G}>0$ for $ s> \frac{( \Lambda_i,\varrho)}{(\Lambda_i,\Lambda_i)}$. On the critical strip $0<  s< \frac{( \Lambda_i,\varrho)}{(\Lambda_i,\Lambda_i)}$, the Eisenstein series is integrable and therefore 
\be
0=\int_\mathcal{M}d\mu\,\Delta E_{s\Lambda_i}^{G}(\mathcal{V}) =2  (s  \Lambda_i -\varrho , s \Lambda_i)  \int_\mathcal{M}d\mu\,E_{s\Lambda_i}^{G}(\mathcal{V}) \; , 
\ee
such that $E_{s\Lambda_i}^{G}(\mathcal{V})$ must be negative at its global minimum.

We find therefore that when the moduli space $K\backslash G$ admits a symmetric point $\mathcal{V}_0$ with a unique invariant quadratic polynomial in $p$, the symmetric point $\mathcal{V}_0$ is a local minimum of any absolutely convergent Eisenstein series. Moreover, $\mathcal{V}_0$ is a local minimum of any integrable automorphic function that is negative at $\mathcal{V}_0$. 

When there are two $G^{\mathcal{V}_0}(\mathds{Z})$-invariant quadratic polynomials we need to compute the two independent eigenvalues of the Hessian matrix at $\mathcal{V}_0$. For this purpose we shall define a two-dimensional subspace of $SO(N) \backslash SL(N)$ that includes the relevant symmetric points. There also exist symmetric points with more invariant polynomials, but we find that they never correspond to global mimima and shall disregard them. 

We will now discuss the cases of $SL(N)$ and $SO(N,N)$.

\subsection{$SL(N)$ symmetric points}
The symmetric points in  $SO(N) \backslash SL(N)$ are invariant under maximal  finite irreducible subgroups of $PSL(N,\mathds{Z})$. The maximal  finite irreducible subgroups of $GL(N,\mathds{Z})$ have been classified for all $N\le 10$ \cite{AutoL4,AutoL5,AutoL9}. For $N\le 4$ all the maximal finite subgroups of $PSL(N,\mathds{Z})$ are stabilisers of an even-bilinear form  $C_{L}$ of Cartan type. One gets for irreducible groups
\bea N=1 : \qquad\qquad&& A_1\; ,  \\
N=2 : \qquad\qquad&& A_2  \;, \quad 2A_1\; ,  \\
N=3 : \qquad\qquad&& A_3 \;, \quad A_3^* \;, \quad 3A_1 \; , \\
N=4 : \qquad\qquad&& D_4\;, \quad A_4 \;, \quad A_4^* \;, \quad 4A_1 \; , \quad 2 A_2\; , \quad A_2\times  A_2 \; .
\eea
The reducible maximal subgroups stabilise  the reducible bilinear forms $A_1{+}A_2$, $A_1{+}A_3$, $A_1{+}A_3^*$. One easily checks that reducible bilinear forms that are extrema are always saddle points. To prove this, one considers coordinates associated to the maximal parabolic subgroup $P_k = \mathds{R}_+\times SL(k)\times SL(N-k)\ltimes \mathds{R}^{k\times(N-k)}\subset SL(N) $ in which 
\be H[n,m] = y^{2\frac{k-N}{N}} H_k[n+X m] +y^{2\frac{k}{N}} H_{N-k}[m] \ee
with $n\in \mathds{Z}^k$, $m\in \mathds{Z}^{N-k}$ and $X\in \mathds{R}^{k\times (N-k)}$. The Fourier expansion of the Epstein series in the maximal parabolic $P_k$ can be obtained by Poisson summation and reads 
\bea
&& \hspace{-2mm}\Ep^N_s(H) = y^{2\frac{N-k}{N} s}\Ep^k_s(H_k)  +\pi^{\frac k2} \frac{\Gamma(s-\frac k2)}{\Gamma(s)} y^{2\frac{k}{N}(\frac{N}{2}-s)} \Ep^{N-k}_{s-\frac{k}{2}}(H_{N-k})  \\
&&\quad  +\frac{2\pi^s}{\Gamma(s)} \sum^\prime_{n\in\mathds{Z}^k}\sum_{m\in\mathds{Z}^{N-k}}^\prime \hspace{-2mm} y^{\frac{N-2k}{N} s + \frac{k}{2} }\frac{H_k^{-1}[n]^{\frac{s}{2}-\frac{k}{4}}}{ H_{N-k}[m]^{\frac{s}{2}-\frac{k}{4}}}  K_{s-\frac{k}{2}}\Bigl( 2\pi  y \sqrt{ H_k^{-1}[n] H_{N-k}[m]} \Bigr)\cos(2\pi n^\intercal X m )\; . \nonumber
\eea
The second derivative with respect to $X$ at $X=0$ is negative definite
\bea&&  {\rm d}^2 \Ep^N_s(H) |_{X=0}\\
&=&- \frac{2\pi^s}{\Gamma(s)} \sum^\prime_{n\in\mathds{Z}^k}\sum_{m\in\mathds{Z}^{N-k}}^\prime y^{\frac{N-2k}{N} s + \frac{k}{2} }\frac{H_k^{-1}[n]^{\frac{s}{2}-\frac{k}{4}}}{ H_{N-k}[m]^{\frac{s}{2}-\frac{k}{4}}}  K_{s-\frac{k}{2}}\Bigl( 2\pi  y \sqrt{ H_k^{-1}[n] H_{N-k}[m]} \Bigr)(2\pi n^\intercal {\rm d}X m )^2 \; ,\nonumber \eea
as the absolutely convergent sum of positive terms. Therefore any extremum at $X=0$ is a saddle point. We will therefore only analyse symmetric points for which the bilinear form $H$ is irreducible. 

For $N=5$ the irreducible bilinear forms invariant under a maximal finite subgroup of $PSL(5,\mathds{Z})$ are \cite{AutoL5}
\be N=5 : \qquad\qquad D_5\; , \quad D_5^*\; , \quad A_5\;, \quad A_5^*\;, \quad A_5^{+2} \;, \quad (A_5^{+2})^*= A_5^{+3} \;  , \ee
where $A_5^{+2} $ is the ${\rm S}_6$-invariant lattice $A_5\oplus (A_5 {+} \Lambda_2) \oplus (A_5{+}\Lambda_4)$ that includes the second fundamental weight $\Lambda_2$, and $A_5^{+3} =A_5\oplus (A_5 {+} \Lambda_3) $. All these points lie on dimension zero boundaries of Grenier's fundamental domain, as we show in Appendix \ref{Grenierdomainboundaries}.

It will be convenient to introduce the following parametrisation of $\mathcal{V}\in SO(N) \backslash SL(N) $ in the maximal parabolic subgroup $P_{N-1}$
\be
\mathcal{V}= y^{-\frac{1}{N}}\left(\begin{array}{cc} \; N^{\frac1{2(N-1)}} \mathcal{V}_1 \; & \; N^{\frac1{2(N-1)}}  \mathcal{V}_1 {\bf x}\; \\ \; 0\; & \; \frac{y}{\sqrt{N}} \end{array}\right) \; , \label{VSLNCN} \ee
where $ \mathcal{V}_1 \in SO(N-1)\backslash SL(N-1)$, ${\bf x} \in \mathds{R}^{N-1}$, and $y\in \mathds{R}_+$ has been rescaled. When $ \mathcal{V}_1$ is at the $A_{N{-}1}$ symmetric point 
\be N^{\frac1{N-1}}  \mathcal{V}_1^\intercal  \mathcal{V}_1 = C_{A_{N{-}1}} \; . \label{VVeqC} \ee 
In this paper we define the $A_N$ point in the fundamental domain as the symmetric matrix 
\be
C_{A_N} =\begin{pmatrix}2&1&\dots&1\\1&2&\ddots&\vdots\\\vdots&\ddots&\ddots&1\\1&\dots&1&2\end{pmatrix}\; , 
\label{H}
\ee
or equivalently 
\be C_{A_N}[n] = 2 \sum_{i=1}^N n_i^2 + \sum_{i\ne j} n_i n_j \; . \label{ANCartan}  \ee
In this way the symmetry under the subgroup S$_N\subset {\rm S}_{N+1}$ is manifest. One gets the $A_N$, $D_N$, $E_N$ representatives by fixing $\mathcal{V}_1$ at the $A_{N{-}1}$ lattice point, and all $x_i$ components equal such that 
\bea y^{\frac{2}{N}} H = C_{A_{N}} \qquad  &\mbox{for}& \qquad x_i=\frac{1}{N}\, , \; y  =\sqrt{N+1 }\; , \\
  y^{\frac{2}{N}} H = C_{D_{N}} \qquad  &\mbox{for}& \qquad x_i=\frac{2}{N}\, , \; y  =2 \; , \label{DnfromAn} \\
 y^{\frac{2}{N}}  H =  C_{E_{N}} \qquad  &\mbox{for}& \qquad x_i=\frac{3}{N}\, , \; y  =\sqrt{9-N}\;. \label{CNfromyx}
\eea
One moreover checks  that the Gram matrix $C_{\! A_{5}^{+2}}$ and $C_{\! A_{5}^{+3}}$ of $A_5^{+2}$ and $A_5^{+3}$ are given by \footnote{This is not the bilinear form given in \cite{AutoL5}, the latter is equivalent to $3 C_{\! A_{5}^{+2}}$ up to $SL(5,\mathds{Z})$.}
\bea  y^{\frac{2}{5}} H = C_{\! A_{5}^{+2}}  \qquad  &\mbox{for}& \qquad x_i=\frac{2}{5}\, , \; y  =\sqrt{\frac{2}{3} }\; , \label{F5fromyx} \\
 y^{\frac{2}{5}} H = C_{\! A_{5}^{+3}}  \qquad  &\mbox{for}& \qquad x_i=\frac{2}{5}\, , \; y  =\sqrt{\frac{3}{2} } \; .  \eea
This ansatz for $y>0$ and a single $x\in [0,\frac12]$ therefore describes many symmetric points. In particular it covers all the symmetric points (up to inversion) for $SL(3,\mathds{Z})$ and $SL(5,\mathds{Z})$,  the groups relevant  in string theory. The dual lattice points can be studied in the same way with the inverse matrix or using the functional relation \eqref{FunctionalEpstein}. This representative of the $A_N$ symmetric point is in the Grenier domain, but not the others, see Appendix  \ref{Grenierdomainboundaries}. 

The densest sphere packing criterion can be checked using the density \eqref{densityLattice}  
\be \rho(C_{A_N}) = \frac{\bigl( \frac{\pi}{2} \bigr)^{\frac{N}{2}} }{\sqrt{N+1} \Gamma(\frac{N}{2}+1)} \underset{N\ge 4}{<} \rho(C_{D_N}) = \frac{\bigl( \frac{\pi}{2} \bigr)^{\frac{N}{2}} }{2 \Gamma(\frac{N}{2}+1)} \underset{N=6,7,8}{<} \rho(C_{E_N}) = \frac{\bigl( \frac{\pi}{2} \bigr)^{\frac{N}{2}} }{\sqrt{9-N} \Gamma(\frac{N}{2}+1)}   \ee
in agreement  with the list \eqref{ADEbest}. The dual lattices $D_N^*$ and $A_N^*$ are always less dense than the lattices $D_N$ and $A_N$, but $E_N^*$ is denser than $D_N$ and $A_N$ for $N=6,7,8$. For $N=5$ one finds 
\be \rho(C_{A_5}) =\frac{\pi^2}{15 \sqrt{3}} <  \rho(C_{\! A_{5}^{+2}}) = \frac{4 \sqrt{2} \pi^2}{135} < \rho(C_{D_5}) =  \frac{\pi^2}{15 \sqrt{2}}\; ,  \ee
so the densest sphere lattice packing is indeed the lattice $D_5$, and $A_{5}^{+2}$ is denser than $A_5$. The dual lattices are less dense, with 
\be  \rho(C_{A_5^*}) =\frac{5\sqrt{5} \pi^2}{432} < \rho(C_{D_5^*}) =  \frac{\pi^2}{30} <  \rho(C_{\! A_{5}^{+3}}) = \frac{3 \pi^2}{80} \; .  \ee

On the surface where $H(y,x,C_{A_{N-1}})$ is parametrised by $y>0$ and $- \frac12 \le x\le \frac{1}{2}$ as in \eqref{VSLNCN}, one  writes the explicit formula for the $SL(N)$ Epstein series as
\begin{multline} \Ep^N_s(y,x,C_{A_{N-1}}) = \frac{y^{\frac{2s}{N}}}{N^{\frac{s}{N-1}}}  \Ep^{N-1}_s(C_{A_{N-1}}) +  \frac{2\pi^{\frac{N-1}{2}} \Gamma(s-\frac{N-1}{2})\zeta(2s-N+1)  }{\Gamma(s)} \frac{y^{(N-1)( 1- \frac{2s}{N})}}{N^{\frac{N}{2}-s}}   \\
+ \frac{4\pi^s}{\Gamma(s)} \frac{y^{\frac{N-1}{2} - \frac{N-2}{N}s}}{\sqrt{N}} \sum_{n \in \mathds{Z}^{N-1}}^\prime   \sigma_{N-1-2s}(n) \frac{ K_{s-\frac{N-1}{2}}\bigl( 2\pi y  \sqrt{  C_{A_{N-1}}^{-1}[n] /N } \bigr)}{\bigl( N C_{A_{N-1}}^{-1}[n] \bigr)^{\frac{N-1}{4}-\frac{s}{2}}  }  \cos\Bigl( 2\pi x \sum_{i=1}^{N-1} n_i \Bigr)  \label{EpsteinExpansion}\; .  \end{multline}
It is defined by induction in $N$. For $N=2$ the Epstein series evaluated at $\tau = \frac12 + i \frac{\sqrt{3}}{2}$ can be written as the zeta function over $\mathds{Q}(\sqrt{-3})$ \cite{MinimaSL2R}\footnote{The expression using Hurwitz zeta function is obtained as the Euler product $$ \sum^\prime_{z\in \mathds{Z}(e^{\frac{2\pi i}{3}})}    \frac{1}{|z|^{2s}} =  6 \frac{1}{1-3^{-s}} \prod_{p = 1\, {\rm mod} \, 3} \left( \frac{1}{1-p^{-s}}\right)^2 \prod_{p = 2\, {\rm mod}\,  3} \frac{1}{1-p^{-2s}} \; . $$}
\be \EpN{2}_s(C_{A_{2}}) = \sum^\prime_{z\in \mathds{Z}(e^{\frac{2\pi i}{3}})}  \frac{( \frac{\sqrt{3}}{2})^s }{|z|^{2s}} = \frac{6}{12^{\frac{s}{2}}} \zeta(s) \bigl( \zeta(s,\tfrac13) -\zeta(s,\tfrac23)\bigr) \;  . \ee

For $SL(N)$ the elements $p \in  \mathfrak{p} = \mathfrak{g}\ominus \mathfrak{k}$ are symmetric-traceless $N\times N$ matrices, and the Killing--Cartan form is normalised such that 
\be  \kappa_{ab} p^a p^b = 2 \tr  p^2 = 2 \sum_{i,j=1}^N p_{ij}^2 \; . \ee
We will now study the different symmetric points.

\subsubsection{The $A_N$ symmetric point}
\label{ANsymmetricPoint} 
The automorphism group of the $A_N$ lattice is $\mathds{Z}_2\ltimes {\rm S}_{N+1}$, but only the alternating group Alt$_{N+1}$ embeds inside $PSL(N,\mathds{Z})$ for $N$ even and  S$_{N+1}\subset PSL(N,\mathds{Z})$ for $N$ odd. The Eisenstein series  are nonetheless invariant under the action of  $PGL(N,\mathds{Z})$, so their symmetry group is always S$_{N+1}$. The automorphism group $\mathds{Z}_2\ltimes {\rm S}_{N+1}$ is realised on the lattice vectors such that $\mathds{Z}_2$ acts as 
\be n_i \mapsto - n_i \; , \ee
and S$_{N+1}$ acts as the permutation of the $N$ elements $n_i$ and a fictitious $N{+}1$th element $-\sum_{j=1}^Nn_j$, i.e. $\sigma\in {\rm S}_N$ acts as
\be
\sigma : n_i\mapsto n_{\sigma(i)}\;  , 
\ee
and  
\be
\sigma_{i,N+1} : n_i\mapsto -\sum_{j=1}^Nn_{j}\; ,\quad n_{j\neq1}\mapsto n_j\; . 
\ee
It is useful to split this group into ${\rm S}_N$ and
\be
{\rm S}_{N+1}/{\rm S}_N=\mathds{Z}_{N+1}
\label{splitting the symmetric group}
\ee
where $\mathds{Z}_{N+1}$ is the cyclic group generated by the transformation
\be
\sigma_{f} : n_{i\neq N}\mapsto n_{i+1}\; ,\quad n_N\mapsto -\sum_{i=1}^Nn_{i}\; . 
\label{sigma_f}
\ee
To determine the S$_{N+1}$-invariant quadratic polynomials, it is convenient to define 
\be \tilde{p} =(N+1)^{\frac{1}{N}}  \mathcal{V}_0^{\intercal} p\mathcal{V}_0 \; , \ee for which the traceless condition reads 
\be \tr C_{A_N}^{-1} \tilde{p} = \frac{1}{N+1}\biggl(N\sum_{i=1}^N \tilde{p}_{ii}-\sum_{i\ne j}\tilde{p}_{ij}\biggr)=0\; . \ee
For any  invariant polynomial  $f_{\mathcal{V}_0}(p)$, one defines the polynomial 
\be
\tilde f_{\mathcal{V}_0}(\tilde p)=f_{\mathcal{V}_0}\bigl((N+1)^{-\frac{1}{N}} \mathcal{V}_0^{-1\,\intercal}\tilde p\mathcal{V}^{-1}_0\bigr)\; ,  \label{tildepPol} 
\ee
which is invariant under 
\be \tilde f_{\mathcal{V}_0}(\tilde p) = \tilde f_{\mathcal{V}_0}(\gamma^\intercal \tilde p \gamma )\; , \ee
for any  $\gamma\in G^{A_N}(\mathds{Z})$ representing S$_{N+1}$. The symmetric matrix  $\tilde{p}$ simply transforms under S$_{N+1}$ as a traceless bilinear form in the $n_i$'s, i.e. under  $\sigma\in {\rm S}_N$ as
\be
\sigma : \tilde{p}_{ij}\mapsto \tilde{p}_{\sigma^{-1}(i)\sigma^{-1}(j)}\; , 
\ee
and under $\sigma_f$ as
\be\begin{split}
\sigma_f &: \tilde{p}_{11}\mapsto \tilde{p}_{NN}, \\
\sigma_f &: \tilde{p}_{1j}\mapsto \tilde{p}_{NN}-\tilde{p}_{j-1N}, 
\end{split}\hspace{4mm}
\begin{split}
 &\tilde{p}_{i\neq 1i\neq 1}\mapsto \tilde{p}_{NN}+\tilde{p}_{i-1i-1}-2\tilde{p}_{i-1N}\; , \\
&\tilde{p}_{i\neq 1j\neq1}\mapsto \tilde{p}_{NN}+\tilde{p}_{i-1j-1}-\tilde{p}_{i-1N}-\tilde{p}_{j-1N}\; .  \end{split} \label{tpAN} 
\ee
The vector $n$ transforms in the standard representation of S$_{N+1}$ associated to the partition $(N,1)$. The symmetric tensor product decomposes into irreducible representations as \footnote{Recall that $(N)$ is the trivial representation, $(N,1)$ is the basic representation of dimension $N$, and $(N-1,2)$ is the symmetric tensor irreducible representation of dimension $\frac{(N+1)(N-2)}{2}$.}
\be (N,1) \otimes_S (N,1) = (N) \oplus (N,1)\oplus(N-1,2) \; , \ee
such that $\tilde{p}$ splits into the irreducible components $q \in (N,1)$ and $\tilde{p}^\perp \in (N-1,2)$. The vector $q$ in the standard representation  is defined as 
\be
q_i\equiv \tilde p_{ii}-\frac{2}{N-1}\sum_{\substack{j=1\\j\neq i}}^N\tilde p_{ij}\; , \label{qinp} 
\ee
and transforms by permutation under S$_N$ and as 
 \be \sigma_f : q_1 \mapsto -q_N\; , \quad q_{i\neq 1} \mapsto q_{i-1} -q_N\;, \ee
 under $\sigma_f$. The $ (N-1,2)$ tensor $\tilde{p}^\perp$ is defined as 
 \be \tilde{p}^\perp_{ii} = \tilde{p}_{ij} - q_i + \frac{2}{N+1} \sum_{j=1}^N q_j \; , \quad   \tilde{p}^\perp_{i\neq j} = \tilde{p}_{ij}  + \frac{1}{N+1} \sum_{k=1}^N q_k \;  ,   \ee
 and transforms under S$_{N+1}$ as $\tilde{p}$ in \eqref{tpAN}. 
It follows that there are two independent invariant quadratic polynomials in $\tilde{p}$, one quadratic in $q$ and one quadratic in $ \tilde{p}^\perp$, such that the Killing--Cartan form splits into 
\be \frac12  \kappa_{ab} p^a p^b=  \tr C_{A_N}^{-1} \, \tilde{p} \, C_{A_N}^{-1} \, \tilde{p}  = \frac{N-1}{N+1} C_{A_N}^{-1}[q] + \tr C_{A_N}^{-1} \tilde{p}^\perp C_{A_N}^{-1} \tilde{p}^\perp\; .  \ee
For completeness, we give the explicit proof that there are only two invariant quadratic polynomials in Appendix \ref{InvQuaPolAN}. 

To study the Taylor expansion of Epstein series at the symmetric point $A_N$, we use the coset representative \eqref{VSLNCN} and its vielbeins 
\be \mathcal{P} = \frac12 {\rm d} \mathcal{V} \mathcal{V}^{-1}  + \frac12 ({\rm d} \mathcal{V} \mathcal{V}^{-1})^\intercal =\begin{pmatrix}\mathcal{P}_1-\frac{1}{N}\frac{{\rm d}y}{y}&\frac{N^{\frac{N}{2(N-1)}}}{2y}\mathcal{V}_1\,{\rm d}x\\\frac{N^{\frac{N}{2(N-1)}}}{2y}\mathcal{V}_1 \,{\rm d}x&\frac{N-1}{N}\frac{{\rm d}y}{y}\end{pmatrix}  \; . \ee
At the symmetric point we take $\mathcal{V}_1$ constant and such that $N^{\frac{1}{N-1}} \mathcal{V}^\intercal_1 \mathcal{V}_1= C_{A_{N{-}1}}$ and we define the pull-back momentum such that  $\mathcal{P}_{1*}=0$. One computes then 
\bea \widetilde{ \mathcal{P}}_* &\equiv& (N+1)^{ \frac{1}{N}}   \mathcal{V}_0^\intercal \mathcal{P}_* \mathcal{V}_0 \\
&=& \left( \begin{array}{cc} \, - \frac{1}{N} C_{A_{N{-}1}} \frac{\dd y}{y}\; & \; \frac{\sqrt{N+1} }{2y} C_{A_{N{-}1}} \dd x -  \frac{1}{N} C_{A_{N{-}1}}x_*  \frac{\dd y}{y} \\
\frac{\sqrt{N+1} }{2y}  \dd x^\intercal C_{A_{N{-}1}} - \frac{1}{N} x_*^\intercal C_{A_{N{-}1}}  \frac{\dd y}{y} \; &\; \frac{N-1}{N} \frac{ \dd y}{y} + \frac{\sqrt{N+1} }{y}  x_*^\intercal C_{A_{N{-}1}} \dd x\end{array} \right)\nonumber\; ,  \eea
where $x_{* i} = \frac{1}{N}$. This gives the standard representation components of $ \widetilde{ \mathcal{P}}_* $ that are defined from $\widetilde{\mathcal{P}}$ as in \eqref{qinp} and take the explicit form 
\be \mathcal{Q}_{i\ne N *} = -  \frac{\sqrt{N+1}}{N-1} \Bigl( C_{A_{N{-}1}} \frac{\dd x}{y} \Bigr)_i \; , \quad \mathcal{Q}_{N*} =   \Bigl( \frac{N+1}{N} \frac{ \dd y}{y} - \frac{\sqrt{N+1}}{N-1} x^\intercal_* C_{A_{N{-}1}} \frac{\dd x}{y} \Bigr) \; ,  \ee
and one checks that the $ (N-1,2)$ components $\widetilde{ \mathcal{P}}^\perp_* $  only depend on $\dd x$
\be \frac{y}{\sqrt{N+1}} \widetilde{ \mathcal{P}}^\perp = \left(\begin{array}{cc} - \frac{1}{N-1}  {\rm diag}[ C_{A_{N{-}1}} \dd x] \; &\; \tfrac12 C_{A_{N{-}1}} \dd x \\ \;   \tfrac12 \dd x^\intercal C_{A_{N{-}1}}  \; & - \frac{N}{N-1} x_*^\intercal C_{A_{N{-}1}} \dd x \end{array}\right) -\frac{1}{N-1} C_{A_N} ( x_*^\intercal C_{A_{N{-}1}} \dd x)  \; .  \ee
 A direct computation gives 
\bea C_{A_N}^{-1} [\mathcal{Q}_*] &=& \frac{N+1}{N-1}  \Bigl( \frac{N-1}{N} \frac{\dd y^2}{y^2} + \frac{1}{N-1} \frac{ C_{A_{N{-}1}}[\dd x]}{y^2} \Bigr)\; , \CR
2\tr C_{A_N}^{-1} \widetilde{ \mathcal{P}}^\perp_* C_{A_N}^{-1} \widetilde{ \mathcal{P}}^\perp_*  &=& \Bigl( N-\frac{2}{N-1} \Bigr)\frac{ C_{A_{N{-}1}}[\dd x]}{y^2}  \; . \eea
We define the pull-back of an automorphic function 
\be f_*(y,x) = f(y, \mathcal{V}_1 = \mathcal{V}_{1*},x_i = x) \; , \ee
to the two-dimensional subspace parametrised by $y$ and $x$. The second differential of the pull-back function must be consistent with the S$_{N+1}$ symmetry, and therefore 
\bea
&& \dd^2 f_*(y ,x)\Big|_{y = \sqrt{N+1},x = \frac{1}{N}} =f_{yy}\dd y^2+f_{xx}\dd x^2+2f_{yx}\dd y\dd x\nonumber\\
&=&a_+\frac{2}{N+1}\left(\frac{N-1}{N}\dd y^2+N \dd x^2\right)+a_-\frac{(N^2(N-1)-2N)}{N+1}\dd x^2\; , 
\eea
for two coefficients $a_+$ and $a_-$ determined by the second derivatives of the automorphic function at the symmetric point. In particular 
\be
f_{yy}=2\frac{N-1}{(N+1) N}a_+,\quad f_{xx}=\frac{2N}{N+1}a_++\frac{(N^2(N-1)-2N)}{N+1}a_-, \quad f_{yx}=0
\ee
Hence the condition for the symmetric point $A_N$ to be a local minimum of the automorphic function $f$ is that $a_+>0$ and $a_->0$, which is equivalent to the condition that  
\be
f_{xx}>\frac{N^2}{N-1}f_{yy}>0\; . \label{FminAN} 
\ee
It is therefore sufficient to study the pull-back function $f_*(y,x)$ to determine if the symmetric point $A_N$ is a local minimum of the automorphic function $f(y,\mathcal{V}_1,x_i)$ in $\mathcal{M}$. 

We have carried out this computation numerically for the Epstein series and found that the $A_N$ symmetric point is always a local minimum for $s$ large enough, but the eigenvalue $a_-$ becomes negative for small $s$ for $N\ge 4$, giving a saddle unstable along the corresponding $\frac{(N+1)(N-2)}{2}$-dimensional hypersurface. It is always a local minimum for $N=3$, and is in fact the global minimum for $s>\frac{3}{4}$ \cite{MinimaSL3}. We find that it is a local minimum for $s>1$ for $SL(4)$, and for $s\gtrsim 3.16603$ for $SL(5)$. For $SL(6)$ and $SL(7)$ one finds similarly that it is not a local minimum for $s< s_{A_N}$ with $s_{A_N}$ slightly above  the critical value $\frac{N}{2}$. Most importantly for us, it is not a local minimum of the Epstein function $\EpN{5}_{\frac{3}{2}}$ that defines the leading Wilson coefficient in type II string theory on $T^3$. 

\subsubsection{The $D_N$ symmetric point}

The $D_N$ symmetric point is realised with the ansatz \eqref{CNfromyx} as the bilinear form
\be
C_{D_N}[n]=2\sum_{i=1}^{N-1} n_i^2+4n_{N}^2+2 \sum_{i<j<N }n_in_j+4n_{N}\sum_{i=1}^{N-1} n_i\; .  
\ee
It is easy to show that it is related to the identity matrix by a $GL(N,\mathds{Q})$ transformation that determines 
\be
n_{i\ne N} = m_i\; , \qquad n_{N}=-\frac{1}{2}\sum_{i=1}^{N} m_i \; , \label{IdentityDN}
\ee
for $N$ integers $m_i$ such that $\sum_i m_i = 0 $ mod $2$. One then gets
\be C_{D_N}[m]=\sum_{i=1}^{N} m_i^2 \; ,
\ee
which reproduces the standard construction of the $D_N$ root lattice in Euclidean space. In this basis the automorphisms S$_N \ltimes \mathds{Z}_2^N $ of the lattice $D_N$ are simply realised as the $SL(N,\mathds{Z})$ transformations 
\begin{align}
    \sigma &: m_i \mapsto m_{\sigma(i)}\; , \nonumber\\
    \varpi_i &: m_i \mapsto -m_i \; , \quad m_{j \ne i} \mapsto m_j\; ,
\end{align}
that preserve the condition that $\sum_i m_i = 0 $ mod $2$. For $N=4$, one has moreover the triality automorphisms. This determines the action of the automorphism group on the momentum $p \in  \mathfrak{p}$ for generic $N$ as
\begin{align}
    \sigma &: p_{ij} \mapsto p_{\sigma(i)\sigma(j)}\; , \nonumber\\
    \varpi_i &: p_{ij} \mapsto - p_{ij}  \; , \quad p_{ii}\mapsto p_{ii} \; , \quad p_{jk} \mapsto p_{jk} \; , \quad \forall j,k \ne i  \; .
\end{align}
Note that the transformation $\prod_{i=1}^{N} \varpi_i$ flips the sign of the whole vector $\vec{m}$ and therefore acts trivially on  $p$, so that only ${\rm S}_N \ltimes \mathds{Z}^{N-1}_2  \subset PGL(N,\mathds{Z})$ acts on the moduli space. 

The components $p_{ij}$ split into the $N-1$ independent $p_{ii}$ that transform in the standard representation of ${\rm S}_N$ and are invariant under $\mathds{Z}_2^{N-1}$, and the $p_{ij}$ for $i\ne j$ that transform under the reducible representation $N\oplus (N-1,1)\oplus (N-2,2)$ of ${\rm S}_N$, but mix together under $\mathds{Z}_2^{N-1}$. One straightforwardly checks that the only invariant quadratic polynomials are 
\be \sum_{i=1}^N p_{ii}^2\; , \quad \sum_{i\ne j} p_{ij}^2\; . \ee
For $N=4$ the triality symmetry implies that there is a unique invariant quadratic polynomial, while for generic $N$ 
one straightforwardly computes that 
\bea \sum_{i=1}^N \mathcal{P}_{ii *}^2 &=& \tr \begin{pmatrix} -\frac{1}{N}\frac{{\rm d}y}{y_*}& 0 \\ 0 &\frac{N-1}{N}\frac{{\rm d}y}{_*y}\end{pmatrix}^2 = \frac{N-1}{N}\frac{{\rm d}y^2}{y^2_*}  \; , \CR
\sum_{i\ne j}  \mathcal{P}_{ij}^2 &=& \tr \begin{pmatrix} 0 &\frac{N^{\frac{N}{2(N-1)}}}{2y_*}\mathcal{V}_{1*}\,{\rm d}x\\\frac{N^{\frac{N}{2(N-1)}}}{2y_*}\mathcal{V}_{1*} \,{\rm d}x&0 \end{pmatrix}^2 = \frac{N}{2 y_*^2}C_{A_{N-1}}[\dd x] = \frac{N^2(N-1)}{2y_*^2} \dd x^2    \; . \quad \eea
Hence one has a local minimum of an automorphic function $f$ at the $D_N$ symmetric point if and only if 
\be
f_{xx}>0 \; , \qquad f_{yy}>0\; ,
\ee
with 
\bea
\dd^2 f_*(y ,x)\Big|_{y = 2,x = \frac{2}{N}} &=&f_{yy}\dd y^2+f_{xx}\dd x^2+2f_{yx}\dd y\dd x\nonumber\\
&=&a_+\frac{N-1}{N}\dd y^2 +a_-\frac{N^2(N-1)}{2}\dd x^2\; , 
\eea
and $a_+$ and $a_-$ the two eigenvalues of the Hessian of $f$ at the symmetric point. One finds by numerical evaluation  that the $D_N$ symmetric point is indeed a local minimum of the Epstein series $\Ep_s^N$ for all $s$ and $N\le 7$. We do not expect this to be true for arbitrary large $N$, but this is not relevant for the study of the type II string theory effective action. We find in particular that among all symmetric points, the $D_N$ symmetric point gives the minimum value of the Epstein series $\Ep_s^N$ for $N=4,5$ and $s \ge \frac{N}{4}$,\footnote{Note that $D_4^*$ and $D_4$ define the same unimodular matrix $H$, so the $D_4$ symmetric point is the lowest value for all $s>0$.} leading to the conjecture that $D_N$ is the global minimum. 

\subsubsection{Other symmetric points}
In this subsection we shall briefly describe the other symmetric points. 

\vskip 2mm

\noindent {\bf $\bullet$ The dual symmetric points}\\
Let us first observe that we do not need to describe the dual symmetric point for the lattices $A_N^*$ and $D_N^*$ separately. By construction $C_{L^*}= C_{L}^{-1}$, and the inverse of $H = \mathcal{V}^\intercal \mathcal{V}$ with the ansatz \eqref{VSLNCN} allows to describe all dual lattices. One easily extends all the results using the functional relation \eqref{FunctionalEpstein}. As a consistency check we have evaluated numerically the function for both the lattice and its dual. The analysis of the polynomials in $p \in \mathfrak{p}$ invariant under the automorphisms of the lattice proceeds in the same way for the dual lattices, and gives by construction the same number of invariant polynomials. 

For example for $A_N^*$ one defines 
\be \tilde{p} =(N+1)^{-\frac{1}{N}}  \mathcal{V}_0^{\intercal} p\mathcal{V}_0 \; , \ee 
for which the traceless condition reads 
\be \tr C_{A_N} \tilde{p} =0\; . \ee
The action of the Weyl group is then such that ${\rm S}_N$ acts by permutation of the indices and $\sigma_f$ acts on $\tilde{p}$ as
\be
\sigma_f : \tilde{p}_{i\ne N j\ne N}\mapsto \tilde{p}_{i+1\, j+1}\; , \quad
\sigma_f : \tilde{p}_{N j\ne N}\mapsto - \sum_{k=1}^N \tilde{p}_{N k}\; ,  \quad \sigma_f : \tilde{p}_{NN}\mapsto - \sum_{k=1}^N \tilde{p}_{kk}\; . \label{ANdual} \ee
It decomposes into the vector $q$ in the irreducible representation $(N,1)$  
\be q_i = \tilde{p}_{ii}\; , \ee
and the orthogonal component $\tilde{p}^\perp$ in $(N-1,2)$ that has zero diagonal entries $\tilde{p}_{ii}^\perp=0$ and
\be \tilde{p}_{ij}^\perp = \tilde{p}_{ij} + \frac{1}{N-1} \bigl( \tilde{p}_{ii} + \tilde{p}_{jj}\bigr) \; , \ee
for $i\ne j$.  One obtains then two independent quadratic polynomials in $q$ and $\tilde{p}^\perp$ respectively, such that 
\be \frac12  \kappa_{ab} p^a p^b=  \tr C_{A_N} \, \tilde{p} \, C_{A_N}\, \tilde{p}  = \frac{N-1}{N+1} C_{A_N}[q] + \tr C_{A_N} \tilde{p}^\perp C_{A_N} \tilde{p}^\perp\; .  \ee
For large values of $s$, the Epstein series evaluated at the dual symmetric points $D_N^*$ and $A_N^*$ is larger, as expected from the sphere packing density. By the functional relation \eqref{FunctionalEpstein} this must be the opposite for $s< \frac{N}{4}$, and in particular a symmetric point and its dual give the same value at $s=\frac{N}{4}$. Accordingly, there is always a value of $s_{\rm c} >\frac{N}{4}$ where $\Ep_{s_{\rm c}}^N(D_N^*) = \Ep_{s_{\rm c}}^N(A_N)$ and $\Ep_{s}^N(D_N^*) < \Ep_{s}^N(A_N)$ for $s<s_{\rm c}$. The numerical evaluation for $N\le 5$ shows that there is only one such transition, and that $\Ep_{s}^N(D_N^*) - \Ep_{s}^N(A_N)$ and $\Ep_{s}^N(D_N) - \Ep_{s}^N(D^*_N)$ have only one zero at $s=s_{\rm c}$ and $s=\frac{N}{4}$, respectively.

\vskip 2mm 
\noindent {\bf $\bullet$ The symmetric point $A_{5}^{+2}$}\\
The symmetric point $A_{5}^{+2}$ admits the same symmetry S$_6\subset SL(5,\mathds{Z})$ as  the point $A_5$ and its dual $A_5^*$. This is easy to prove using the relation
\be C_{\! A_{5}^{+2}} = \upgamma^\intercal C_{A_5}^{-1} \upgamma \ee
where $\upgamma$ the $GL(5,\mathds{Q})$ matrix defined such that 
\be \vec{m} = \upgamma \, \vec{n} \ee
is the change of basis \eqref{IdentityDN}. In particular $C_{D_5} = \upgamma^\intercal \upgamma$. One checks that for any  $g \in {\rm S}_6 \subset SL(5,\mathds{Z})$,  $\upgamma g \upgamma^{-1} \in SL(5,\mathds{Z})$, such that the automorphism group of $A_{5}^{+2}$ is conjugate to the one of $A_5^*$ in $SL(5,\mathds{Q})$  \cite{AutoL5}. It follows that the coset $p\in \mathfrak{p}$ decomposes into the two irreducible representations $(5,1)$ and $(4,2)$ of ${\rm S}_6$. To finds the explicit polynomial it is convenient to introduce  
\be \widetilde{\mathcal{P}}_* = \bigl( \tfrac{2}{3}\bigr)^{\frac{1}{5}} \upgamma^{-1} \mathcal{V}_*^\intercal \mathcal{P}_*  \mathcal{V}_* \upgamma^{-1 \intercal} \; , \ee
which transforms under ${\rm S}_6$ as $\tilde{p}$ in \eqref{ANdual}. Repeating the computations of Section \ref{ANsymmetricPoint} for the ansatz \eqref{VSLNCN},\eqref{F5fromyx} one obtains the following invariants polynomials 
\be  \frac{4}{3} C_{A_5} [\mathcal{Q}_*] =  \frac{12}{5} \frac{\dd y^2}{y^2} + 15 \dd x^2 \; , \qquad 2\tr C_{A_5} \widetilde{ \mathcal{P}}^\perp_* C_{A_5} \widetilde{ \mathcal{P}}^\perp_*  = 135 \dd x^2  \; .\ee
One obtains that the condition for the Hessian of the function $f$ to be positive definite is \eqref{FminAN} for $N=5$, as for the $A_5$ symmetric point. We have checked these conditions numerically and found that the symmetric point $A_{5}^{+2}$ is a local minimum of the Epstein series $\EpN{5}_s$ for $s\gtrsim 2.8849$. Below this value the second derivative with respect to $y$ is negative and the saddle is unstable along $5$ directions. 

In the critical strip $\frac{5}{4}<s<\frac{5}{2}$ we find that 
\be \EpN{5}_s(D_5) <  \EpN{5}_s(D_5^*) <\EpN{5}_s(A_{5}^{+2})< \EpN{5}_s(A_{5}^{+3})<\EpN{5}_s(A_5)<\EpN{5}_s(A_5^*) \; , \ee
and only $D_5$ and $D_5^*$ are local minima.

\vskip 2mm 

\noindent {\bf $\bullet$  The $E_6$, $E_7$, $E_8$ symmetric points}\\
We shall be very brief about these cases since they are not relevant to the analysis of the type II string theory low energy effective action. One finds in these three cases that the Weyl group of $E_N$ is large enough to impose that there is a unique invariant quadratic polynomial in $p$. It is therefore sufficient to check that the Epstein series is negative in the critical strip to ensure that the $E_N$ symmetric point is a local minimum for all values of $s$. One finds indeed by numerical evaluation that the $E_N$ symmetric points are local minima for all values of $s$, and among all symmetric points we have checked they give the smallest value of the Epstein series $\Ep_s^N$ for all values of $s\ge \frac{N}{4}$.\footnote{We have only checked symmetric points of Cartan type $A_N$, $D_N$, $E_N$ and their dual. This exhausts all maximal irreducible  symmetry groups for $N=7$, but not for $N=6,8$ \cite{AutoL5}.}
In the critical strip $\frac{N}{4}<s<\frac{N}{2}$ we find that 
\be \Ep_s^N(E_N) \underset{N=6,7}{<}  \Ep_s^N(E_N^*) <\Ep_s^N(D_N)\underset{N=6,7}{<}  \Ep_s^N(D_N^*)<\Ep_s^N(A_N)<\Ep_s^N(A_N^*) \; . \ee

\vskip 2mm 

\noindent {\bf $\bullet$  The $A_d \times A_{\frac{N}{d}}$ symmetric points for $d|N$}\\
For $N$ not prime one can also have symmetric points associated to irreducible bilinear form that are tensor products of lower dimensional bilinear forms. For example for $N=4$ one has 
\be C_{A_2\times A_2} = \left( \begin{array}{cccc} \, 4\, &\, 2\, &\, 2\, &\, 1\, \\ 2&4&1& 2\\ 2&1&4&2\\ 1&2&2&4\end{array}\right) \; . \ee
More generally the density of the sphere packing of $A_d \times A_{\frac{N}{d}}$ for $d$ dividing $N$ is 
\be \rho(C_{A_d\times A_{\frac{N}{d}}}) = \frac{ \pi^{\frac{N}{2}} }{(1+d)^{\frac{N}{2d}} ( 1+\frac{N}{d})^{\frac{d}{2}} \Gamma(\frac{N}{2}+1)} \; . \ee
For $N\le 8$ the possible lattices are less dense than $E_N$, $D_N$ and $A_N$, but one checks that in the critical strip 
\be \EpN{4}_s(D_4) < \EpN{4}_s(A_2\times A_2)<  \EpN{4}_s(A_4) \; . \ee
So the ordering is reversed with the $A_4$ lattice at low values of $s$. Nevertheless, the symmetric point $D_4$ remain the global minimum of $ \EpN{4}_s$ for all value of $s>0$.

\subsection{$SO(N,N)$ symmetric points}

We now consider the symmetric space $(SO(N)\times SO(N))\backslash SO(N,N)$ with the discrete group $SO(N,N,\mathds{Z})$ preserving the split signature even  lattice $\II_{N,N}$. We introduce the vector representation Epstein series
\be \Ep_s^{N,N}(H) = \sum^\prime_{\substack{q \in \II_{N,N}\\ (q,q) = 0}} \frac{1}{H[q]^s} = 2\zeta(2s) E^{\scalebox{0.6}{$SO(N,N)$}}_{s\Lambda_1} \; , \label{EpNN} \ee
which is proportional to the maximal parabolic Eisenstein series of infinitesimal character $\lambda=2s\Lambda_1-\rho$. It is absolutely convergent for Re$[s]> N-1$, and admits an analytic continuation to a meromorphic function of $s\in \mathds{C}$. We will be interested in particular in the value $s = \frac{N-2}{2}$, for which the Epstein series is in the minimal representation.\footnote{This can be understood from the Langlands functional identity $\xi(N-2) E^{\scalebox{0.6}{$SO(N,N)$}}_{\frac{N-2}{2} \Lambda_1} = \xi(2) E^{\scalebox{0.6}{$SO(N,N)$}}_{ \Lambda_N}= \xi(2) E^{\scalebox{0.6}{$SO(N,N)$}}_{ \Lambda_{N-1}}$ for $N\ge 3$ which shows that this function can be realised either as a vector representation Epstein series or a spinor representation Epstein series.}  It will be convenient to consider the representation of the symmetric $SO(N,N)$ matrix $H$ in the $P_N = GL(N,\mathds{R})\ltimes \wedge^2 \mathds{R}^N$ parabolic subgroup such that for any vector $q = (m,n)$ with $m,n\in \mathds{Z}^N$, 
\be
H[q]=G^{-1}[m+Bn]+G[n]\; , 
\ee
and the split signature bilinear form is
\be (q,q) = 2 m \cdot n \; . \ee
In type II string theory on $T^d$ the $SO(d,d)$ symmetric matrix $H$ is parametrised by the Narain moduli, with the torus metric $G$ and the Kalb--Ramond two-form $B$ in string length. Then $m$ is the vector of Kaluza--Klein modes and $n$ the vector of winding numbers. For the U-duality group $Spin(5,5,\mathds{Z})$ one can view $G$ as the M-theory metric on $T^5$ and $B$ as the Hodge dual of the three-form potential on $T^5$.

To define symmetric points in $SO(N,N)$ we start with the assumption that $G$ is itself a symmetric point of $SO(N)\backslash GL(N)$. From the previous section we find that $G$ is then proportional to the Gram matrix $C_L$ of an even  lattice $L$, and the relevant solution will turn out to be
\be G = \frac{1}{2} C_L \; , \qquad G+B  = 0 \; {\rm mod} \; \mathds{Z}\; . \label{LatticeHAnsatz} \ee
One finds in this case that the charges $q_\pm$ 
\be  q_\pm = \frac{1}{2} G^{-1} \bigl( m + (B \pm G) n \bigr)  \; , \label{LeftRightProject} \ee
satisfy 
\be q_+-q_- \in \mathds{Z}^N\; , \ee
while $q_\pm$ are by construction in the dual lattice $L^*= C_L^{-1} \mathds{Z}^N$. It follows that this bilinear form describes the isomorphism 
\be \II_{N,N} \cong \bigoplus_{\mu \in L^* \! / L } \Bigl( ( L + \mu ) \oplus ( L + \mu)[-1] \Bigr) \; , \ee
or equivalently for the Narain theta series 
\be \sum_{q\in \II_{\scalebox{0.5}{$N,N$}}} e^{- \pi \tau_2 H[q] +  i \pi \tau_1 (q,q)} = \sum_{\mu \in L^* \! / L }\;  \biggl| \sum_{n\in \mathds{Z}^N} e^{ i \pi  \tau C_L[n+\mu ]} \biggr|^2 \; . \ee
For any pair of elements $\gamma_\pm \in {\rm Aut}(L)$ such that 
\be \gamma_+ \mu = \gamma_- \mu \; {\rm mod} \; L\; , \label{gammapm} \ee
one has an automorphism $O(N,N,\mathds{Z})$ of the split signature lattice $\II_{N,N}$ that preserves the symmetric matrix $H$.  For Cartan type lattices $A_N,D_N,E_N$, the automorphism group is the product of the outer automorphisms of the Dynkin diagram and the Weyl group
\be {\rm Aut}(L) = {\rm Out}(L) \ltimes W(L)\; , \ee
and the Weyl group $W(L)$ preserves all weights $\mu \in L^*/L$, so one has the automorphism group 
\be {\rm Aut}(H) = {\rm Out}(L)  \ltimes W(L) \times W(L) \; . \label{AutH}  \ee
The simplest example of a symmetric point is for $N=1$, in which case $G = y^2$ and 
\be  \Ep_s^{1,1}(H) =2 \zeta(2s) \bigl( y^{2s} + y^{-2s}  \bigr) \; . \ee
Its global minimum at $y=1$ can be interpreted as the $SU(2)$ self-dual radius in string theory on $S^1$. The $A,D,E$ points described above correspond more generally to the points of enhanced gauge symmetry in string theory. For $N=2$ one recovers the minimum of the product of two $SL(2)$ Epstein series at $T = U = \frac12 + i \frac{  \sqrt{3}}{2}$ for the symmetric matrix $H$ associated to the lattice  $L=A_2$. We will see below that for $N=3$ the symmetric matrix associated to $L=A_3$ also reproduces the minimum of the $SL(4)$ Epstein series at the symmetric point $D_4$. 

Assuming $H$ to be determined by a lattice Gram matrix as in \eqref{LatticeHAnsatz}, one can compute the limit at large $s$ from the theta series 
\bea  \Ep_s^{N,N}(H)  &=& \frac{\pi^s}{\Gamma(s)} \int_0^\infty \frac{d\tau_2}{\tau_2} \tau_2^s \int_{-\frac12}^{\frac12} d\tau_1  \sum^\prime_{q\in \II_{\scalebox{0.5}{$N,N$}}} e^{- \pi \tau_2 H[q] +  i \pi \tau_1 (q,q)} \CR
&= & \sum_{\mu \in L^*\! / L} \sum^\prime_{\substack{m,n\in \mathds{Z}^N\\ C_L[n+\mu] = C_L[m+\mu]}} \frac{1}{(2 C_L[n+\mu])^s} \; ,  \eea
where the prime removes the point $m=n=0$ for $\mu=0$ only. At large $s$, the leading term in $\Ep_s^{N,N}(H)  $ is proportional to the smallest length of a vector in $L^*$ to the power $-2s$, and one obtains a minimum for the even lattice with the largest possible minimal length of  a vector in $L^*$. Although this is not exactly the same criterion as for the densest lattice sphere packing, it gives the same $ADE$ classification \eqref{ADEbest} for $N\le 8$. 

For small values of $s$ we need instead to use the Fourier expansion of the Epstein series \footnote{In our convention the $SL(N)$ Epstein series only depend on the unimodular bilinear form so that $\Ep^N_s(G) = \Ep^N_s( \det G^{- \frac{1}{N}} G))$, and in particular  $\det G^{\frac{s}{N}} \Ep^N_s(G^{-1})  = \sum_{n\in \mathds{Z}^N}^\prime \frac{1}{G^{-1}[n]^s}$.}
\begin{multline} \Ep_s^{N,N}(H)   = \det G^{\frac{s}{N}} \Ep^N_s(G^{-1}) + \frac{ \pi^{\frac{N-1}{2}} \Gamma(s - \frac{N-1}{2})}{\Gamma(s) } \frac{\zeta(2s-N+1)}{\zeta(2s-N+2)} \det G^{\frac{ N-1-s}{N}} \Ep^N_{s-\frac{N-2}{2}}(G) \label{VectorEpsteinExpand} \\
+ \frac{4\pi^s}{\Gamma(s)} \sqrt{\det G} \sum^\prime_{\substack{Q\in \wedge^2 \mathds{Z}^N\\ Q\wedge Q=0}}  \frac{ \sigma_{N-1-2s}(Q)}{{\rm gcd}(Q)^{\frac{N-2}{2}-s}} E^{\scalebox{0.6}{$SL(2)$}}_{s-\frac{N-2}{2}}(U_Q) \frac{K_{s- \frac{N-1}{2}}(2\pi \sqrt{G[Q]})}{\sqrt[4]{G[Q]}}  e^{\pi i \tr B Q} \end{multline} 
where the sum is over all non-zero rank one antisymmetric integer matrices $Q$, the bilinear form is defined as 
\be G[Q] = - \frac12 \tr G Q G Q \; ,\ee
and $E^{\scalebox{0.6}{$SL(2)$}}_{s}(U_Q)$ is the $SL(2)$ real analytic Eisenstein series evaluated on the $SL(2)$ subgroup of the stabiliser of $Q$.\footnote{Normalised such that $\EpN{2}_s(U) = 2 \zeta(2s) E^{\scalebox{0.6}{$SL(2)$}}_{s}(U)$.} Note that the stabiliser of $Q$ in $SL(N)$ is $SL(2)\times SL(N-2)\ltimes \mathds{R}^{2\times (N-2)}$. 

For the value $s=\frac{N-2}{2}$, this expression simplifies drastically to 
\be \Ep_{\frac{N-2}{2}}^{N,N}(H)   = \det G^{\frac{N-2}{2N}} \Ep^N_{\frac{N-2}{2}}(G^{-1}) + \frac{ \pi^{\frac{N}{2}}  \sqrt{ \det G }  }{\Gamma(\frac{N-2}{2}) }   \Biggl( \frac{1}{3} + 2\hspace{-1.3mm} \sum^\prime_{\substack{Q\in \wedge^2 \mathds{Z}^N\\ Q\wedge Q=0}}  \sigma_{1}(Q) \frac{e^{- 2\pi \sqrt{ G[Q]}} }{\pi \sqrt{G[Q]}}  e^{\pi i \tr B Q} \Biggr) \; . \label{EpSONNMin} \ee 
We will use this expression to evaluate numerically the Epstein series for $N=5$.

We find evidence that the Epstein series in the vector representation \eqref{EpNN} admits its global minimum for all $s>0$ at the symmetric point where $G = \frac{1}{2} C_L$ and $G + B = 0$ mod $\mathds{Z}$ for the Cartan type best packing lattices \eqref{ADEbest}.

\subsubsection{Symmetric points as minima}
In order to describe the symmetries of the polynomials in the coset derivatives we need to define a coset representative $\mathcal{V}\in SO(N,N)$. We introduce two vielbeins basis $V_\pm$ for the same metric $G  =V^\intercal_\pm V_\pm$ such that $\mathcal{V}$ reads 
\be \mathcal{V} = \left(\begin{array}{cc} \frac{1}{\sqrt{2}} V^{-1\intercal}_+ &  \frac{1}{\sqrt{2}} V^{-1\intercal}_+ (G+B) \\  \frac{1}{\sqrt{2}} V^{-1\intercal}_-  &  \frac{1}{\sqrt{2}} V^{-1\intercal}_- (-G+B)  \end{array}\right) \; .\label{SOcoset}  \ee
It transforms under left action of   $k_\pm \in SO(N)$ and right action of $\gamma \in O(N,N,\mathds{Z})$ as  
\be \mathcal{V} \rightarrow \left(\begin{array}{cc} k_+ & 0 \\ 0 & k_-   \end{array}\right) \mathcal{V}  \gamma\; . \ee 
One can of course set $V_+ = V_- = V$, but it is convenient to keep them different to make manifest the covariance under $SO(N)\times SO(N)$. 
The coset differential is then 
\be\frac12 \dd \mathcal{V} \mathcal{V}^{-1} + \frac12  \mathcal{V}^{-1\intercal } \dd \mathcal{V}^\intercal =  \left(\begin{array}{cc} 0 & \mathcal{P}^\intercal \\ \mathcal{P}  & 0 \end{array}\right) \ee
with 
\be \mathcal{P} =- \frac12 V_-^{-1\intercal} \bigl( \dd G  +\dd B\bigr)  V_+^{-1} \; . \ee
We now consider a point \eqref{LatticeHAnsatz} for $L$ of Cartan type, that we write 
\be \mathcal{V}_0 =  \left(\begin{array}{cc} \sqrt{2} V_{0 +}& 0 \\ 0 &\sqrt{2} V_{0-}    \end{array}\right) \left(\begin{array}{cc} C_L^{-1}   &C_L^{-1}   (G+B)_0 \\  C_L^{-1}    &  C_L^{-1}  (-G+B)_0  \end{array}\right) \; . \ee
Using the charges $q_\pm $ that transform under the two copies of the Weyl group at a symmetric point \eqref{LeftRightProject}, one obtains that 
\be \mathcal{V}_0 Q = \left(\begin{array}{c}\sqrt{2} V_{0+} q_+  \\  \sqrt{2} V_{0-} q_- \end{array}\right)\; .  \ee
It follows that under an automorphism in \eqref{AutH} realised as a $\gamma\in O(N,N,\mathds{Z})$ one has 
\be \mathcal{V}_0 \gamma = \left(\begin{array}{cc} k_+(\gamma) & 0 \\ 0 & k_-(\gamma)   \end{array}\right) \mathcal{V}_0 =  \left(\begin{array}{cc} \sqrt{2} V_{0+}  \gamma_+ & 0 \\ 0 &\sqrt{2} V_{0-} \gamma_-     \end{array}\right)  \left(\begin{array}{cc} C_L^{-1}   &C_L^{-1}   (G+B)_0 \\  C_L^{-1}    &  C_L^{-1}  (-G+B)_0  \end{array}\right)  \ee
where the $\gamma_\pm $ satisfy \eqref{gammapm}. In particular, the $\gamma_\pm$ can be two independent elements of the Weyl group of the Cartan type lattice $L$. 

It appears therefore convenient to expand the automorphic function of interest with respect to the pull-back momentum  
\be \mathcal{P}_* =-  \bigl( \dd G  +\dd B\bigr)_*\; . \ee
Similarly as in section \ref{ANsymmetricPoint}, we introduce the variable 
\be \tilde{p} = 2 V_{0-}^\intercal  {p} V_{0+}  \; .\ee 
For any  invariant polynomial  $f_{H_0}(p)$ of $p\in \mathfrak{so}(N,N) \ominus (\mathfrak{so}(N)\oplus \mathfrak{so}(N))$, one defines the polynomial 
\be
\tilde f_{\mathcal{V}_0}(\tilde p)=f_{\mathcal{V}_0 }\bigl(  2V_{0-}^\intercal  {p} V_{0+}  \bigr)\; ,  \label{tildepPolSO} 
\ee
that is invariant under 
\be \tilde f_{\mathcal{V}_0}(\gamma_- \tilde p \gamma_+) = \tilde f_{\mathcal{V}_0}(\tilde p ) \; .  \ee
One then finds for all irreducible Cartan type lattices $L$, that the unique invariant quadratic polynomial is the Killing--Cartan form 
\be 2 \tr p {p}^\intercal  = 2 \tr C_L^{-1} \tilde{p}\,  C_L^{-1} \tilde{p}^\intercal \; . \label{KCSO} \ee
Let us prove this for $A_N$ and $D_N$.
\vskip 2mm

\noindent {\bf $\bullet$ $A_N$ symmetric point}\\
For $A_N$, the momentum $\tilde{p}$ transforms in the tensor product of two standard representations $(N,1)$ associated to the two copies of S$_{N+1}$. It is therefore in an irreducible representation of S$_{N+1} \times $ S$_{N+1}$ and there is obviously no linear  invariant. Introducing the indices $i=1$ to $N$ for one side and $\hat{\imath} = 1 $ to $N$ for the other, one obtains the  action of S$_{N} \times $ S$_{N}$
\be
\sigma \times \hat{\sigma} : \tilde{p}_{i\hat{\jmath}}\mapsto \tilde{p}_{\sigma^{-1}(i)\hat{\sigma}^{ -1}(\hat{\jmath})}\; , 
\ee
while the left $\sigma_f$ acts as
\be
\sigma_f \times 1 : \tilde{p}_{1\hat{\jmath}}\mapsto \tilde{p}_{N\hat{\jmath}}\; , \qquad  \sigma_f \times 1 : \tilde{p}_{i\neq 1 \hat{\jmath}}\mapsto \tilde{p}_{N\hat{\jmath}}-\tilde{p}_{i-1\hat{\jmath}}\; , 
\ee
and identically for the right  $\hat{\sigma}_f$. For the quadratic polynomial one can use the decomposition into irreducible representations of the tensor product of two standard representations 
\be (N,1) \otimes (N,1) = (N) \oplus (N,1)\oplus(N-1,2) \oplus (N-1,1,1) \; , \ee
to conclude that the quadratic polynomials in $\tilde{p}$ decompose into the irreducible representations of S$_{N+1} \times $ S$_{N+1}$  
\be  \bigl[ (N+1) \oplus (N,1)\oplus(N-1,2) \bigr] \otimes  \bigl[ (N+1) \oplus (N,1)\oplus(N-1,2) \bigr] \oplus (N-1,1,1) \otimes (N-1,1,1) \; . \ee
It follows that there is single quadratic invariant polynomial \eqref{KCSO}. According to the discussion of section \ref{TaylorExpansion}, we conclude that the $A_N$ symmetric point is a local minimum of any $SO(N,N)$ Eisenstein series in the domain of absolute convergence. We have checked numerically that $\Ep_s^{5,5}(H)$ is negative at the critical value $s = \frac{3}{2}$ relevant in the string theory effective action, and the symmetric point $A_5$ is therefore a local minimum. 

\vskip 2mm

\noindent {\bf $\bullet$ $D_N$ symmetric point}\\
For $D_N$, the momentum ${p}$ transforms in the tensor product of two vector representations of  S$_N \ltimes \mathds{Z}^{N-1}_2 $. It is convenient to define the action of the two copies of S$_N \ltimes \mathds{Z}^{N}_2 $
\begin{align}
    \sigma \times \hat{\sigma} &: p_{i\hat{\jmath}} \mapsto p_{\sigma(i)\hat{\sigma}(\hat{\jmath})}\; , \nonumber\\
    \varpi_i \times 1 &: p_{i\hat{\jmath}} \mapsto - p_{i\hat{\jmath}}  \; , \quad p_{j\hat{\jmath}}\mapsto p_{j\hat{\jmath}}  \; , \quad \forall j \ne i  \; ,
\end{align}
and  identically for the right  $\hat{\varpi}_{\hat{\jmath}}$, and then take the subgroup of elements with an even number of $  \varpi_i $ and $\hat{\varpi}_{\hat{\jmath}}$. It follows directly that the only invariant quadratic polynomial is 
\be \sum_{i=1}^N \sum_{\hat{\jmath}=1}^N p_{i\hat{\jmath}} p_{i\hat{\jmath}} = \tr p p^\intercal \; , \ee
which is the Killing--Cartan form \eqref{KCSO}. One finds therefore that the $D_N$ symmetric point is a local minimum of  any $SO(N,N)$ Eisenstein series in the domain of absolute convergence. We have checked numerically that $\Ep_s^{5,5}(H)$ is negative at the critical value $s = \frac{3}{2}$ relevant in the string theory effective action, and the symmetric point $D_5$ is therefore a local minimum. It is a lower value than the $A_5$ symmetric point and we conjecture that it is the global minimum of the minimal Epstein series.

\subsubsection{$SO(3,3)$ and $SL(4)$}
Because of the homomorphism Spin$_0(3,3)  = SL(4)$ it is relevant to compare the results we have obtained for $SO(N,N)$ and $SL(N)$ in this case. It appears that the vector representation Epstein series of $SO(3,3)$ and the Epstein series of $SL(4)$ are related at the special $s$ value
\be \pi \Ep_{\frac12}^{3,3}(H) = \EpN{4}_1(H)\; , \label{EpsSO33SL4} \ee
with the identification of the $SL(4)$ matrix 
\be
\mathcal{V}= y^{-\frac{1}{4}}\left(\begin{array}{cc} \; 2^{\frac1{3}} \mathcal{V}_1 \; & \; 2^{\frac1{3}}  \mathcal{V}_1 x\; \\ \; 0\; & \; \frac{y}{2} \end{array}\right)= \frac{1}{\sqrt{\det \! V}} \left(\begin{array}{cc} \; V  \; & \;   V  x\; \\ \; 0\; & \; \det \! V   \end{array}\right) \; , \ee
and $V_+=V_-=V$ and $B_{ij} = \varepsilon_{ijk} x_k$ in \eqref{SOcoset}. In particular for \eqref{CNfromyx} one has $G = \frac{y}{4} C_{A_3}$, and one can check the functional relation \eqref{EpsSO33SL4} using \eqref{EpSONNMin} and \eqref{EpsteinExpansion}
\bea  \pi \Ep_{\frac12}^{3,3}(H)   &=& \pi \det G^{\frac{1}{6}} \EpN{3}_{\frac12}(G^{-1}) +  \pi^2 \sqrt{ \det G }   \Biggl( \frac{1}{3} + 2\hspace{-1.3mm} \sum^\prime_{Q\in \wedge^2 \mathds{Z}^3}  \sigma_{1}(Q) \frac{e^{- 2\pi \sqrt{ G[Q]}} }{\pi \sqrt{G[Q]}}  e^{\pi i \tr B Q} \Biggr) \CR
&=& \frac{y^{\frac{1}{2}}}{4^{\frac{1}{3}}}  \EpN{3}_1(C_{A_{3}}) + \frac{\pi^2}{3} \frac{y^{\frac32} }{4} +  \pi \sqrt{y} \sum_{n \in \mathds{Z}^{3}}^\prime   \sigma_{1}(n) \frac{e^{-\pi y  \sqrt{  C_{A_{3}}^{-1}[n]  }}}{\sqrt{ C_{A_{3}}^{-1}[n] } }  \cos\Bigl( 2\pi x \cdot n \Bigr) 
\; . \eea 
For the special points \eqref{CNfromyx} of the $SL(4)$ Epstein series we have 
\be G = \frac{y}{4} \begin{pmatrix}2\; &\, 1\, &\; 1 \\ 1\;  &\, 2\,  &\; 1 \\ 1 \; &\, 1\,  &\; 2 \end{pmatrix}\; ,\quad B=x \begin{pmatrix}0&1&-1 \\-1&0&1 \\ 1 &-1 &0\end{pmatrix} \; , \ee 
so that the $SL(4)$ $D_4$ symmetric point corresponds to the $SO(3,3)$ $A_3$ symmetric point, 
\be G = \frac{1}{2}  C_{A_3} \; ,\quad B+G = \begin{pmatrix}1\; &\, 1\, &\; 0 \\  0\;  &\, 1\, &\; 1 \\ 1\;  &\,  0\,   &\; 1\end{pmatrix} \; , \ee 
while the $SL(4)$ $A_4$ symmetric point gives 
\be G = \frac{\sqrt{5}}{4} \begin{pmatrix}2\; &\, 1\, &\; 1 \\ 1\;  &\, 2\,  &\; 1 \\ 1 \; &\, 1\,  &\; 2 \end{pmatrix}\; ,\quad B=\frac{1}{4} \begin{pmatrix}0&1&-1 \\-1&0&1 \\ 1 &-1 &0\end{pmatrix} \; . \ee 
We find consistently that the conjectured global minimum of the $SL(4)$ Epstein series  at the $D_4$ symmetric point agrees with the conjectured global minimum of the $SO(3,3)$ Epstein series at the $A_3$ symmetric point. One finds that they have the same automorphism groups 
\be
\text{Aut}(\II_{3,3}^{A_3})=\mathds{Z}_2\ltimes({\rm S}_4\times {\rm S}_4) \cong 
\text{Aut}(D_4)={\rm S}_3\ltimes({\rm S}_4 \ltimes \mathds{Z}^3_2 )\; ,
\ee
because of the triality automorphism of $D_4$, explaining that there is a unique invariant quadratic polynomial in this case. 

\section{Fixing the logarithmic ambiguity in eight dimensions}
\label{Renormalisation}
The different terms in the low energy expansion of the two-graviton amplitude \eqref{LowEnergyExpansion} are not well defined individually in eight dimensions because the 1-loop box integral 
\be I_4(s,t) = \int\frac{d^{D}p}{(2\pi)^{D}}\frac{1}{p^2(p-k_1)^2(p-k_1-k_2)^2(p+k_4)^2} \ee
diverges logarithmically and so do the Epstein series defining $\mathcal{E}_\gra{0}{0}(\varphi)$. To determine unambiguously the amplitude we must analyse the low energy limit of the one-loop string theory amplitude \cite{Green:1981yb}
\be \mathcal{A}^{\scalebox{0.6}{1-loop}} = 2\pi \alpha^{\prime \, 3} g_{\scalebox{0.6}{$8$}}^{\; 2} \int_{\mathcal{F}} \frac{{\rm d}^2\tau}{\tau_2^{\, 2}} \Gamma_{\II_{2,2}}\prod_{a=1}^4 \int_\Sigma\hspace{-1.8mm}  \frac{{\rm d}^2z_a}{\tau_2}   \tau_2 \delta^\ord{2}(z_4) e^{-\frac{\alpha^\prime}{2} \! \sum_{a>b}\! {G}(z_a-z_b) k_a \cdot  k_b} \label{1loopstring}\; , \ee
where 
\be
\Gamma_{\II_{d,d}}(\tau)=\tau_2^\frac{d}{2}\sum_{q\in \II_{\scalebox{0.5}{$d,d$}}} e^{- \pi \tau_2 H[q] +  i \pi \tau_1 (q,q)}\; ,
\ee
and $G(x)$ is the torus Green function. Following \cite{Green:1999pv}, we split the $SL(2)$ fundamental domain into the truncated fundamental domain 
 \be \mathcal{F}_L  = \bigl\{ \tau_2 < L, \, - \tfrac12 \le \tau_1 \le \tfrac12 ,\,  |\tau| > 1 \bigr\} \ee 
and the complementary region for which $\tau_2>L$ and  $ - \tfrac12 \le \tau_1 \le \tfrac12$. At leading order one obtains 
\begin{multline} \frac{\mathcal{A}^{\scalebox{0.6}{1-loop}}}{\lP^{\; 6}} = 2\pi \int_{\mathcal{F}_L} \hspace{-1.3mm} \frac{{\rm d}^2\tau}{\tau_2^{\, 2}} \Gamma_{\II_{2,2}}  + 4\pi \int_{L}^\infty  \hspace{-0.5mm} \frac{{\rm d}\tau_2}{\tau_2}   \int_0^1 \hspace{-2mm} {\rm d}x_3 \int_0^{x_3} \hspace{-3mm} {\rm d}x_2\int_0^{x_2} \hspace{-3mm} {\rm d}x_1    e^{\pi \tau_2   \alpha^\prime[ (x_2-x_1)(1-x_3) s + x_1 (x_3-x_2) t]}  + \circlearrowleft \\+ \mathcal{O}(s^2) \; ,\end{multline}
where $\circlearrowleft$ represents the two cyclic permutations of the Mandelstam variables. To compute the two terms separately it is more convenient to introduce a dimensional regularisation $d = 2 +2 \epsilon$. This can be achieved with the insertion of $\tau_2^\epsilon$ in both terms. One can then remove the $L$-dependent terms that cancel out and use instead 
\be \lim_{\epsilon\rightarrow 0_+} \Bigl( 2\pi \int_{\mathcal{F}} \hspace{-1.3mm} \frac{{\rm d}^2\tau}{\tau_2^{\, 2}} \tau_2^\epsilon \Gamma_{\II_{2,2}}  + 4\pi \int_{0}^\infty  \hspace{-0.5mm} \frac{{\rm d}\tau_2}{\tau_2} \tau_2^\epsilon  \int_0^1 \hspace{-2mm} {\rm d}x_3 \int_0^{x_3} \hspace{-3mm} {\rm d}x_2\int_0^{x_2} \hspace{-3mm} {\rm d}x_1    e^{\pi \tau_2   \alpha^\prime[ (x_2-x_1)(1-x_3) s + x_1 (x_3-x_2) t]}+ \circlearrowleft  \Bigr) \ee
For the one-loop supergravity amplitude one gets the dimensional regularisation
\bea && (4\pi)^5  (4\pi^2 \alpha')^{-\epsilon}   \int\frac{d^{8-2\epsilon}p}{(2\pi)^{8-2\epsilon}}\frac{1}{p^2(p-k_1)^2(p-k_1-k_2)^2(p+k_4)^2}\CR
&=& 4\pi \int_{0}^\infty  \hspace{-0.5mm} \frac{{\rm d}\tau_2}{\tau_2}  \tau_2^\epsilon \int_0^1 \hspace{-2mm} {\rm d}x_3 \int_0^{x_3} \hspace{-3mm} {\rm d}x_2\int_0^{x_2} \hspace{-3mm} {\rm d}x_1    e^{\pi \tau_2   \alpha^\prime[ (x_2-x_1)(1-x_3) s + x_1 (x_3-x_2) t]}   \CR
&=& -8\pi^{1-\epsilon}\frac{\Gamma( \epsilon-2)\Gamma(3-\epsilon)^2}{\Gamma(5-2\epsilon)}\int_0^1dx\left((-\alpha's)^{-\epsilon} \frac{(1-x)^{1-\epsilon}}{\left(1+\frac{t}{s}\right)-1}+(-\alpha't)^{-\epsilon}  \frac{(1-x)^{1-\epsilon} }{\left(1+\frac{s}{t}\right)-1}\right) \CR
&=& \frac{2\pi}{3} \left(\frac{1}{\epsilon} + \frac{11}{3} - \gamma_{\scalebox{0.6}{E}} - {\rm ln} \, \pi -\frac{s}{s+t}\ln{\left(-\alpha^\prime s\right)}- \frac{t}{s+t}\ln{\left(-\alpha^\prime t\right)}-\frac12 \frac{s t}{(s+t)^2}\Bigl(\ln{\Bigl(\frac{t}{s}\Bigr)}^2+\pi^2\Bigr)\right)\CR
&& + \mathcal{O}(\epsilon) \; . \eea
We define accordingly the renormalised 1-loop box integral in $D=8$ dimensions 
\be \hat I_{4,\mu}(s,t) =- \frac{1}{6 (4\pi)^4}\left( \frac{s}{s+t}\ln{\left(-  s/\mu^2\right)}+ \frac{t}{s+t}\ln{\left(-t/\mu^2 \right)}+\frac12 \frac{s t}{(s+t)^2}\Bigl(\ln{\Bigl(\frac{t}{s}\Bigr)}^2+\pi^2\Bigr)\right)\; ,\label{RenI4}  \ee
where we introduced a renomalisation scale $\mu$. Note that $\hat{I}_{4,\mu}(s,t) = \hat{I}_{4,1}(s/\mu^2,t/\mu^2)$.

For the perturbative Wilson coefficient one can replace $\tau_2^\epsilon$ by the real analytic Eisenstein series $E_\epsilon(\tau)$ without modifying the limit. One computes as in   \cite{Green:2010wi} that 
\bea && 2\pi \int_{\mathcal{F}} \hspace{-1.3mm} \frac{{\rm d}^2\tau}{\tau_2^{\, 2}} E_\epsilon(\tau) \Gamma_{\II_{2,2}}  \CR
&=& 4\pi \xi(2\epsilon) E_\epsilon(T) E_\epsilon(U)\CR
&=&  - \frac{2\pi}{\epsilon} + 2\pi \bigl( \gamma_{\scalebox{0.6}{E}}- {\rm ln}(4\pi) \bigr) - 2\pi\,  {\rm ln}(U_2\lvert\eta(U)\lvert^4) - 2\pi\,  {\rm ln}(T_2\lvert\eta(T)\lvert^4)+ \mathcal{O}(\epsilon)\; ,  \eea
where $\eta$ is the Dedekind eta function. One obtains in total 
\bea  \frac{\mathcal{A}^{\scalebox{0.6}{1-loop}}}{\lP^{\; 6}} &=& - 2\pi\,  {\rm ln}(U_2\lvert\eta(U)\lvert^4) - 2\pi\,  {\rm ln}(T_2\lvert\eta(T)\lvert^4)+\frac{22\pi}{3}- 4\pi\ln(2\pi)\CR
&& +(4\pi)^5(\hat I_{4,1}(\alpha's,\alpha't)+\hat I_{4,1}(\alpha't,\alpha'u)+\hat I_{4,1}(\alpha'u,\alpha's)) + \mathcal{O}(s^2) \CR
&=& - 2\pi\,  {\rm ln}(U_2\lvert\eta(U)\lvert^4) - 2\pi\,  {\rm ln}(T_2\lvert\eta(T)\lvert^4) + \frac{4\pi}{3} {\rm ln}\, g_{\scalebox{0.6}{$8$}} +\frac{22\pi}{3}- 4\pi\ln(2\pi \lP\mu ) \CR
&& +(4\pi)^5\bigl(\hat I_{4,\mu}(s, t)+\hat I_{4,\mu}(t, u)+\hat I_{4,\mu}(u, s)\bigr) + \mathcal{O}(s^2) \eea
where we used $\alpha^{\prime\, 3} g_{\scalebox{0.6}{$8$}}^{\; 2} = \lP^{\; 6}$ in the last step.

We  use the same convention as in \cite{Green:2010wi} for the renormalised Eisenstein series, such that we define the $SL(2)$ Epstein series 
\be
\Epreg^{\!\! 2}_1(U)=\lim_{\epsilon\rightarrow0}\left(\EpN{2}_{1+\epsilon}(U)-\frac{\pi}{\epsilon}-2\pi(\gamma_{\scalebox{0.6}{E}}-\ln{2})\right)=-\pi\ln{(U_2\lvert\eta(U)\lvert^4)}\; , 
\ee
and the $SL(3)$ Epstein series 
\bea
\Epreg^{\!\!3}_\frac{3}{2}(H)&=&\lim_{\epsilon\rightarrow0}\left(\EpN{3}_{\frac{3}{2}+\epsilon}(H)-\frac{2\pi}{\epsilon}-4\pi(\gamma_{\scalebox{0.6}{E}} -1)\right) \CR
&=& \frac{2\zeta(3)}{g_{\scalebox{0.6}{$8$}}^{\; 2}}-2\pi\ln{(T_2\lvert\eta(T)\lvert^4)}+\frac{4\pi}{3}\ln  g_{\scalebox{0.6}{$8$}} \CR
&& \qquad  +\frac{8\pi}{g_{\scalebox{0.6}{$8$}}}   \sum^\prime_{m,n\in \mathds{Z}} \frac{ \sqrt{T_2}\,  \sigma_2(m,n)}{  |m + T n|} K_1\Bigl(\tfrac{2\pi}{g_{\scalebox{0.6}{$8$}} \sqrt{T_2} }  |m + T n|\Bigr) e^{ 2\pi i ( c_0 m + c_2 n )}  \; . 
\eea
With these definitions we have
\be  \mathcal{A}(s,t,u,\varphi)  = \frac{64}{s t u} + (4\pi)^{5} \lP^{\; 6} \bigl(\hat I_{4,\mu}( s, t)+\hat I_{4,\mu}( t, u)+\hat I_{4,\mu}( u,s)\bigr)  + \lP^{\; 6} \mathcal{E}_{\gra{0}{0},\mu}(\varphi)  + \mathcal{O}( s^2) \ee
and
\be \mathcal{E}_{\gra{0}{0},\mu}(\varphi)  = \Epreg^{\!\!3}_\frac{3}{2}(H) + 2 \Epreg^{\!\! 2}_1(U)  + \frac{22\pi}{3}- 4\pi\ln(2\pi \lP\mu ) \; . \ee
Note that the logarithm  $4\pi\ln(2\pi \lP ) $ can be absorbed in the logarithm of the type IIB torus volume $V_2 = (2\pi \lP)^2 T_2 g_{\scalebox{0.6}{$8$}}^{-2/3} $, equivalently $V_2 = (2\pi \lP)^2 U_2 g_{\scalebox{0.6}{$8$}}^{-2/3} $ in type IIA.

\section{Numerical approximations}

In this section we explain the numerical checks we have carried out to establish the conjectured global minima of the Epstein series. We concentrate on the two cases of $SL(5)$ and $SO(5,5)$ that constitute the main new results. The global minimum of the $SL(3)$ Epstein series was obtained as the symmetric point $A_3$ for all $s>\frac{3}{4}$, including the renormalised value at $s=\frac{3}{2}$ by minimal subtraction, in \cite{MinimaSL3}. 

The moduli space $SO(5)\backslash SL(5)$ has 14 dimensions and  $\bigl(\hspace{-0.1mm} SO(5)\times SO(5)\hspace{-0.1mm}\bigr)\backslash SO(5,5)$ 25 dimensions, and it is rather difficult to study the value of the Epstein series systematically over the whole moduli space in these two cases. Nonetheless, the Epstein series are  very regular functions, that decrease somewhat monotonically from the cusps at infinity to the symmetric points in the interior.

\subsection{The $SL(5)$ Epstein series}
We consider several $SL(5,\mathds{Z})$ equivalent realisations of the $D_5$ symmetric point $H = 2^{- \frac{2}{5}} C_{D_5}$ with 
\be C_{D_5} = \left( \begin{array}{ccccc} \; 2\; &\; 1\; &\; 1\; &\; 1\; &\;  2\; \\ \; 1\; &\; 2\; &\; 1\; &\; 1\; &\;  2\; \\ \; 1\; &\; 1\; &\; 2\; &\; 1\; &\;  2\; \\ \; 1\; &\; 1\; &\; 1\; &\; 2\; &\;  2\; \\ \; 2\; &\; 2\; &\; 2\; &\; 2\; &\;  4\; \end{array} \right) \; , \quad  C^\prime_{D_5} = \left( \begin{array}{ccccc} \; 2\; &\; 1\; &\; 1\; &\; 2\; &\;  1\; \\ \; 1\; &\; 2\; &\; 1\; &\; 2\; &\;  1\; \\ \; 1\; &\; 1\; &\; 2\; &\; 2\; &\;  1\; \\ \; 2\; &\; 2\; &\; 2\; &\; 4\; &\;  2\; \\ \; 1\; &\; 1\; &\; 1\; &\; 2\; &\;  2\; \end{array} \right) \; , \quad C^{\prime\prime}_{D_5} = \left( \begin{array}{ccccc} \; 2\; &\; 1\; &\; 0\; &\; 0\; &\;  0\; \\ \; 1\; &\; 2\; &\; 0\; &\; 0\; &\;  1\; \\ \; 0\; &\; 0\; &\; 2\; &\; 0\; &\;  1\; \\ \; 0\; &\; 0\; &\; 0\; &\; 2\; &\;  1\; \\ \; 0\; &\; 1\; &\; 1\; &\; 1\; &\;  2\; \end{array} \right) \; . \ee
They are determined by iterated  inclusions of lattices 
\bea && A_2 \subset A_3\subset A_4 \subset D_5 \; , \CR
&& A_2 \subset A_3\subset D_4 \subset D_5 \; , \CR
&& A_2 \subset A_2+A_1\subset A_2 + 2 A_1 \subset D_5 \; . 
\eea
The first case is described with the ansatz \eqref{VSLNCN} and the expansion of the Epstein series \eqref{EpsteinExpansion} with $N=5$ as a function of $y$ and $x$. We use \eqref{EpsteinExpansion}  iteratively and truncate the sum over $n_i \in \mathds{Z}$ for $N=3$ between $-50$ and $50$, the sum over $n_i \in \mathds{Z}$ for $N=4$ between $-20$ and $20$ and the sum over $n_i\in \mathds{Z}$ for $N=5$ between $-10$ and $10$. We checked that increasing the ranges of the Fourier modes $n_i$ does not change the result within the approximation. We analyse the values of $s$ between $\frac{5}{4}$ and $5$. For large values of $s$ the Fourier expansion converges more slowly because the Bessel function does not decrease fast enough for the truncated expansion \eqref{EpsteinExpansion} to give a good approximation.

The obtained function of $y$ and $x$ admits only three local minima, corresponding to the symmetric points $A_5$, $A_5^{+2}$ and $D_5$, with $D_5$ being the global minimum.  Only $D_5$ is a local minimum of the function of $H \in SO(5) \backslash SL(5)$ for low values of $s$. 

\begin{figure}[h!]
    \centering
    \subfloat[]{{\includegraphics[width=8.0cm]{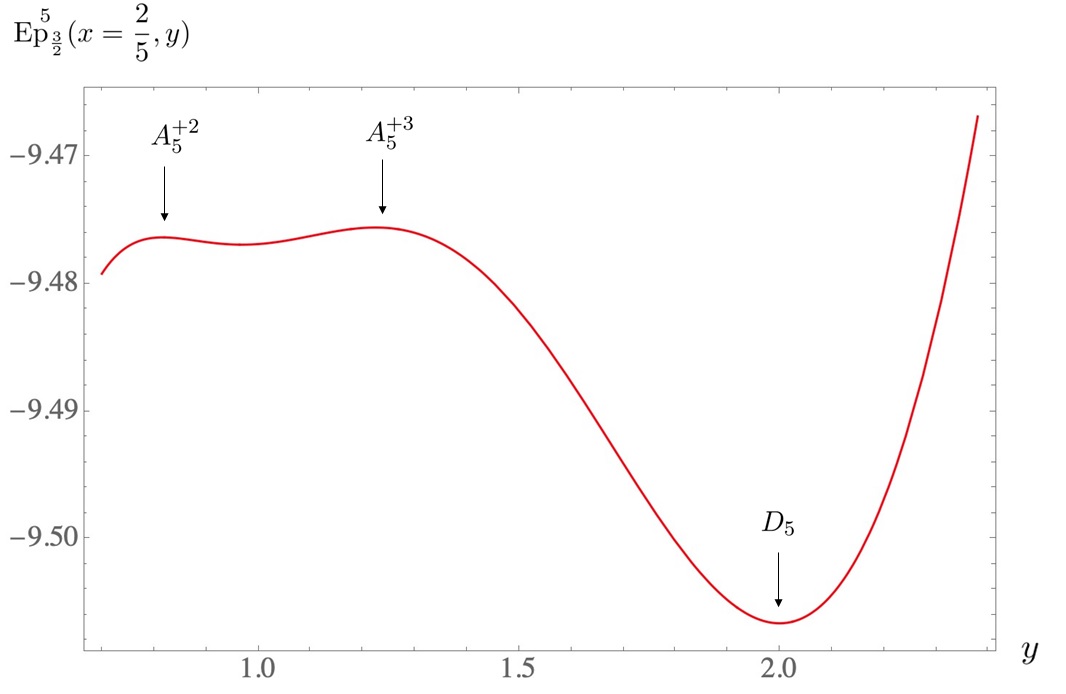} }}
    \subfloat[]{{\includegraphics[width=7.8cm]{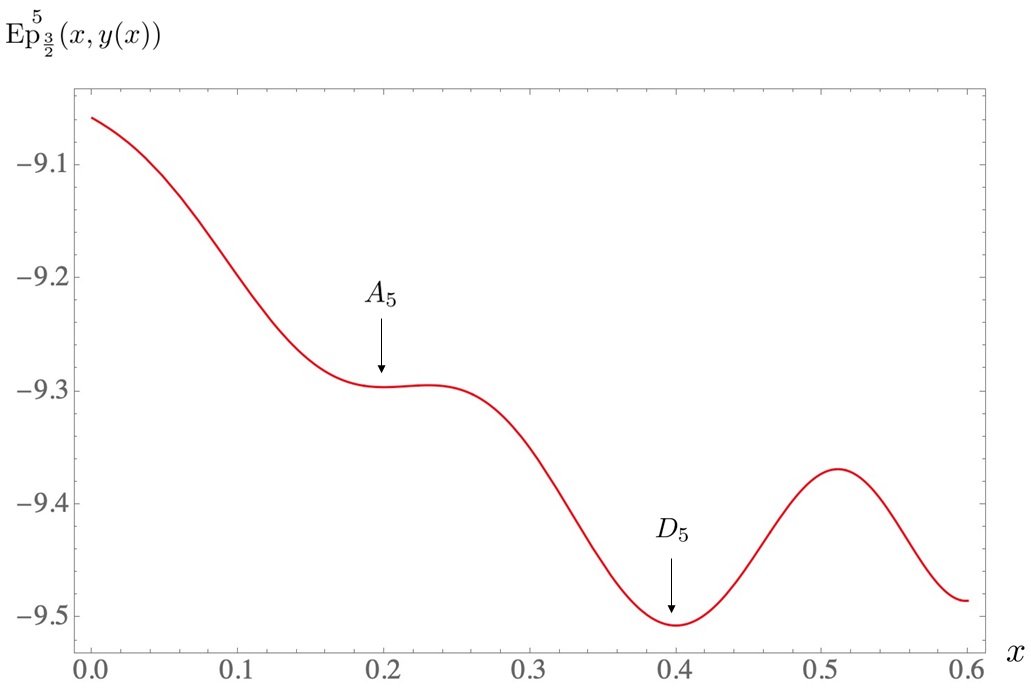} }}
    \caption{Pullback of the $SL(5)$ Epstein series along the two dimensional $(x,y)$ surface defined by \eqref{VSLNCN} for $s=\frac32$. (a) Slice $x=\frac25$ containing $A_5^{+2}$ , $A_5^{+3}$  and $D_5$, only $D_5$ is a local minimum. (b) Slice $y=-5(\sqrt{6}-2)x+2(\sqrt{6}-1)$ containing $A_5$ and $D_5$, along this slice $A_5$ looks like a local minimum but it is a saddle point in $SO(5) \backslash SL(5)$. $D_5$ is the global minimum on the surface.}
    \label{Plots}
\end{figure}

One also checks the pull-back of the function on different surfaces associated to the parametrisation \eqref{EpsteinExpansion} for $N=4$ as $\Ep^5_s(2,\frac{2}{5},H) = \frac{2^{\frac{2s}{5}}}{5^{\frac{s}{4}}}  \Ep^{4}_s(y,x,A_3) + \dots $ and similarly for $N=3$ and find again that the $D_5$ point is the global minimum on each surface. 

\vskip 2mm

In the second case we consider the $SL(N)$ coset representative 
\be
\mathcal{V}= y^{-\frac{1}{N}}\left(\begin{array}{cc} \; 2^{\frac1{N-1}} \mathcal{V}_1 \; & \; 2^{\frac1{N-1}}  \mathcal{V}_1 {\bf x}\; \\ \; 0\; & \; y/2 \end{array}\right) \; , \label{VSLNCNDN} \ee
with $\mathcal{V}_1$ at the $D_{N-1}$ symmetric point and ${\bf x}$ restricted to the last component $x_{\scalebox{0.6}{$N{-}1$}}$. One then obtains the Fourier expansion of the Epstein series
\begin{multline} \Ep^N_s(y,x,C_{D_{N-1}}) = \frac{y^{\frac{2s}{N}}}{4^{\frac{s}{N-1}}}  \Ep^{N-1}_s(C_{D_{N-1}}) +  \frac{2\pi^{\frac{N-1}{2}} \Gamma(s-\frac{N-1}{2})\zeta(2s-N+1)  }{\Gamma(s)} \frac{y^{(N-1)( 1- \frac{2s}{N})}}{4^{\frac{N}{2}-s}}   \\
+ \frac{2\pi^s}{\Gamma(s)} y^{\frac{N-1}{2} - \frac{N-2}{N}s} \hspace{-1mm} \sum_{n \in \mathds{Z}^{N-1}}^\prime  \hspace{-1mm}  \sigma_{N-1-2s}(n) \frac{ K_{s-\frac{N-1}{2}}\bigl( \pi y  \sqrt{  C_{D_{N-1}}^{-1}[n]  } \bigr)}{\bigl( 4 C_{D_{N-1}}^{-1}[n] \bigr)^{\frac{N-1}{4}-\frac{s}{2}}  }  \cos\bigl( 2\pi x  n_{\scalebox{0.6}{$N{-}1$}} \bigr)  \label{EpsteinExpansionDN} \end{multline}
which defines another surface parametrised by $(x,y)$  in the moduli space. In this case one obtains for $N=5$ that the only local minimum is at $y=2$ and $x=\frac{1}{2}$ corresponding to the $D_5$ symmetric point. 

\vskip 2mm

In the third case we consider the $SL(N)$ coset representative 
\be \mathcal{V}= y^{-\frac{1}{N}}\left(\begin{array}{cc} \; (4N{-}8)^{\frac1{2(N-1)}} \mathcal{V}_1 \; & \;(4N{-}8)^{\frac1{2(N-1)}}  \mathcal{V}_1 {\bf x}\; \\ \; 0\; & \; \frac{y}{2\sqrt{N-2}}  \end{array}\right) \; , \label{VSLNCNReduced} \ee
with $\mathcal{V}_1$ at the $A_{N-3}{+}2A_1$ symmetric point and ${\bf x} = ( - \frac{2}{N-2},\dots , 2\frac{N-3}{N-2}, 1,1)x$.  One obtains again for $N=5$ that the only local minimum is at $y=2$ and $x=\frac{1}{2}$ corresponding to the $D_5$ symmetric point.

\vskip 2mm

We obtain numerically the value of the Epstein series
\be \EpN{5}_\frac32(D_5) \approx -9.50663\; . \ee
The numerical value is stable under modification of the truncation up to fifteen digits.

\subsection{The $SO(5,5)$ vector representation Epstein series}
The vector representation Epstein series at generic $s$ is much harder to approximate because of the Eisenstein series appearing in the Fourier coefficients in \eqref{VectorEpsteinExpand}. Therefore we shall only analyse the minimal Epstein series  \eqref{EpSONNMin}  at $s=\frac{3}{2}$ 
\be \Ep_{\frac{3}{2}}^{5,5}(H)   = \det G^{\frac{3}{10}} \EpN{5}_{\frac{3}{2}}(G^{-1}) + \pi \sqrt{ \det G }    \Biggl( \frac{\pi}{6} + \sum^\prime_{\substack{Q\in \wedge^2 \mathds{Z}^5\\ Q\wedge Q=0}}  \sigma_{1}(Q) \frac{e^{- 2\pi \sqrt{ G[Q]}} }{\sqrt{G[Q]}}  e^{\pi i \tr B Q} \Biggr) \; , \label{EpSO55Min} \ee 
that defines the leading Wilson coefficient in the string theory effective action. 

One difficulty is to generate an appropriate large set of rank two antisymmetric integer matrices $Q \in \wedge^2 \mathds{Z}^5$. To obtain a good approximation, we must include all the M2-instanton charges $Q$ with a Euclidean action below some threshold
\be S[Q] = 2\pi \sqrt{- \tfrac12 \tr G Q G Q} \le \Lambda \; , \ee
without including two many charges $Q$ with a strictly greater Euclidean action. The incomplete sets of charges for a given action greater than $\Lambda $ do not a priori spoil the approximation, but their inclusion increases the evaluation time of the function without providing a better approximation. 

To compute a sample of charges we generate the set of rank two matrices with entries $|Q_{ij}|\le 6$ and only keep the charges with a Euclidean action evaluated at the $D_5$ symmetric point bounded by $\Lambda = 2\pi \sqrt{6}$. The values of the action is $S = \pi \sqrt{1+4 n}$ or $S = 2\pi \sqrt{n}$ for all integers $n\ge 1$. We have checked up to $S= 2\pi \sqrt{5}$ that the set of charges we have obtained  define complete orbits of the Weyl group action $W(D_5)$. They all satisfy $|Q_{ij}|\le 4$, so we believe they are the complete set of charges for $\Lambda = 2\pi \sqrt{5}$. One also checks that this set of charges gives all charges of action bounded by $\Lambda = 2\pi \sqrt{6}$ at the $A_5$ symmetric point with entries $|Q_{ij}|\le 5$. The values of the action is then $S = 2\pi \sqrt{n}$ or $S = \pi \sqrt{3+4 n}$ for all integers $n\ge 1$.  We expect therefore this set of charges to provide a good approximation of the Epstein series on the fundamental domain. 

We consider two surfaces, the first parametrised by 
\be G = \frac{y}{2} \left( \begin{array}{ccccc} \; 2\; &\; 1\; &\; 1\; &\; 1\; &\;  2\; \\ \; 1\; &\; 2\; &\; 1\; &\; 1\; &\;  2\; \\ \; 1\; &\; 1\; &\; 2\; &\; 1\; &\;  2\; \\ \; 1\; &\; 1\; &\; 1\; &\; 2\; &\;  2\; \\ \; 2\; &\; 2\; &\; 2\; &\; 2\; &\;  4\; \end{array} \right) \; , \qquad B = x \left( \begin{array}{ccccc} \; 0\; &\; 1\; &\; 1\; &\; 1\; &\;  2\; \\ -1\; &\; 0\; &\; 1\; &\; 1\; &\;  2\; \\ - 1\; &-1\; &\; 0\; &\; 1\; &\;  2\; \\ -1\; &- 1\; &- 1\; &\; 0\; &\;  2\; \\ - 2\; &- 2\; &- 2\; &- 2\; &\;  0\; \end{array} \right)  \; , \ee
only includes a global minimum at the $D_5$ symmetric point at $y=1$ and $x=\frac12$. 

The second is parametrised by 
\be G = \frac{1}{2} \left( \begin{array}{ccccc} \; 2\; &\; 1\; &\; 1\; &\; 1\; &\;  5x\; \\ \; 1\; &\; 2\; &\; 1\; &\; 1\; &\;  5x\; \\ \; 1\; &\; 1\; &\; 2\; &\; 1\; &\;  5x \; \\ \; 1\; &\; 1\; &\; 1\; &\; 2\; &\;  5x \; \\ \; 5x \; &\; 5x \; &\; 5x \; &\; 5x \; & \frac{y^2}{5}{+}20 x^2 \end{array} \right) \; , \qquad B = \frac12 \left( \begin{array}{ccccc} \; 0\; &\; 1\; &\; 1\; &\; 1\; &\;  5x \; \\ -1\; &\; 0\; &\; 1\; &\; 1\; &\;  5x \; \\ - 1\; &-1\; &\; 0\; &\; 1\; &\;  5x \; \\ -1\; &- 1\; &- 1\; &\; 0\; &\;  5x \; \\ - 5x \; &- 5x \; &- 5x \; &- 5x \; &\;  0\; \end{array} \right)  \; . \ee
On this surface we find that the $D_5$ symmetric point is a global minimum at $y =2$ and $x=\frac{2}{5}$. One also finds the local minimum at the $A_5$ symmetric point at $y = \sqrt{6}$ and $x=\frac{1}{5}$.  

The numerical approximation gives the minimum of the Epstein series  
\be \Ep_{\frac{3}{2}}^{5,5}(D_5) \approx - 3.4447 \pm 0.0002\; .  \ee 

\appendix
\section{Grenier domain boundaries}
\label{Grenierdomainboundaries}
Consider the $A_N$ symmetric point in $SO(N)\backslash SL(N)$ where $H=(N{+}1)^{-\frac{1}{N}} C_{A_N}$ is proportional to the $A_N$ Cartan matrix.  Its representative \eqref{ANCartan} in  $SO(N)\backslash SL(N)/ SL(N,\mathds{Z})$ is in the Grenier domain. To exhibit that this is the case, one writes $ C_{A_N}$ as
\be C_{A_N}[n] = \sum_{k=0}^{N-1} \frac{k+2}{k+1} \biggl( n_{k+1} + \frac{1}{k+1} \sum_{i=k+2}^N n_i \biggr)^2 \ee
and identifies the bilinear forms  
\be Y_k[n] = \sum_{i=k+1}^N n_i^2 + \frac{2}{k+2} \sum_{k+1\le i<j\le N} n_i n_j \ee
in the notation of Section \ref{GrenierDomain}. One finds from this formula that $x_{k+1,i} = \frac{1}{k+1} \in [ 0, \frac12]$ and for each $Y_k[n]$, the minimal length vectors are length one, consistently with the definition of the Grenier domain. This point is on the boundary of the Grenier domain since the inequality  
\be Y_0[n] \ge 1 \ee
is saturated for  all $A_N$ root, i.e. 
\be n_i = \pm \delta_{ij} \; , \qquad n_i = \delta_{ij} - \delta_{ik} \; , \ee
for any $j$ and $k$. For $Y_k[n]$ with $k\ge 1$, the only vectors of minimal length $1$ are $n_i = \delta_{ij}$ for all $j\ge k+1$. In fact it is a boundary of dimension zero. To prove this one can write a generic point in $SO(N)\backslash SL(N)$ using the matrix \footnote{Take care that the coordinates $y_i$ and $x_{ij}$ are not the same as in \eqref{Vyx}.}
\be Y_0[n] = n_1^2 + \sum_{i=2}^N y_i^2 n_i^2 + \sum_{i<j} x_{ij} n_i n_j \; . \ee
Now, solving the condition that $Y_0[n]=1$ for all roots of  $A_N$, one obtains 
\be y_i^2=1\; , \qquad y_i^2 + y_j^2 -x_{ij} = 1\; ,  \ee
which determines $y_i = x_{ij} = 1$ and fixes the $A_N$ point $H = (N+1)^{- \frac{1}{N}} C_{A_N}$. 

\vskip 5mm

One can consider similarly the symmetric point $D_{N{+}1}$ in $SO(N)\backslash SL(N)$. Starting form \eqref{DnfromAn} and shifting $n_{2}\rightarrow n_2-n_{N+1}$ one obtain the equivalent representative of the $D_{N{+}1}$ Cartan matrix 
\be C^\prime_{D_{N{+}1}} = 2 \sum_{i=1}^{N+1} n_i^2 +  2\hspace{-5mm} \sum_{\substack{i<j\\ (i,j) \ne (2,N+1)}} \hspace{-5mm}  n_i n_j  \; ,  \ee
where the second sum is over all pairs except $n_2 n_{N+1}$. One checks that $H = 2^{-\frac{2}{N+1}} C^\prime_{D_{N+1}} $ is in the Grenier domain with the triangular form
\bea Y_0 &=&  \Bigl( n_1 + \frac12 \sum_{i=2}^{N+1} n_i\Bigr)^2 + \frac{3}{4}   \Bigl( n_2 + \frac13 \sum_{i=3}^{N}n_i - \frac{1}{3} n_{N+1} \Bigr)^2 \CR
 && + \frac{1}{2} \sum_{k=2}^{N-1} \frac{k+2}{k+1} \biggl( n_{k+1} + \frac{1}{k+1} \sum_{i=k+2}^N n_i  + \frac{2}{k+2} n_{N+1}\biggr)^2 + \frac{4}{N+1} n_{N+1}^2 \; , \eea
 and the sub-components 
\bea Y_0 &=&\sum_{i=1}^{N+1} n_i^2 +  \hspace{-5mm} \sum_{\substack{i<j\\ (i,j) \ne (2,N+1)}} \hspace{-5mm}  n_i n_j \CR
 Y_1 &=& \sum_{i=2}^{N+1} n_i^2 + \frac{2}{3} \hspace{-5mm} \sum_{\substack{2\le i<j\\ (i,j) \ne (2,N+1)}} \hspace{-5mm} n_i n_j - \frac{2}{3} n_2 n_{N+1}  \; , \CR
 Y_{k\ge 2} &=& \sum_{i = k+1}^{N+1}n_i^2  +\frac{2}{k+2} \hspace{-1mm} \sum_{k+1\le i<j\le N } \hspace{-2mm} n_i n_j +\frac{4}{k+2} n_{N+1} \sum_{i=k+1}^N n_i\; .    \eea
The triangular form of $Y_0$ exhibits that all $x_{ij} \in [ 0,\frac{1}{2}] $ but $x_{2,N+1}= -\frac13 \in [- \frac12,\frac12]$. The sub-components all have minimum length vectors of length one, and $H$ is therefore in the Grenier domain.  This point sits on the boundary because the inequalities $Y_0[n]\ge 1$ are saturated for all $D_{N+1}$ roots, i.e. 
\bea n_i &=&\pm  \delta_{i,j} \; , \qquad n_i = \delta_{i,j} - \delta_{i,k} \quad \mbox{for} \; (j,k) \ne (2,N+1) \; {\rm or} \; (N+1,2) \; ,\CR
n_i &=& \pm ( \delta_{i,j} - \delta_{i,2} - \delta_{i,N+1} )  \quad\mbox{for} \; j \ne 2,N+1 \; , \CR
n_i &=& \pm ( \delta_{i,j} +\delta_{i,k} - \delta_{i,2} - \delta_{i,N+1} )  \quad\mbox{for} \; j,k \ne 2,N+1 \; . 
\eea
For $Y_k$ with $k=1$ and $k\ge 3$, the only vectors of length $1$ are $n_i = \delta_{i,j}$ for all $j\ge k+1$. For $k=2$ one gets all vectors $n_i = \delta_{i,j}$ with $j\ge 3$ and the vectors $n_i = \pm \delta_{i,j} \mp \delta_{i,N+1}$ for $3\le i \le N$. Writing the equalities $Y_0[n]=1$ for all $D_{N+1}$ roots for the generic matrix 
\be Y_0[n] = n_1^2 + \sum_{i=2}^N y_i^2 n_i^2 + \sum_{i<j} x_{ij} n_i n_j \ee
one obtains 
\bea && y_i^2 =1  \; , \qquad  y_i^2 + y_j^2 -x_{ij}= 1 \quad \mbox{for} \; (i,j) \ne (2,N+1)  \; ,\\
&& y_i^2 + y_2^2 + y_{N+1}^2 - x_{i2} -x_{i\, N+1}+x_{2\, N+1}= 1 \quad\mbox{for} \; i \ne 2,N+1 \; , \CR
&& y_i^2 +y_j^2 +  y_2^2 + y_{N+1}^2 +x_{ij} - x_{i2} -x_{i\, N+1}- x_{j2} -x_{j\, N+1}+x_{2\, N+1}= 1  \quad\mbox{for} \; i,j \ne 2,N+1 \; . \nonumber
\eea
This implies that $y_i=1$ and $x_{ij}=1$ except for $x_{2\, N+1}=0$ and therefore $H = 2^{-\frac{2}{N+1}} C^\prime_{D_{N+1}} $. 

\vskip 5mm

In general one does not expect reducible matrices to correspond to a zero dimensional boundary of the Grenier domain. Take as an example $H= (2N+2)^{- \frac{1}{N+1}} C_{A_N +A_1}$ with 
\be C_{A_N+A_1}[n] = 2 \sum_{i=1}^{N+1} n_i + 2 \sum_{i<j\le N} n_i n_j \; . \ee
One checks similarly that $H$ is in the Grenier domain, but the saturated inequalities still allow for a dimension $N$ boundary parametrised by 
\be Y_0[n] =  \sum_{i=1}^{N+1} n_i +  \sum_{i<j\le N} n_i n_j  + \sum_{i=1}^N x_i n_{N+1} n_i \; , \ee
provided they satisfy 
\be \sum_{i=1}^k x_i^2 - \frac{2}{k+1} \sum_{i<j\le k} x_i x_j \le 1\; , \ee
for all $k$ between $1$ and $N$.

\vskip 4mm

For the lattice  $ A_{5}^{+2}$, the matrix \eqref{F5fromyx}
\be C_{\! A_{5}^{+2}}= \left( \begin{array}{ccccc} \; 2\; &\; 1\; &\; 1\; &\; 1\; &\;  2\; \\ \; 1\; &\; 2\; &\; 1\; &\; 1\; &\;  2\; \\ \; 1\; &\; 1\; &\; 2\; &\; 1\; &\;  2\; \\ \; 1\; &\; 1\; &\; 1\; &\; 2\; &\;  2\; \\ \; 2\; &\; 2\; &\; 2\; &\; 2\; &\;  \frac{10}{3} \; \end{array} \right) \;  \ee
is not  in the Grenier domain. But using the $SL(5,\mathds{Z})$ matrix 
\be \gamma = \left( \begin{array}{ccccc} \; 0\; &\; 0\; &\; 0\; &\; 1\; &\;  1\; \\ \; 1\; &\; 1\; &\; 0\; &\; 0\; &\; 1 \; \\ \;0 \; &\; 1\; &\; 1\; &\; 0\; &\; 1 \; \\ \; 0\; &\; 1\; &\; 0\; &\; 0\; &\;  0\; \\ -1\; &-2\; &-1\; &-1\; &-2 \; \end{array} \right) \; , \ee
one obtains that  
\be C^\prime_{\! A_{5}^{+2}} = \gamma^\intercal  C_{\! A_{5}^{+2}} \gamma \ee
is. This is manifest in the upper triangular form  
\begin{multline} C^\prime_{\! A_{5}^{+2}}[n] = \frac{4}{3} \Biggl(  \bigl(n_1 + \tfrac12 n_2 +\tfrac14  n_3  + \tfrac14 n_4 + \tfrac12  n_5\bigr)^2\\ + 
   \frac{3}{4} \biggl( \bigl(n_2 + \tfrac12 n_3 - \tfrac12 n_4 \bigr)^2 + \bigl(n_3 + \tfrac12 n_4 + \tfrac12 n_5\bigr)^2 + 
      \frac34 \Bigl( \bigl(n_4 + \tfrac13 n_5\bigr)^2 + \tfrac89 n_5^{\; 2}\Bigr) \biggr)\Biggr) \; .  \end{multline}
This bilinear form admits 30 vectors of norm square $\frac{4}{3}$ in $\mathds{Z}^5$, the 15 weights in the Weyl orbit of the highest weights $\Lambda_2$ and the 15 in the Weyl orbit of $\Lambda_4$. The corresponding conditions $Y_0[n]=1$ determine $C^\prime_{\! A_{5}^{+2}}$, and the lattice    $ A_{5}^{+2}$ is therefore at a dimension-zero boundary of the Grenier domain.  

\section{Invariant polynomials at the $A_N$ symmetric point}
\label{InvQuaPolAN}

We prove here that there are indeed only two quadratic invariant polynomials at the $A_N$ symmetric point by constructing them explicitly. We can easily see that for any polynomial $F(\tilde p)$, its average on ${\rm S}_{N+1}$ defined by

\be
\langle F(\tilde p)\rangle_{{\rm S}_{N+1}}=\frac{1}{\lvert {\rm S}_{N+1}\lvert}\sum_{\gamma\in {\rm S}_{N+1}}F(\gamma^{\intercal}\tilde p\gamma)\; ,
\ee
is an ${\rm S}_{N+1}$-invariant polynomial. Therefore by calculating the $\langle \tilde p_{ij}\rangle_{{\rm S}_{N+1}}$ and $\langle \tilde p_{ij}\tilde p_{rs}\rangle_{{\rm S}_{N+1}}$ we can generate all independent ${\rm S}_{N+1}$-invariant linear and quadratic polynomials. We start with the linear polynomials where the calculations are less tedious but are completely analogous to the quadratic case. We can see that\footnote{Actually it is not obvious that we don't need to take another average on ${\rm S}_N$ in the end but it turns out that this is indeed not the case}

\be
\langle \tilde p_{ij}\rangle_{{\rm S}_{N+1}}=\langle\langle \tilde p_{ij}\rangle_{{\rm S}_{N}}\rangle_{\mathds{Z}_{N+1}}
\ee
It is easy to see that the average on ${\rm S}_N$ doesnâ€™t depend on $i,j$ explicitly however the result will differ if $i=j$ or $i\neq j$. Therefore there are naively only two possible linear invariant polynomials. We can easily compute the two independent averages on ${\rm S}_N$.

\begin{align}
\langle \tilde p_{ii}\rangle_{{\rm S}_{N}}&=\frac{1}{N!}\sum_{\sigma\in {\rm S}_{N}}\tilde p_{\sigma^{-1}(i)\sigma^{-1}(i)}=\frac{1}{N}\sum_{i=1}^N\tilde p_{ii}
\\\langle \tilde p_{ij}\rangle_{{\rm S}_{N}}&=\frac{1}{N!}\sum_{\sigma\in {\rm S}_{N}}\tilde p_{\sigma^{-1}(i)\sigma^{-1}(j)}=\frac{2}{N(N-1)}\sum_{i<j}\tilde p_{ij}
\end{align}
where $i\neq j$ in the second line. Hence all that remains is to calculate $\langle \tilde p_{ij}\rangle_{\mathds{Z}_{N+1}}$. If we use the convention that the indices $i,j$ of $\tilde p_{ij}$ are taken modulo $N+1$ we can write an arbitrary number of iterations of the transformation $\sigma_f$ in the simplified form

\be
\sigma_f^k.\tilde p_{ij}=\tilde p_{N-k+1N-k+1}+\tilde p_{i-kj-k}-\tilde p_{i-kN-k+1}-\tilde p_{j-kN-k+1}
\ee

for all $k\geq0$. We find that

\be
    \langle \tilde p_{ij}\rangle_{\mathds{Z}_{N+1}}=\frac{1}{N+1}\left(\sum_{k=1}^N\tilde p_{kk}+\sum_{k=1}^{N+1}\tilde p_{k+ik+j}-\sum_{k=1}^N\tilde p_{k+ik}-\sum_{k=1}^N\tilde p_{k+jk}\right)
\ee

for any $i,j$. Hence we find that

\begin{align}
\langle \tilde p_{ii}\rangle_{{\rm S}_{N+1}}&=\frac{1}{N}\sum_{i=1}^N\langle \tilde p_{ii}\rangle_{\mathds{Z}_{N+1}}=\frac{2}{N(N+1)}\left(N\sum_{i=1}^N\tilde p_{ii}-2\sum_{i<j}\tilde p_{ij}\right)=0\\
\langle \tilde p_{ij}\rangle_{{\rm S}_{N+1}}&=\frac{2}{N(N-1)}\sum_{i<j}\langle \tilde p_{ij}\rangle_{\mathds{Z}_{N+1}}=\frac{1}{N(N+1)}\left(N\sum_{i=1}^N\tilde p_{ii}-2\sum_{i<j}\tilde p_{ij}\right)=0
\end{align}
where we have used the traceless condition and where $i\neq j$ in the second line. Therefore there are no linear ${\rm S}_{N+1}$-invariant polynomials of $\tilde p$. This means that $C_{A_N}$ as defined in \eqref{H} is an extremum of any automorphic function on $SO(N) \backslash SL(N)$.\\

We will now apply the same reasoning to show that there are only two independent quadratic ${\rm S}_{N+1}$-invariant polynomials of $\tilde p$. We can show that

\begin{align}
\sigma_f^k.\tilde p_{ij}\tilde p_{rs}=& \sigma_f^k.\tilde p_{ij}\,\sigma_f^k.\tilde p_{rs}\nonumber\\
=&(\tilde p_{N-k+1N-k+1})^2\nonumber\\
&-\tilde p_{N-k+1N-k+1}(\tilde p_{i-kN-k+1}+\tilde p_{j-kN-k+1}+\tilde p_{r-kN-k+1}+\tilde p_{s-kN-k+1})\nonumber\\
&+\tilde p_{N-k+1N-k+1}(\tilde p_{i-kj-k}+\tilde p_{r-ks-k})+(\tilde p_{i-kN-k+1}+\tilde p_{j-kN-k+1})(\tilde p_{r-kN-k+1}+\tilde p_{s-kN-k+1})\nonumber\\
&-\tilde p_{i-kj-k}(\tilde p_{r-kN-k+1}+\tilde p_{s-kN-k+1})-\tilde p_{r-ks-k}(\tilde p_{i-kN-k+1}+\tilde p_{j-kN-k+1})\nonumber\\
&+\tilde p_{i-kj-k}\tilde p_{r-ks-k}
\end{align}

For all $i,j,r,s$. Let us now calculate $\langle \tilde p_{ij}\tilde p_{rs}\rangle_{\mathds{Z}_{N+1}}$

\begin{align}
\langle \tilde p_{ij}\tilde p_{rs}\rangle_{\mathds{Z}_{N+1}}=&\frac{1}{N+1}\left(\sum_{k=1}^N(\tilde p_{kk})^2-\sum_{k=1}^N\tilde p_{kk}(\tilde p_{ik+ik}+\tilde p_{k+jk}+\tilde p_{k+rk}+\tilde p_{k+sk})\right.\nonumber\\
&+\sum_{k=1}^N\tilde p_{kk}(\tilde p_{k+ik+j}+\tilde p_{k+rk+s})+\sum_{k=1}^N(\tilde p_{k+ik}+\tilde p_{k+jk})(\tilde p_{k+rk}+\tilde p_{k+sk})\nonumber\\
&-\sum_{k=1}^N\tilde p_{k+ik+j}(\tilde p_{k+rk}+\tilde p_{k+sk})-\sum_{k=1}^N\tilde p_{k+rk+s}(\tilde p_{k+ik}+\tilde p_{k+jk})\nonumber\\
&\left.+\sum_{k=1}^{N+1}\tilde p_{k+ik+j}\tilde p_{k+rk+s}\right)
\end{align}

for all $i,j,r,s$. We can now calculate the average on ${\rm S}_{N+1}$ in an analogous way to the linear case. Just like the linear case we can see that the average on ${\rm S}_N$ will not depend on $i,j,r,s$ explicitly, however the result will differ depending on if some indices are equal to each other. One can see that there are 7 different averages to compute: $\langle (\tilde p_{ii})^2\rangle_{{\rm S}_{N+1}}$, $\langle \tilde p_{ii}\tilde p_{ij}\rangle_{{\rm S}_{N+1}}$, $\langle \tilde p_{ii}\tilde p_{jj}\rangle_{{\rm S}_{N+1}}$, $\langle (\tilde p_{ij})^2\rangle_{{\rm S}_{N+1}}$, $\langle \tilde p_{ii}\tilde p_{jr}\rangle_{{\rm S}_{N+1}}$, $\langle \tilde p_{ij}\tilde p_{ir}\rangle_{{\rm S}_{N+1}}$ and $\langle \tilde p_{ij}\tilde p_{rs}\rangle_{{\rm S}_{N+1}}$ where all indices are assumed to be different. We can decompose each of them on a basis of ${\rm S}_N$ invariant polynomials as

\begin{align}
&a_1\sum_{i=1}^N(\tilde p_{ii})^2+a_2\sum_{i<j}\tilde p_{ii}\tilde p_{jj}+a_3\sum_{i\neq j}\tilde p_{ii}\tilde p_{ij}+a_4\sum_{i< j}(\tilde p_{ij})^2\nonumber\\&+a_5\sum_{i=1}^N\sum_{\substack{j< l\\j\neq i\\l\neq i}}\tilde p_{ii}\tilde p_{jl}+a_6\sum_{i=1}^N\sum_{\substack{j< l\\j\neq i\\l\neq i}}\tilde p_{ij}\tilde p_{il}+a_7\sum_{\substack{i\neq j\neq m\neq l\\i<j\\m<l\\i<m}}\tilde p_{ij}\tilde p_{ml}
\end{align}
For $N>2$ we can check that only three out of the seven averages are actually independent and they are linearly dependent with the Killing-Cartan form

\begin{align}
\tr C_{A_N}^{-1} \, \tilde{p} \, C_{A_N}^{-1} \, \tilde{p}=&\frac{4}{(N+1)^2}\Biggl(N^2\sum_{i=1}^N(\tilde p_{ii})^2+2\sum_{i<j}p_{ii}\tilde p_{jj}\Biggr.\nonumber\\
&-4N\sum_{i\neq j}\tilde p_{ii}\tilde p_{ij}+2(N^2+1)\sum_{i< j}(\tilde p_{ij})^2\nonumber\\
&\Biggl.+4\sum_{i=1}^N\sum_{\substack{j< l\\j\neq i\\l\neq i}}\tilde p_{ii}\tilde p_{jl}-4(N-1)\sum_{i=1}^N\sum_{\substack{j< l\\j\neq i\\l\neq i}}\tilde p_{ij}\tilde p_{il}+8\sum_{\substack{i\neq j\neq m\neq l\\i<j\\m<l\\i<m}}\tilde p_{ij}\tilde p_{ml}\Biggr)
\end{align}
as well as

\begin{align}
\text{tr}(C_{A_N}^{-1}&\tilde p)^2=\frac{4}{(N+1)^2}\Biggl(N^2\sum_{i=1}^N(\tilde p_{ii})^2+2N^2\sum_{i<j}\tilde p_{ii}\tilde p_{jj}-4N\sum_{i\neq j}\tilde p_{ii}\tilde p_{ij}\Biggr.\nonumber\\
&\Biggl.+4\sum_{i< j}(\tilde p_{ij})^2-4N\sum_{i=1}^N\sum_{\substack{j< l\\j\neq i\\l\neq i}}\tilde p_{ii}\tilde p_{jl}+8\sum_{i=1}^N\sum_{\substack{j< l\\j\neq i\\l\neq i}}\tilde p_{ij}\tilde p_{il}+8\sum_{\substack{i\neq j\neq m\neq l\\i<j\\m<l\\i<m}}\tilde p_{ij}\tilde p_{ml}\Biggr)=0
\end{align}
which vanishes once the traceless condition is imposed. This proves that there are two independent non trivial quadratic ${\rm S}_{N+1}$-invariant polynomials of $\tilde p$. The other non trivial independent polynomial is given by one of the averages, we give the simplest one

\be
\langle (\tilde p_{ii})^2\rangle_{{\rm S}_{N+1}}=\frac{2}{(N+1)N}\left(N\sum_{i=1}^N(\tilde p_{ii})^2-4\sum_{i\neq j}\tilde p_{ii}\tilde p_{ij}+2\sum_{i< j}\tilde p_{ii}\tilde p_{jj}+4\sum_{i< j}(\tilde p_{ij})^2\right)
\ee
In the case $N=2$ only the first four averages exist and they are all proportional to the Killing-Cartan form once the traceless condition is imposed. Therefore we can recover the well known result that $A_2$ is a local minimum (and indeed the global minimum) for the $SL(2)$ Eisenstein series.


\begin{thebibliography}{10}

\bibitem{Palti:2019pca}
E.~Palti, ``{The Swampland: Introduction and Review},''
  \href{http://dx.doi.org/10.1002/prop.201900037}{{\em Fortsch. Phys.}
  {\bfseries 67} no.~6, (2019) 1900037},
  \href{http://arxiv.org/abs/1903.06239}{{\ttfamily arXiv:1903.06239
  [hep-th]}}.

\bibitem{Gross:1986mw}
D.~J. Gross and J.~H. Sloan, ``{The Quartic Effective Action for the Heterotic
  String},''
\href{http://dx.doi.org/10.1016/0550-3213(87)90465-2}{{\em Nucl. Phys.}
  {\bfseries B291} (1987) 41--89}.

\bibitem{Green:2008uj}
M.~B. Green, J.~G. Russo, and P.~Vanhove, ``{Low energy expansion of the
  four-particle genus-one amplitude in type II superstring theory},''
  \href{http://dx.doi.org/10.1088/1126-6708/2008/02/020}{{\em JHEP} {\bfseries
  0802} (2008) 020},
\href{http://arxiv.org/abs/0801.0322}{{\ttfamily arXiv:0801.0322 [hep-th]}}.

\bibitem{Green:2008bf}
M.~B. Green, J.~G. Russo, and P.~Vanhove, ``{Modular properties of two-loop
  maximal supergravity and connections with string theory},''
  \href{http://dx.doi.org/10.1088/1126-6708/2008/07/126}{{\em JHEP} {\bfseries
  0807} (2008) 126},
\href{http://arxiv.org/abs/0807.0389}{{\ttfamily arXiv:0807.0389 [hep-th]}}.

\bibitem{Green:1981yb}
M.~B. Green and J.~H. Schwarz, ``{Supersymmetrical String Theories},''
\href{http://dx.doi.org/10.1016/0370-2693(82)91110-8}{{\em Phys. Lett.}
  {\bfseries B109} (1982) 444--448}.

\bibitem{Green:1997tv}
M.~B. Green and M.~Gutperle, ``{Effects of D-instantons},''
  \href{http://dx.doi.org/10.1016/S0550-3213(97)00269-1}{{\em Nucl. Phys.}
  {\bfseries B498} (1997) 195--227},
\href{http://arxiv.org/abs/hep-th/9701093}{{\ttfamily arXiv:hep-th/9701093}}.

\bibitem{Green:1997di}
M.~B. Green and P.~Vanhove, ``{D-instantons, strings and M-theory},''
  \href{http://dx.doi.org/10.1016/S0370-2693(97)00785-5}{{\em Phys. Lett.}
  {\bfseries B408} (1997) 122--134},
\href{http://arxiv.org/abs/hep-th/9704145}{{\ttfamily arXiv:hep-th/9704145}}.

\bibitem{Berkovits:1997pj}
N.~Berkovits, ``{Construction of $R^4$ terms in $\mathcal{N}=2$ $D = 8$
  superspace},'' \href{http://dx.doi.org/10.1016/S0550-3213(97)00817-1}{{\em
  Nucl. Phys. B} {\bfseries 514} (1998) 191--203},
  \href{http://arxiv.org/abs/hep-th/9709116}{{\ttfamily arXiv:hep-th/9709116}}.

\bibitem{Pioline:1998mn}
B.~Pioline, ``{A note on non-perturbative $R^4$ couplings},''
  \href{http://dx.doi.org/10.1016/S0370-2693(98)00554-1}{{\em Phys. Lett.}
  {\bfseries B431} (1998) 73--76},
\href{http://arxiv.org/abs/hep-th/9804023}{{\ttfamily arXiv:hep-th/9804023}}.

\bibitem{Green:1998by}
M.~B. Green and S.~Sethi, ``{Supersymmetry constraints on type IIB
  supergravity},'' \href{http://dx.doi.org/10.1103/PhysRevD.59.046006}{{\em
  Phys. Rev.} {\bfseries D59} (1999) 046006},
\href{http://arxiv.org/abs/hep-th/9808061}{{\ttfamily arXiv:hep-th/9808061}}.

\bibitem{Obers:1999um}
N.~A. Obers and B.~Pioline, ``{Eisenstein series and string thresholds},''
  \href{http://dx.doi.org/10.1007/s002200050022}{{\em Commun. Math. Phys.}
  {\bfseries 209} (2000) 275--324},
\href{http://arxiv.org/abs/hep-th/9903113}{{\ttfamily arXiv:hep-th/9903113}}.

\bibitem{Green:1999pv}
M.~B. Green and P.~Vanhove, ``{The low energy expansion of the one-loop type II
  superstring amplitude},''
  \href{http://dx.doi.org/10.1103/PhysRevD.61.104011}{{\em Phys. Rev.}
  {\bfseries D61} (2000) 104011},
\href{http://arxiv.org/abs/hep-th/9910056}{{\ttfamily arXiv:hep-th/9910056}}.

\bibitem{Kazhdan:2001nx}
D.~Kazhdan, B.~Pioline, and A.~Waldron, ``{M}inimal representations, spherical
  vectors, and exceptional theta series. {I}'' {\em Commun. Math. Phys.}
  {\bfseries 226} (2002) 1--40,
\href{http://arxiv.org/abs/hep-th/0107222}{{\ttfamily hep-th/0107222}}.

\bibitem{Basu:2008cf}
A.~Basu and S.~Sethi, ``{Recursion Relations from Space-time Supersymmetry},''
  \href{http://dx.doi.org/10.1088/1126-6708/2008/09/081}{{\em JHEP} {\bfseries
  09} (2008) 081},
\href{http://arxiv.org/abs/0808.1250}{{\ttfamily arXiv:0808.1250 [hep-th]}}.

\bibitem{Green:2005ba}
M.~B. Green and P.~Vanhove, ``{Duality and higher derivative terms in M
  theory},'' \href{http://dx.doi.org/10.1088/1126-6708/2006/01/093}{{\em JHEP}
  {\bfseries 0601} (2006) 093},
\href{http://arxiv.org/abs/hep-th/0510027}{{\ttfamily arXiv:hep-th/0510027
  [hep-th]}}.

\bibitem{Pioline:2010kb}
B.~Pioline, ``{$R^4$ couplings and automorphic unipotent representations},''
  \href{http://dx.doi.org/10.1007/JHEP03(2010)116}{{\em JHEP} {\bfseries 03}
  (2010) 116},
\href{http://arxiv.org/abs/1001.3647}{{\ttfamily arXiv:1001.3647 [hep-th]}}.

\bibitem{Green:2011vz}
M.~B. Green, S.~D. Miller, and P.~Vanhove, ``{Small representations, string
  instantons, and Fourier modes of Eisenstein series},''
  \href{http://dx.doi.org/10.1016/j.jnt.2013.05.018}{{\em J. Number Theor.}
  {\bfseries 146} (2015) 187--309},
\href{http://arxiv.org/abs/1111.2983}{{\ttfamily arXiv:1111.2983 [hep-th]}}.

\bibitem{Bossard:2014lra}
G.~Bossard and V.~Verschinin, ``{Minimal unitary representations from
  supersymmetry},'' \href{http://dx.doi.org/10.1007/JHEP10(2014)008}{{\em JHEP}
  {\bfseries 1410} (2014) 008},
\href{http://arxiv.org/abs/1406.5527}{{\ttfamily arXiv:1406.5527 [hep-th]}}.

\bibitem{Bossard:2014aea}
G.~Bossard and V.~Verschinin, ``{$\mathcal{E} \nabla^4 R^4$ type invariants and
  their gradient expansion},''
  \href{http://dx.doi.org/10.1007/JHEP03(2015)089}{{\em JHEP} {\bfseries 03}
  (2015) 089},
\href{http://arxiv.org/abs/1411.3373}{{\ttfamily arXiv:1411.3373 [hep-th]}}.

\bibitem{Gustafsson:2014iva}
H.~P.~A. Gustafsson, A.~Kleinschmidt, and D.~Persson, ``{Small automorphic
  representations and degenerate Whittaker vectors},''
\href{http://arxiv.org/abs/1412.5625}{{\ttfamily arXiv:1412.5625 [math.NT]}}.

\bibitem{Bossard:2015uga}
G.~Bossard and V.~Verschinin, ``{The two $\nabla^{6}$ R$^{4}$ type invariants
  and their higher order generalisation},''
  \href{http://dx.doi.org/10.1007/JHEP07(2015)154}{{\em JHEP} {\bfseries 07}
  (2015) 154},
\href{http://arxiv.org/abs/1503.04230}{{\ttfamily arXiv:1503.04230 [hep-th]}}.

\bibitem{Gourevitch:2019knu}
D.~Gourevitch, H.~P.~A. Gustafsson, A.~Kleinschmidt, D.~Persson, and S.~Sahi,
  ``{Fourier coefficients of minimal and next-to-minimal automorphic
  representations of simply-laced groups},''
  \href{http://dx.doi.org/10.4153/S0008414X20000711}{{\em Can. J. Math.}
  {\bfseries 74} no.~1, (2022) 122--169},
  \href{http://arxiv.org/abs/1908.08296}{{\ttfamily arXiv:1908.08296
  [math.NT]}}.

\bibitem{Paulos:2016but}
M.~F. Paulos, J.~Penedones, J.~Toledo, B.~C. van Rees, and P.~Vieira, ``{The
  S-matrix bootstrap II: two dimensional amplitudes},''
  \href{http://dx.doi.org/10.1007/JHEP11(2017)143}{{\em JHEP} {\bfseries 11}
  (2017) 143}, \href{http://arxiv.org/abs/1607.06110}{{\ttfamily
  arXiv:1607.06110 [hep-th]}}.

\bibitem{Paulos:2017fhb}
M.~F. Paulos, J.~Penedones, J.~Toledo, B.~C. van Rees, and P.~Vieira, ``{The
  S-matrix bootstrap. Part III: higher dimensional amplitudes},''
  \href{http://dx.doi.org/10.1007/JHEP12(2019)040}{{\em JHEP} {\bfseries 12}
  (2019) 040}, \href{http://arxiv.org/abs/1708.06765}{{\ttfamily
  arXiv:1708.06765 [hep-th]}}.

\bibitem{EliasMiro:2019kyf}
J.~Elias~Mir\'o, A.~L. Guerrieri, A.~Hebbar, J.~a. Penedones, and P.~Vieira,
  ``{Flux Tube S-matrix Bootstrap},''
  \href{http://dx.doi.org/10.1103/PhysRevLett.123.221602}{{\em Phys. Rev.
  Lett.} {\bfseries 123} no.~22, (2019) 221602},
  \href{http://arxiv.org/abs/1906.08098}{{\ttfamily arXiv:1906.08098
  [hep-th]}}.

\bibitem{Guerrieri:2021ivu}
A.~Guerrieri, J.~Penedones, and P.~Vieira, ``{Where Is String Theory in the
  Space of Scattering Amplitudes?},''
  \href{http://dx.doi.org/10.1103/PhysRevLett.127.081601}{{\em Phys. Rev.
  Lett.} {\bfseries 127} no.~8, (2021) 081601},
  \href{http://arxiv.org/abs/2102.02847}{{\ttfamily arXiv:2102.02847
  [hep-th]}}.

\bibitem{Guerrieri:2022sod}
A.~Guerrieri, H.~Murali, J.~Penedones, and P.~Vieira, ``{Where is M-theory in
  the space of scattering amplitudes?},''
  \href{doi:10.1007/JHEP06(2023)064}{{\em JHEP} {\textbf{06}}
  (2023) 064}, 
  \href{http://arxiv.org/abs/2212.00151}{{\ttfamily arXiv:2212.00151
  [hep-th]}}.

\bibitem{Coleman:1967ad}
S.~R. Coleman and J.~Mandula, ``{All Possible Symmetries of the S Matrix},''
  \href{http://dx.doi.org/10.1103/PhysRev.159.1251}{{\em Phys. Rev.} {\bfseries
  159} (1967) 1251--1256}.

\bibitem{Aks:1965}
S. O. Aks, ``{Proof that scattering implies production in quantum field theory},''
\href{doi:10.1063/1.1704305}{{\em J. Math. Phys. } {\textbf{6}}, no.4, 516-532 (1965)}.

\bibitem{Antunes:2023irg}
A.~Ant\'{o}nes, M.~S.~Costa and J.~Pereira,
``Exploring Inelasticity in the S-Matrix Bootstrap,''
\href{http://arxiv.org/abs/2301.13219}{{\ttfamily arXiv:2301.13219
  [hep-th]}}.


\bibitem{Duff:1995wd}
M.~J. Duff, J.~T. Liu, and R.~Minasian, ``{Eleven-dimensional origin of
  string-string duality: A One loop test},''
  \href{http://dx.doi.org/10.1016/0550-3213(95)00368-3}{{\em Nucl. Phys. B}
  {\bfseries 452} (1995) 261--282},
  \href{http://arxiv.org/abs/hep-th/9506126}{{\ttfamily arXiv:hep-th/9506126}}.

\bibitem{Witten:1996hc}
E.~Witten, ``{Five-brane effective action in M theory},''
  \href{http://dx.doi.org/10.1016/S0393-0440(97)80160-X}{{\em J. Geom. Phys.}
  {\bfseries 22} (1997) 103--133},
  \href{http://arxiv.org/abs/hep-th/9610234}{{\ttfamily arXiv:hep-th/9610234}}.

\bibitem{Kim:2019vuc}
H.-C. Kim, G.~Shiu, and C.~Vafa, ``{Branes and the Swampland},''
  \href{http://dx.doi.org/10.1103/PhysRevD.100.066006}{{\em Phys. Rev. D}
  {\bfseries 100} no.~6, (2019) 066006},
  \href{http://arxiv.org/abs/1905.08261}{{\ttfamily arXiv:1905.08261
  [hep-th]}}.

\bibitem{Guerrieri:2020bto}
A.~L. Guerrieri, J.~Penedones, and P.~Vieira, ``{S-matrix bootstrap for
  effective field theories: massless pions},''
  \href{http://dx.doi.org/10.1007/JHEP06(2021)088}{{\em JHEP} {\bfseries 06}
  (2021) 088}, \href{http://arxiv.org/abs/2011.02802}{{\ttfamily
  arXiv:2011.02802 [hep-th]}}.

\bibitem{Kiritsis:1997em}
E.~Kiritsis and B.~Pioline, ``{On $R^4$ threshold corrections in type IIB
  string theory and (p,q) string instantons},''
  \href{http://dx.doi.org/10.1016/S0550-3213(97)00645-7}{{\em Nucl. Phys.}
  {\bfseries B508} (1997) 509--534},
\href{http://arxiv.org/abs/hep-th/9707018}{{\ttfamily arXiv:hep-th/9707018}}.

\bibitem{Green:2010wi}
M.~B. Green, J.~G. Russo, and P.~Vanhove, ``{Automorphic properties of low
  energy string amplitudes in various dimensions},''
  \href{http://dx.doi.org/10.1103/PhysRevD.81.086008}{{\em Phys.Rev.}
  {\bfseries D81} (2010) 086008},
\href{http://arxiv.org/abs/1001.2535}{{\ttfamily arXiv:1001.2535 [hep-th]}}.

\bibitem{MinimaSL2R}
R.~Rankin, ``{A minimum problem for the Epstein zeta function},''
  \href{http://dx.doi.org/10.1017/S2040618500035668}{{\em Glasg. Math. Assoc.}
  no.~1, (1953) 149--158}.

\bibitem{MinimaSL3}
P.~Sarnak and A.~Str\"{o}mbergsson, ``{Minima of Epstein's Zeta function and
  heights of flat tori},''
  \href{http://dx.doi.org/10.1007/s00222-005-0488-2}{{\em Invent. math}
  no.~165, (2006) 115--151}.

\bibitem{MinimaLatticePacking}
S.~Ryshkov, ``{On the question of final $\zeta$-optimality of lattices
  providing the closest lattice packing of $n$-dimensional spheres},'' {\em
  Sib. Math. J.} no.~14, (1974) 743â€“750.

\bibitem{Caron-Huot:2022ugt}
S.~Caron-Huot, Y.-Z. Li, J.~Parra-Martinez, and D.~Simmons-Duffin, ``{Causality
  constraints on corrections to Einstein gravity},''
  \href{http://dx.doi.org/10.1007/JHEP05(2023)122}{{\em JHEP} {\bfseries 05}
  (2023) 122}, \href{http://arxiv.org/abs/2201.06602}{{\ttfamily
  arXiv:2201.06602 [hep-th]}}.

\bibitem{Green:2010sp}
M.~B. Green, J.~G. Russo, and P.~Vanhove, ``{String theory dualities and
  supergravity divergences},''
  \href{http://dx.doi.org/10.1007/JHEP06(2010)075}{{\em JHEP} {\bfseries 1006}
  (2010) 075},
\href{http://arxiv.org/abs/1002.3805}{{\ttfamily arXiv:1002.3805 [hep-th]}}.

\bibitem{Bossard:2015oxa}
G.~Bossard and A.~Kleinschmidt, ``{Supergravity divergences, supersymmetry and
  automorphic forms},'' \href{http://dx.doi.org/10.1007/JHEP08(2015)102}{{\em
  JHEP} {\bfseries 08} (2015) 102},
\href{http://arxiv.org/abs/1506.00657}{{\ttfamily arXiv:1506.00657 [hep-th]}}.

\bibitem{Hull:1994ys}
C.~M. Hull and P.~K. Townsend, ``Unity of superstring dualities,'' {\em Nucl.
  Phys.} {\bfseries B438} (1995) 109--137,
\href{http://arxiv.org/abs/hep-th/9410167}{{\ttfamily hep-th/9410167}}.

\bibitem{zbMATH02628371}
H.~M.~A. {Speiser} and H.~{Weyl}, ``{Gesammelte Abhandlungen von {\it Hermann
  Minkowski}. Unter Mitwirkung von {\it Andreas Speiser} und {\it Hermann Weyl}
  herausgegeben von {\it David Hilbert}.}'' {Leipzig u. Berlin: B. G. Teubner.
  gr. $8^\circ$}, 1911.

\bibitem{zbMATH04146017}
D.~{Grenier}, ``{Fundamental domains for the general linear group},''
  \href{http://dx.doi.org/10.2140/pjm.1988.132.293}{{\em {Pac. J. Math.}}
  {\bfseries 132} no.~2, (1988) 293--317}.

\bibitem{Peterson:1983}
D.~H. Peterson and K.~Kac, ``Infinite flag varieties and conjugacy theorems,''
  \href{http://dx.doi.org/10.1002/prop.200810570}{{\em Proc. Nat. Acad. Sci.
  U.S.A.} {\bfseries 80} (1983) 1778--1782}.

\bibitem{MinimaSL2C}
J.~Cassels, ``{On a problem of Rankin about the Epstein zeta function},''
  \href{http://dx.doi.org/10.1017/S2040618500033906}{{\em Glasg. Math. Assoc.}
  no.~4, (1963) 73â€“80}.

\bibitem{Epstein2}
P.~Epstein, ``{Theorie allgemeiner Zetafunctionen, II.},'' {\em Math. Ann.}
  no.~63, (1907) 205â€“216.

\bibitem{ConwaySloane}
J.~Conway and N.~Sloane, {\em Sphere Packings, Lattices and Groups}.
\newblock Springer, New York, 1999.

\bibitem{MinimaEpsteinTerras}
A.~Terras, ``{On the minima of quadratic forms and the behaviour of the Epstein
  and Dedekind Zeta functions},'' {\em Journal of Number Theory} no.~12, (1980)
  258--272.

\bibitem{SpherePacking}
H.~Cohn, A.~Kumar, S.~Miller, D.~Radchenko, and M.~Viazovska, ``{On the minima
  of quadratic forms and the behaviour of the Epstein and Dedekind Zeta
  functions},'' \href{http://dx.doi.org/10.4007/annals.2022.196.3.3}{{\em
  Annals of mathematics} {\bfseries 196} no.~3, (2022) 983--1082},
  \href{http://arxiv.org/abs/1902.05438}{{\ttfamily arXiv:1902.05438
  [math.MG]}}.

\bibitem{AutoL4}
E.~Dade, ``{The maximal finite groups of $4\times4$ integral matrices},''
  \href{http://dx.doi.org/10.1215/ijm/1256067584}{{\em Illinois J. Math.}
  {\bfseries 9} no.~1, (1965) 99}.

\bibitem{AutoL5}
W.~Plesken and M.~Pohst, ``{On Maximal Finite Irreducible Subgroups of $GL(n,
  \mathds{Z})$: I. The five and seven dimensional cases.},'' {\em Mathematics
  of Computation} {\bfseries 31} no.~138, (1977) 536--551.

\bibitem{AutoL9}
W.~Plesken and M.~Pohst, ``{On Maximal Finite Irreducible Subgroups of $GL(n,
  \mathds{Z})$: V. The eight-dimensional case and a complete description of
  dimensions less than ten.},'' \href{http://dx.doi.org/10.2307/2006235}{{\em
  Mathematics of Computation} {\bfseries 34} no.~149, (1980) 277}.

\end{thebibliography}

\providecommand{\href}[2]{#2}\begingroup\raggedright\endgroup

\end{document}